\documentclass[12pt]{iopart}

\usepackage{epsfig}

\def\la{\mathrel{\mathchoice {\vcenter{\offinterlineskip\halign{\hfil
$\displaystyle##$\hfil\cr<\cr\sim\cr}}}
{\vcenter{\offinterlineskip\halign{\hfil$\textstyle##$\hfil\cr
<\cr\sim\cr}}}
{\vcenter{\offinterlineskip\halign{\hfil$\scriptstyle##$\hfil\cr
<\cr\sim\cr}}}
{\vcenter{\offinterlineskip\halign{\hfil$\scriptscriptstyle##$\hfil\cr
<\cr\sim\cr}}}}}

\def\ga{\mathrel{\mathchoice {\vcenter{\offinterlineskip\halign{\hfil
$\displaystyle##$\hfil\cr>\cr\sim\cr}}}
{\vcenter{\offinterlineskip\halign{\hfil$\textstyle##$\hfil\cr
 >\cr\sim\cr}}}
{\vcenter{\offinterlineskip\halign{\hfil$\scriptstyle##$\hfil\cr  
 >\cr\sim\cr}}}
{\vcenter{\offinterlineskip\halign{\hfil$\scriptscriptstyle##$\hfil\cr
 >\cr\sim\cr}}}}}

\def\HII{H{\sc ii}}
\def\HI{H{\sc i}}

\def\CII{C{\sc ii}}
\def\CI{C{\sc i}}

\begin{document}

\title[Molecules in Galaxies]{Molecules in Galaxies}

%\maketitle

\author{Alain Omont}

\address{Institut d'Astrophysique de Paris, CNRS and Universit\'e Pierre et Marie Curie, ~~~~~~98bis Bd Arago, 75014 Paris, France}
\ead{omont@iap.fr}

\begin{abstract} 
The main achievements, current developments and prospects of molecular studies in external galaxies are reviewed. They are put in the context of the results of several decades of studies of molecules in local interstellar medium, their chemistry and their importance for star formation. CO observations have revealed the gross structure of molecular gas in galaxies. Together with other molecules, they are among the best tracers of star formation at galactic scales. Our knowledge about molecular abundances in various local galactic environments is progressing. They trace physical conditions and metallicity, and they are closely related to dust processes and large aromatic molecules. Major recent developments include mega-masers, and molecules in Active Galactic Nuclei; millimetre emission of molecules at very high redshift; and infrared H$_2$ emission as tracer of warm molecular gas, shocks and photodissociation regions. The advent of sensitive giant interferometers from the centimetre to sub-millimetre range,  especially ALMA in the near future in the mm/submm range, will open a new area for molecular studies in galaxies and their use to trace star formation at all distances.
\end{abstract}

\maketitle

{\bf Outline
%\bigskip
\medskip

%$\bullet$ 
1. Introduction

  {\it 1.1 Central role of the interstellar medium (ISM) in the physics, structure, evolution and multi-$\lambda$ emission of galaxies
  
  1.2 Molecular gas is a major component of the ISM
  
  1.3 Landmarks of the history of the discovery and studies of interstellar molecules in the Milky Way
  
  1.4 Main achievements of molecular studies in external galaxies}
\smallskip

 2. Cosmic molecules and local interstellar medium

	{\it 2.1 Physics and various components of the local interstellar molecular gas
	
	2.2 Observational techniques
	
	2.3 Modelling (millimetre) molecular line formation, and diagnostic of physical
conditions and molecular abundances in the ISM

	2.4 Uniqueness of basic processes of interstellar chemistry
	
	2.5 Relation between interstellar and other cosmic molecules}
\smallskip	
	
 3. CO observations and gross structure of molecular gas in the Milky Way

	{\it 3.1 Introduction: atomic and molecular interstellar gas
	
	3.2 Molecular gas in the Milky Way at galactic scale}
\smallskip	

 4. CO observations and gross structure of molecular gas in other galaxies
	
	{\it 4.1 Molecular gas in spiral galaxies from CO surveys of local galaxies
	
	4.2 Molecular gas in other types of galaxies
	
	4.3 OH mega-masers}
\smallskip	
	
 5. Molecules as tracers of star formation at galactic scales

	{\it 5.1 Introduction. Star formation rate
	
	5.2 Star formation and molecular clouds in non-starburst galaxies
	
	5.3 Molecular gas and starbursts in luminous and ultra-luminous infrared galaxies}
\smallskip	
	
 6. Molecular abundances in various local galactic environments

	{\it 6.1 Summary of standard molecular abundances in various regions of the Milky Way,
and models of interstellar chemistry

	6.2 Observed abundance variations in local galaxies: I. The Magellanic Clouds
	
	6.3 Observed abundance variations in local galaxies: II. Nearby starbursts and other galaxies

	6.4 Abundance ratios of isotopic varieties, and inferences for chemical evolution of
galaxies}
\smallskip

 7. Trends in molecular abundances from absorption lines at high redshift

	{\it 7.1 H$_2$ UV absorpion lines in Damped Lyman-$\alpha$ systems of quasars
	
	7.2 Millimetre and radio absorption lines in lensing galaxies and radio-sources}
\smallskip	
	
 8. Molecules, dust and PAHs (large aromatic molecules)

	{\it 8.1 Introduction. Interplay between dust and molecules in galaxies
	
	8.2 Mid-infrared emission of PAHs in galaxies
	
	8.3 Diffuse Interstellar Bands (DIBs) in galaxies
	
	8.4 Molecular infrared spectral features in extragalactic dust}
\smallskip	

 9. Millimetre emission of molecules at very high redshift  

	{\it 9.1 Star formation and ULIRGs at high redshift
	
	9.2 CO studies
	
	9.3 Molecules in the host galaxies of high-z AGN

	9.4 Other species and detailed studies through strong gravitational lensing}
\smallskip	
	
 10. Infrared H$_2$ emission, tracer of warm molecular gas, shocks and
photo-dissociation regions

	{\it 10.1 Basic features and physics of H$_2$ emission, and Milky Way observations
	
	10.2 2\,$\mu$m H$_2$ emission in galaxies
	
	10.3 Mid-IR H$_2$ pure-rotational lines in galaxies
	
	10.4 Prospects for detecting H$_2$ in forming galaxies}
\smallskip	
	
 11. Molecules and Active Galactic Nuclei (AGN)

	{\it 11.1 Interplay between super-massive black holes and their host galaxy
	
	11.2 Molecules in the central regions and fueling the AGN
	
	11.3 H$_2$O mega-masers and AGN molecular disks }
\smallskip

 12. Prospects

	{\it 12.1 Waiting for ALMA: ongoing studies and submillimetre breakthroughs

	12.2 The ALMA revolution
	
	12.3 Accompanying- and post-ALMA: JWST, extremely large telescopes and SKA}
\smallskip
	
 13. General conclusion
\medskip

	References
\medskip

	List of Main Abbreviations}

\clearpage

\section{General Introduction} 

\subsection{Central role of the interstellar medium (ISM) in the physics, structure, evolution and multi-$\lambda$ emission of galaxies}

	The interstellar medium (ISM) encompasses only a modest fraction of the mass of a galaxy, typically a few percents of the stellar mass, less than 1\% of the dark matter mass, of a large spiral galaxy such as the Milky Way. However, it is an essential component of spiral and irregular galaxies (see various books and reviews such as Spitzer 1968, 1998, Balian, Encrenaz \& Lequeux 1975, Dyson \& Williams 1980, Kaler 1997, Lequeux 2005, Tielens 2005). Most importantly, it is the material and the framework of star formation. Gravitational forces are constantly at work to squeeze and fragment its condensations; slowly starting at the scale of giant molecular clouds and ending in the final collapse of individual stars. Star formation is known as a highly complex process, yielding a number of intermediate stages and byproducts which correspond to various components of the ISM such as cold and hot condensations, accretion disks, bipolar flows and jets, energetic stellar winds and shocks, etc. These mechanisms introduce strong feedback actions upon the surrounding ISM, which in turn determine the global properties of star formation and account for its low efficiency.
	
  The symbiosis between stars and the ISM, together with stellar evolution, governs the overall and chemical evolution of galaxies. The properties, chemical composition and global mass of the ISM depend on its interaction with stars not only through the engulfment of material into new stars and the violent action of massive young stars through winds and supernova blasts, but also through the milder winds of the less massive stars, especially during the final stages of their evolution when they disperse their matter in the AGB and planetary nebula phases. 

	The extremely low density of the interstellar gas, $\sim$ 1-10$^{3}$ cm$^{-3}$, outside of the final star-forming condensations, leads to peculiar physical and chemical properties, as concerns time constants, temperature, heating and cooling, chemical reactions, dust and nanoparticles, and the associated radiative transfer which plays a major role in the energy balance and radiative multi-$\lambda$ emission of galaxies (see e.g.\,\,Tielens 2005 and references therein). The peculiarity and richness of the ISM is also a consequence of the variety of the violent processes which permeate it, especially in its lowest density phases: UV radiation, stellar winds, supernova explosions, X-rays and $\gamma$-rays, cosmic rays, inducing ionization, turbulence, magnetic fields, shocks, etc. Such violent processes are fundamentally hostile to molecules. However, the latter find efficient shelter within the densest parts of the ISM which are much less affected.
	
	Indeed, because of its dispersed nature, the ISM is a highly self-interacting and dissipating medium. It is thus at the origin of the formation of the disks of the spiral galaxies, and is subjected to strong interactions in galaxy collisions and merging, leading to major starbursts and feeding the central black hole. The ISM is also constantly fed by accretion of extragalactic gas and it plays a major role in the formation of galactic bars. It is strongly affected and compressed by spiral structures, leading to the spectacular visualisation of spiral arms through massive star formation. In major starbursts, it may convey huge outflows into the galactic halo, and even in the intergalactic medium, being probably at the origin of the enrichment of the extragalactic gas in heavy elements.
	
	The basic features of the energetics - heating and cooling - and the dynamics of the interstellar medium have been understood for 30-50 years (see the general references above  and references therein, and McKee \& Ostriker 1977). The energy involved in interstellar gas processes is generally only a small fraction of the total energy generated in stars (and possibly the central AGN), while interstellar dust may channel into the far-infrared a substantial fraction, and even the quasi-totality, of the energy generated by star formation (and in some cases the AGN). Heating the gas is thus achieved through relatively marginal processes from UV stellar radiation - mainly through photo-electric ejection of electrons from dust grains (Watson 1972), and from atoms in the vicinity of UV stars -, from cosmic rays, stellar winds, interstellar shocks mainly from supernovae explosions, gravitational energy of molecular clouds, etc. Outside of adiabatic expansion, cooling is mainly achieved through spectral lines of atoms, ions and molecules; the main lines depending mostly on the gas temperature. The interplay between heating, cooling and the gas dynamics may generate special thermodynamical properties, such as  major instabilities in some cases (Field 1965), the special physics of frequent shocks (McKee \& Ostriker 1977, Shull \& McKee 1979, McKee \& Hollenbach 1980), peculiar turbulence (Williams et al.\,2000, Falgarone et al.\,2005a, Elmegreen \& Scalo 2004), etc.
	
 	Mostly through dust grains (and atomic absorption in the far-UV and X-ray ranges), the interstellar medium completely reshapes the spectral energy distribution (SED) of galaxies. In most of them, a good part of the UV and visible radiation from stars is depleted and the corresponding energy comes out in the infrared. The main part of this energy is emitted in the far-infrared by relatively large dust grains, while nanoparticles (mostly Polycyclic Aromatic Hydrocarbons, PAHs,  L\'eger \& Puget 1984) emit spectacular aromatic bands in the mid-infrared.
 
  The tenuous interstellar gas, theatre of very peculiar physical and chemical processes, is indeed an essential ingredient of galaxies. It forms galactic disks and stars, reprocesses a large fraction of their radiation into the infrared, constantly exchanging material with stars and often strongly interacting with them through violent processes.

\subsection{Molecular gas is a major component of the ISM}

	As described e.g. in Tielens (2005), Lequeux (2005) and Dyson \& Williams (1980), molecules constitute a significant fraction of the ISM, residing in the densest regions where star formation takes place. In the ISM, the overall abundance of molecules, i.e. in practice that of H$_2$, is determined by the balance between the destruction processes, mostly by UV, and the formation, generally on dust grains for H$_2$ (see Section 2.4). High gas densities favour molecules, both by accelerating their formation, and protecting them against the main destruction process, UV photodissociation, because of the exponential UV shielding by dust and self-shielding of H$_2$. Therefore, at densities of about 10$^3$ cm$^{-3}$, the interstellar gas is almost entirely molecular, except for about 25\% in mass of He. Of course, the molecules are almost 100\% H$_2$ because of the overwhelming abundance of H nuclei (Table 1), and less than 1\% being in other molecules, mostly CO and H$_2$O (see Section 6). The ISM also contains $\sim$1\% in mass of dust grains intimately mixed with the gas, including most of the refractory elements, such as Fe and Si, and a significant fraction of O and C (see Section 8).  The interstellar molecular gas is generally cold (10-50\,K) because of inefficient heating, mostly by cosmic rays in the absence of UV radiation, and efficient cooling through molecular lines, mostly CO. However, there are also special cases with warm molecular gas, from $\sim$100\,K up to a few 10$^3$ K in strong starbursts, shocks, or in the vicinity of hot stars, AGB stars and AGN.

	The molecular ISM is essential in star formation, from the initial condensation and fragmentation of giant molecular clouds, to the dense accretion disks with molecular flows and pre-planetary disks. The total mass of molecular gas is a determining factor for the global star formation in a galaxy. It is impossible to discuss the complex processes of star formation without dealing with the molecular physics of the interstellar gas. Molecules are particularly essential for cooling the molecular gas to the low temperatures which determine the properties of gravitational collapse and will eventually lead to star formation.
	
	Molecules may also be minor components of the atomic interstellar gas. They have then little influence on the physical properties of the ISM and its evolution. However, they can be interesting probes of the ISM, e.g.\,\,for tracing the intensity of cosmic rays or UV radiation, and importantly contribute to the infrared emission through mid-IR PAH bands. There are even cases where tiny amount of molecules may have a dramatic influence. In particular, in the primordial gas, before the formation of the first stars, abundances of H$_2$ as small as a few 10$^{-4}$  may allow cooling proto-galactic condensations down to a few 10$^2$\,K, allowing early collapse of relatively small masses, perhaps of globular cluster size, which could in turn give birth to the first stars (Section 10.4.2).
	
	Molecular gas -- mostly H$_2$ and He, with $\sim$1\% of other molecules and dust -- is definitely the normal state of the {\it dense} ISM ($\ga$\,10$^3$ cm$^{-3}$). It is essential in the physics of the gravitational collapse at the various scales which eventually lead to the {\it formation of stars and planetary systems}.

\subsection{Landmarks of the history of the discovery and studies of interstellar molecules in the Milky Way (MW)}

	The discovery of the first interstellar molecules occurred as early as 1936-1942, in the context of studies of atomic and ionic absorption lines in the sight-line of bright stars, through the detection of absorption bands of a few diatomic species with abundant atoms and strong optical bands: CH$^+$, CH and CN. It is interesting to note that this list of optical detections has been little increased since this initial discovery. There is a relationship between the rotational excitation of the lower level of these transitions and the CMB 3K radiation temperature; however, this relationship was not properly understood until the CMB was discovered 30 years later.
	
	With the development of interstellar dust studies in the 1940-1950's, various discussions took place about possible catalytic synthesis of molecules on the surface of dust grains, and their likely presence in abundance in the interstellar gas. However, because of the lack of adequate optical lines for most molecules, the exploration of the molecular ISM had to wait for the development of microwave, infrared and especially millimetre technics, and space astronomy for UV detections. After the revolution brought in interstellar studies by the radio astronomy with the first observation of the \HI~21\,cm line, the first organized searches for interstellar molecules were triggered by the success of laboratory microwave spectroscopy. They produced in the 1960's the discovery of the very important interstellar molecules OH, NH$_3$, H$_2$O and H$_2$CO, with strong interstellar maser emission for OH and H$_2$O. Then, because most simple molecules have rotation transitions in the millimetre range, the majority of the detections came with the advent of millimetre radio-astronomy. Indeed, the development of interstellar molecules studies has been always greatly dependent on technical advances: first with radio-astronomy technics in the stream of the enormous success of radio continuum and 21cm observations; and then mostly with the full development of millimetre telescopes and detectors, which was indeed a dedicated effort aiming at interstellar molecules studies, strongly motivated by early microwave detection of complex molecules, such as NH$_3$ and H$_2$CO.
	
	The seventies were the golden age of the discoveries of interstellar molecules. After the early detection of CO, it was soon realized that its stability, abundance and widespread distribution in local interstellar clouds, in the entire MW Galactic disk, circumstellar AGB shells, external galaxies, etc., make it a very good tracer of H$_2$ and thus of the whole molecular ISM. H$_2$ itself is very difficult to detect in the normal cold molecular ISM, because of its lack of allowed electric dipole rotational and vibrational transitions, and the large opacity of molecular clouds to UV radiation (Section 10).
	
	In addition to CO, early millimetre searches quickly discovered tens of other interstellar molecules, some of them which were expected, such as HCN, CH$_3$OH, CS, SiO, C$_2$H, etc., but others turned out to be unexpected such as HCO$^+$, N$_2$H$^+$, HC$_n$N, C$_n$H, often unknown in the laboratory (see e.g.\ Watson 1976, and Section 2). This soon led to a very good understanding of the basic features of interstellar chemistry, in particular the need of dust grain synthesis of H$_2$ and the essential role of ionic reactions driven by cosmic rays (see Section 2.4). 
	
	In parallel to millimetre observations, the first space UV telescopes allowed the exploration of the molecular content of translucent clouds through absorption of UV lines (Section 7). Such clouds are mostly atomic, but contain enough molecules, mostly H$_2$, to be detectable.
	
	The advent in the 1980-1990's of large single-dish telescopes and interferometers operating at millimetre wavelengths opened up a new area in our understanding of the interstellar medium through further discoveries of new interstellar molecules, detailed  studies of star forming regions and their physics and chemistry in the Milky Way, and importantly initiated the first comprehensive molecular studies in external galaxies in the local Universe. However, it was realized only in the nineties that molecules could already be detected with the current equipment in strong starbursts in the most distant Universe. In parallel, ground-based and space infrared astronomy developed comprehensive studies of H$_2$ lines and aromatic mid-IR bands.

\subsection{Main achievements of molecular studies in external galaxies}

 %% Landmarks of the history of the discovery and studies of interstellar molecules in the Milky Way begin with the discovery of the first interstellar molecules which occurred as early as 1936 through the detection of visible absorption bands of a few diatomic species. However, the first organized searches took place only in the 1960s in the microwave range, and more systematically in the seventies with the advent of millimetre radio astronomy. They have continuously developed since, together with UV and IR studies.
  
  As it is reviewed below, many of the main achievements of molecular studies in the last decades took place in observations of external galaxies. Such studies have become a major field for understanding galaxies, their evolution and especially star formation and starbursts at galactic scales.
	
	First, numerous, comprehensive millimetre studies of CO and its isotope varieties have brought detailed information about the amount and distribution of the molecular gas in various kinds of local galaxies, and its relation with star formation and physical conditions. One has been able to infer the role of molecular gas in galaxy evolution, spiral arms, galactic bars and AGN feeding, galaxy collisions, etc., and associated physical processes. The detection of a number of other molecules in local galaxies have allowed comparative studies of molecular abundances in various galactic environments, including isotopic varieties, and inferences on physical and chemical modelling.
	
	Mega-masers in the lines of OH and H$_2$O, with enormous power, have been detected in local infrared luminous and ultra-luminous galaxies (LIRGs and ULIRGs), with discussion of their physics. Detection of H$_2$O mega-masers in disks around AGN lead to a precise determination of the mass of their central super-massive black hole. Several other molecules have also been studied in the vicinity of the central torus around AGN, with continuously improving angular resolution at millimetre and infrared wavelengths.
	
	The absorption of many molecular lines of many species at redshift $\sim$0.5-1, in front of a few radio sources, have provided comparative molecular abundances and inferences on the evolution of physical conditions and atomic abundances in galaxies. Rotational emission of molecules is currently detected in ULIRGs at very high redshift up to z=6.4, often around powerful quasars and radio galaxies. Unique information is deduced about the early formation and starbursts of the most massive galaxies.
	
	Large aromatic compounds (PAHs) have shown to be ubiquitous in all kinds of galaxies. Infrared H$_2$ emission lines begin to currently trace the warm molecular gas in the mid-IR from space, and at high angular resolution in the near-IR from ground-based adaptive optics.
	
	Much more is expected with the prospects of order(s) of magnitude gains of sensitivity expected with worldwide new facilities: mostly with ALMA (Atacama Large Millimeter Array) in the millimetre range; and also from future large space infrared telescopes and  SKA (Square Kilometer Array) in the radio. One will thus study fine details in local galaxies and make comprehensive global studies of standard galaxies at high redshift and detailed studies of IR starbursts galaxies at very high redshift. One may thus expect: i) deep progress in understanding the evolution, formation and merging of various kinds of galaxies at all early epochs when the galaxies and most of their present stars formed, and the connections between AGN and their host galaxies; and also ii) major advances advances in comparative interstellar chemistry in external galaxies.

\section{Cosmic molecules and local interstellar medium}

	The properties of the interstellar gas in various environments share many common features between different galaxies. Most of these properties are well examplified in the Milky Way, and in particular in local interstellar clouds. Therefore, the detailed information that we can get from local interstellar molecules is essential to understand the behaviour of the molecular medium through the Universe. Indeed, the techniques are the same, both for observations, mainly millimetre wave radio astronomy, and for modelling line formation as well as physical and chemical properties.

\subsection{Physics and various components of the local interstellar molecular gas}

	Molecules are mostly found in the `molecular gas' which is one of the four or five major components with characteristic properties of temperature and density, and thus ionization and chemical composition, that one traditionally distinguishes in the ISM of a galaxy like the Milky Way (see e.g.\,\,Table 1 of Lequeux 2005 and Table 1.1 of Tielens 2005): hot coronal intercloud gas, warm ionized and neutral media, cold neutral atomic medium, molecular clouds. Molecules can hardly survive to photodissociation in the tenuous (n$_{\rm H}$ $\sim$ 1-100 cm$^{-3}$), warm (T$_{\rm K}$ $\sim$ 100-1000\,K) atomic gas permeated by the UV radiation from massive stars. They are practically completely absent in the ionized gas (T$_{\rm K}$ $\sim$ 10$^4$\,K) of the diffused ionized medium and the dense \HII~regions around massive stars; as well as in the coronal gas (T$_{\rm K}$ $\sim$ 10$^6$\,K). 
	
	The abundances of the various species, atoms, ions, molecules and dust grains, found in the interstellar medium are controlled first by the typical ''cosmic'' abundances of atomic elements, reproduced in Table 1 and characterized by the overwhelming abundance of H and He, and the relatively large abundance of O, C and N.
	
\begin{table}[ht]                                      
\caption{Atomic solar abundances (Asplund, Grevesse, \& Sauval 2006), ~~~~~~~~~~~~~~~~ representative of ''cosmic'' abundances found in the interstellar medium of galaxies.
}
\begin{center}
\scriptsize
\begin{tabular}{l c c c c c c c c}  
Atom	&	n(X)/n(H) & | &	Atom	&	n(X)/n(H) & | &		Atom	&	n(X)/n(H) &  	\\	
H 	&	1.0E+00 & | &		He 	&	8.5E-02 & | &		O 	&	4.6E-04 &  	\\
C 	&	2.5E-04 & | &		N 	&	6.0E-05 & | &		Mg 	&	3.4E-05 &  	\\	
Si 	&	3.2E-05 & | &		Fe 	&	2.8E-05 & | &		S 	&	1.4E-05 &  	\\	
Al 	&	2.3E-06 & | &		Ca 	&	2.0E-06 & | &		Ni 	&	1.7E-06 &  	\\	
Na	&	1.5E-06	& | &		Cr 	&	4.4E-07 & | &		Cl 	&	3.2E-07 &  	\\	
Mn 	&	2.5E-07 & | &		P 	&	2.3E-07 & | &		K 	&	1.2E-07 &  	\\	
Ti 	&	7.9E-08 & | &		Co 	&	8.3E-08 & | &		F 	&	3.6E-08 &  		
\end{tabular}                
\end{center}                     
\scriptsize
\end{table}

	As described e.g.\ in Tielens (2005), Lequeux (2005) and Dyson \& Williams (1980), the molecular gas must be dense enough (n$_{\rm H}$ $\ga$ 10$^2$-10$^3$ cm$^{-3}$) so that its external layers (A$_{\rm v}$ $\approx$ 1, N$_{\rm H}$ $\approx$ 2\,10$^{21}$ cm$^{-2}$) efficiently shield the interior from UV photodissociation. Most of the molecular gas in the Milky Way is distributed in Giant Molecular Clouds (GMCs), with average densities slightly above 100 cm$^{-3}$, typical sizes of 30-50 pc, typical masses of a few 10$^5$ M$_\odot$. While the external layers of GMCs have a substantial warmer \HI~component possibly with comparable mass, most of their interior is very cold, with temperature not much exceeding 10\,K, as the result of the balance between heating by cosmic rays and cooling by CO rotation lines. Their gross structure is well traced by CO lines (see Sections 3 \& 4), possibly complemented by extinction 
%, A$_{\rm v}$\,$\sim$\,10, 
and $\gamma$ ray emission generated by cosmic rays. Except for helium, most of their inner gas is molecular hydrogen, with about 2\% in mass in sub-micron dust grains, large aromatic molecules (PAH), other molecules and a few atoms (mainly O). The degree of ionization maintained by cosmic rays is extremely low, e.g.\,\,$\sim$10$^{-7}$. However, these molecular clouds are complex structures, with clumpiness at various scales and strong turbulence as attested by the width of several km/s of the molecular lines such as CO, as well as enhanced magnetic field of a few tens $\mu$G, roughly proportional to n$^{0.5}$ (see e.g.\ Fig. 2.6 of Lequeux 2005). They are generally self-gravitating, but remain stable for several 10$^7$ years with the balance of magnetic and turbulent pressure and gravity. Their dense condensations (n\,$\sim$\,10$^3$-10$^5$ cm$^{-3}$, parsec size and masses $\sim$10-10$^3$ M$_\odot$) are particularly interesting as related to the process of star formation and the site of a peculiar chemistry. In the absence of stars already formed, such `dark clouds' are even colder than the ambient molecular medium. However, the presence of young luminous stars may heat the gas of neighbouring condensations up to temperatures $\sim$\,100\,K, changing the chemical processes at work.
	
	The molecular clouds are in constant interaction with the other phases of the interstellar medium and the massive stars that they form. They have thus boundary layers where the molecules are more or less photodissociated by the external UV radiation. Such `Photo-Dissociation Regions' (PDR) may be particularly active in the vicinity of OB associations, generating special physical and chemical conditions. Because of their larger inertia, massive molecular clouds are less permeated by interstellar shocks than the more diffuse phases of the intersellar medium. However, the molecules may survive and even be specially synthesized in the compressed, overdense, hot regions of interstellar shocks propagating in less dense regions, where their lines, e.g.\,\,rovibration lines of H$_2$, may provide basic diagnostic of shocks.
	
Decades of active research on the local ISM have very well documented the physical and chemical properties of molecular clouds in our various Milky Way environment, their rich composition, complex structure and dynamics and their symbiosis with young stars. Milky Way molecular clouds provide us with the best templates to understand the molecular ISM in other galaxies, despite the frequent occurrence of more extreme conditions encountered in earlier and more violent stages of galaxy evolution where extragalactic molecules are currently observed.

\subsection{Observational techniques}

	Because the molecular ISM is cold and opaque to the visible and UV radiation, our best information comes from millimetre radio astronomy and molecular rotational transitions, where the photon energy is of the order of kT$_{\rm K}$. Indeed, as reminded in Section 1.3, the development of interstellar molecular astrophysics has closely followed and often motivated the progress of millimetre techniques. The size of the antennae is limited by the need of high surface accuracy and precise pointing. The first studies were performed in the 1970s by 1-10\,m single dishes such as Columbia 1.2\,m, Texas  5\,m, Bell Lab. 7\,m, NRAO 12\,m, FCRAO 14\,m, etc. They were superseded in the 1980-1990s by the Nobeyama 45\,m and IRAM 30\,m larger dishes, and the millimetre interferometers (BIMA 9 x 6\,m, Nobeyama 6 x 10\,m, OVRO 6 x 10.4\,m, IRAM 6 x 15\,m). 
	
	Millimetre astronomy is strongly affected by atmospheric absorption (and emission), mostly from H$_2$O bands,  in a large part of the spectrum, especially at high frequency. However, it benefits from a number of excellent transmission windows. The latter allow multi-transition studies of practically all interesting molecules except hydrids with a single heavy atom. The 3\,mm window, $\sim$72-116\,GHz, which includes at least one rotation transition of most molecules, has even produced important results in ordinary, low elevation sites especially in the first stages of millimetre radio astronomy. It is a chance for studying the Galactic and extra-galactic molecular medium, that the three first transitions of CO, the most important interstellar molecule, occur at frequencies with good atmospheric transmission, J\,=\,1-0 at 115.271\,GHz, J\,=\,2-1 at 230.538\,GHz, J\,=\,3-2 at 345.796\,GHz. However, the advantage of high altitude, dry sites has become essential, even in the 1.3\,mm window which includes the 2-1 CO line which turns out to be often significantly more sensitive than the 1-0 line. Such excellent sites are mandatory for extending molecular observations to the sub-millimetre range, as shown by the pioneer work of CSO, JCMT and SMA at Mauna Kea. This has justified the choice of the 5000\,m Chajnantor site in Chile for the worldwide mm-submm project ALMA (Section 12).
	
	Millimetre studies fully benefit from the fundamental advantages of radio astronomy for high velocity resolution with the heterodyne techniques, and high angular resolution with multi-dish interferometers. Both are of course essential for studying structure and dynamics of galaxies. They will be fully implemented in ALMA with its 54 x 12-m (+ 12 x 7\,m) dishes, its six initial frequency bands and its baseline up to 14.5 km. The development of millimetre astronomy was long impeded by the difficulty of making high sensibility detectors. However, the progress has been constant in this field so that the ALMA receivers, with supra-conducting junctions (SIS) and Hot Electron Transistors (HET), will approach the quantum noise limit in broad-band receivers. In parallel, the high resolution of the `back-end' spectrometers and the transport and processing of the interferometer signals have fully benefited from the progress of high speed information technology and computers, culminating in the giant ALMA correlator (under construction).
	
	Other wavelength ranges are complementary to the millimetre one for investigating the Galactic and extra-galactic  molecular gas for more specific, but important, goals. There is a natural extension to lower frequencies of the centimetre and decimetre radio range, for three main fields: 1) these frequency ranges include a few very important fine structure lines, mostly of abundant hydrids without important millimetre lines, OH, H$_2$O and NH$_3$, with strong masers for OH and H$_2$O; 2) at very high redshift 3\,mm lines, such as CO(1-0), are shifted to cm wavelengths; 3) the first rotation transitions of very heavy molecules, such as HC$_{\rm n}$N, are located at cm wavelengths; however, their low abundance will not allow much development in extragalactic studies. The Green Bank Telescope (GBT) already studies extragalactic mega-masers and CO(1-0) at very high redshift. The extension of the VLA (EVLA) will allow many more similar studies, waiting for the tremendous sensitivity and VLBI capability of the Square Kilometre Array (SKA) which will make a breakthrough in this field. In the infrared, there is also some extension to the far-infrared (100-200\,$\mu$m) for high-J rotational transitions, mostly H$_2$O. The mid-infrared is important for PAH and dust molecular features (Section 8), while the whole infrared range is essential for H$_2$ lines, although there are highly forbidden (see Section 10). The most important ranges for H$_2$ at zero-redshift are: i) $\sim$5-30\,$\mu$m for rotation lines emitted by the warm molecular medium; such studies will certainly have important developments with future infrared space projects such as JWST, SPICA, SAFIR, etc; and ii) $\sim$2-5\,$\mu$m for rovibration lines emitted in shocks and regions with strong UV radiation, and observable from the ground with very large telescopes and eventually adaptive optics. The allowed UV strong absorption lines of H$_2$ (and other molecules) are much more powerful to trace small column densities of H$_2$ as currently observed by FUSE in the local diffuse gas, and with large telescopes such as VLT, when they are redshifted. However, the number of such detections in extra-galactic lines of sight has remained limited (see Section 7.1).

	In conclusion, studies of interstellar molecules belong mainly to the realm of millimetre radio astronomy. They benefit from the outstanding technical progress in this field and the power of image synthesis and of high-resolution heterodyne spectroscopy with large radio arrays, which will culminate with ALMA. However, millimetre techniques will eventually be complemented in the exploration of the  molecular world of galaxies by other impressively large facilities in wavelength ranges from radio to infrared and visible.

\subsection{Modelling (millimetre) molecular line formation, and diagnostic of physical conditions and molecular abundances in the ISM}

	Most of our information on the molecular interstellar medium comes from the detected intensities and profiles of (millimetre) molecular lines. With a relatively simple modelling, one infers estimates of the main physical parameters, temperature, density, velocity distribution, and the chemical abundances of the observed molecules. Of course in practice, all these quantities are some kind of average over the volume traced by the radiotelescope beam. As the main cooling of the molecular medium is achieved through lines of the most abundant species, mostly CO, such a modelling is also the basis of the evaluation of the cooling rates (see Section 2).
	
\subsubsection{Equations of statistical equilibrium and radiative transfer.} The main goal is to determine the populations of the rotation levels and the line radiation intensities which are coupled through the equations of statistical equilibrium and radiative transfer. As usual, the most simple case is that of local thermodynamical equilibrium (LTE) when the collision transition rates dominate over radiative transitions. The rotational populations have then a Boltzman distribution determined by a single excitation temperature equal to the gas kinetic temperature. However, LTE is never fully realized with the low density of the interstellar medium. For a given rotational transition, LTE is only approached when the density n$_{{\rm H2}}$ exceeds a critical value A/C where A is the Einstein coefficient for spontaneous emission and C is the collisional rate. The A coefficient scales as $\mu^2\nu^3$ where $\mu$ is the dipole moment and $\nu$ the line frequency. In practice the decay rate is lowered by line trapping for optically thick lines so that the critical density is decreased. Typical values thus range from $\approx$ 300 cm$^{-3}$ for CO(1-0), because of the low value of the CO dipole, to $\sim$\,10$^5$\,cm$^{-3}$ for the first transitions of molecules with usual, large dipoles, such as HCN, CS or HCO$^+$. For an optically thick line at LTE, the line intensity J$_\nu$ is just equal to the Planck function B$_\nu$(T$_{\rm K}$).
	
	In the general case, one must solve the coupled equations of statistical equilibrium and radiative transfer. For all usual molecules the Einstein radiation coefficients are well known. Values for the collisional transition rates rotational levels  (or cross-sections, mainly for H$_2$, see e.g.\,\,Flower \& Launay 1985) are provided either by full quantum calculations for the most important simple systems, or, for the others, by various approximations accurate enough to match the uncertainties of the radiative transfer geometry. In most cases, the treatment of radiative transfer is much simplified by the use of the Large Velocity Gradient (LVG) approximation based on local photon trapping and the escape probability method (Sobolev 1960, 1963, Goldreich and Kwan 1974, Scoville and Solomon 1974). It is equivalent to replace the Einstein A coefficient by $\beta$A where  $\beta$ is the `escape probability'. For optically thick lines with line optical depth $\tau_l$ $>>$ 1, $\beta$ is just proportional to $\tau_l$$^{-1}$. The line optical depth $\tau_l$ is generally estimated from a Doppler gaussian profile with the observed line width $\Delta$v. It is worth noting that the Lorentz wings of millimetre molecular lines are always negligible because of the very small values of the Einstein A coefficients.
	
	Despite its success in representing a basic feature of radiative transfer with Doppler broadening, the LVG approximation is clearly limited to account for the complexity of radiative transfer in actual media. The development of computing power allows a more and more generalized use of Monte Carlo methods for modelling radiative transfer, with a degree of complexity for describing the structure of the interstellar medium adapted 
to the available information (see e.g.\ Gon{\c c}alves et al. 2004 and references therein). Such methods are particularly well adapted for computing the emerging intensities and complex line profiles resulting from actual inhomogeneous distributions of densities, 
temperatures, abundances, velocity fields, and the resulting populations of the rotational levels.
	
\subsubsection{Radiotelescope signals: flux density and antenna temperature} In brief, through such modelling, rotational millimetre lines are a powerful tool for tracing molecular abundances, including isotope varieties, and physical conditions in molecular clouds such as density, temperature and velocity fields. However, it is clear that the derived information is some kind of average over the molecular gas included in the radiotelescope beam. This is a serious drawback for extragalactic studies with the large beams of single dish millimetre telescopes. For the 10-30m telescopes discussed in Section 2.2, in most cases the beam diameter ranges from 10'' to 1'. For the nearest galaxies at a distance of $\sim$1\,Mpc, this corresponds to scales of $\sim$50-300\,pc, i.e. at best the typical diameter of a giant molecular cloud. At any redshift z\,$\ga$\,0.5, the corresponding beam-encompassed distances (varying little with z) are rather $\sim$50-300\,kpc, so that there is no hope to derive any other information than global values for a galaxy. Therefore, extragalactic molecular studies really need the increased angular resolution of millimetre interferometers. Current best facilities, such as the IRAM interferometer, 
(Fig.\,7a), 
already reach $\sim$0.3-0.5'' resolution. This provides nice details on nearby galaxies, but, for high z galaxies, hardly allows to distinguish molecular emission of their core from the possible one of their most outer regions. However, ALMA, with better than $\sim$0.1'' resolution and high sensitivity, will provide detailed information on sub-kpc molecular structures at practically any redshift. Comparable or even higher resolution are already achieved by cm/dm very long baseline interferometry (VLBI) with the very strong intensities of extragalactic mega-masers (see below). Such capabilities will be greatly extended by SKA, including for thermal low-J, high-z molecular lines. The strongest ones may already be angularly resolved with the current VLA angular resolution ($\sim$0.15'', see e.g.\,\,Walter et al. 2004, Riechers et al. 2006c).
	
	For optically thin lines, the line flux density S$_\nu$ detected at a given frequency by a radiotelescope (generally expressed in Jansky, 1\,Jy\,=\,10$^{-26}$\,Wm$^{-2}$Hz$^{-1}$) is proportional to the total number of molecules in the upper level u of the transition at the corresponding radial velocity v within the telescope beam. The line profile is just given by the velocity distribution of such molecules, and the line velocity  integrated flux density $S\Delta v$ = $\int$ S$_\nu$(v) dv 
\footnote{$\Delta v$ is the line full width at half-maximum (FWHM) and $S$ is approximately the peak line flux density} 
is proportional to the total number of molecules in level u within the beam (or more exactly the convolution of the molecule distribution with the beam). If the population distribution among the rotational level may be estimated, e.g.\,\,with an approximate rotational excitation temperature, the total number $\mathcal{N}$$_{\rm i}$ of the considered molecules i within the beam may be directly inferred. However, the corresponding abundance of molecules i, $\chi$$_{\rm i}$ = N$_{\rm i}$/N$_{{\rm H2}}$, also requires the knowledge of N$_{{\rm H2}}$ which can only be indirectly derived.
		
	When the line ul is not optically thin, the emission from level u at each position in the cloud must be weighted by the absorption factor exp[-$\tau_{iul}$(v)], where $\tau_{iul}$(v) is the optical depth up to the boundary of the cloud, and depends on the frequency $\nu$ and hence on v. This gives a larger weight to the emission by the outer layers of the cloud, and also modifies the line profile. For very thick lines, the specific intensity J$_{\nu}$ is just B$_\nu$(T$_{\rm K}$) at LTE, and it generalizes as J$_{\nu}$ = B$_\nu$(T$_{{\rm exc}}$) where (E$_{\rm u}$--E$_{\rm l}$)/kT$_{{\rm exc}}$ = --log(n$_{\rm u}$g$_{\rm l}$/n$_{\rm l}$g$_{\rm u}$) is the line excitation temperature. Thus, for thick lines, as often CO lines are, the flux density directly reflects the excitation temperature through S$_\nu$ = $\Omega$J$_{\nu}$ =  $\Omega$B$_\nu$(T$_{{\rm exc}}$), where $\Omega$ is the solid angle subtended by the source (more exactly, the solid angle of the source convolved with the telescope beam) 
\footnote{In the Rayleigh-Jeans regim B$_\nu$(T) = 2kT/$\lambda^2$, and the flux density S$_\nu$ is proportional to T$_{{\rm exc}}$. For an extended source broader than the beam, $\Omega$ is the beam solid angle $\approx$ $\lambda^2$/A$_{\rm a}$ where A$_{\rm a}$ is the antenna surface area. For a source of brightness temperature T, the specific power picked up by the antenna is P$_\nu$ = A$_{\rm a}$S$_\nu$ = 2kT (kT per polarization, Nyquist theorem). 

 The simplicity of this relation has induced radioastronomers to use temperature as unit to measure the signal of radiotelescopes. However, for millimetre and sub-millimetre wavelengths and typical values of T$_{{\rm exc}}$, the Rayleigh-Jeans approximation kT/h$\nu$ $>>$ 1 is generally not valid (kT/h$\nu$ = 0.7 [$\lambda$/1\,mm][T/10\,K]). Nevertheless, antenna temperatures are still in use with the same formal definition 
 T$_{\rm A}$$^{\star}$ = ($\lambda^2$/2k)I$_{\nu}$ $\approx$ A$_{{\rm eff}}$S$_\nu$/2k, 
 where the effective antenna area A$_{{\rm eff}}$ takes into account the antenna imperfection. But T$_{\rm A}$$^{\star}$ is no longer equal to the source brightness temperature T$_{\rm B}$ in the case of an extended source and perfect antenna, but to 
 T$_{\rm B}$$^{\star}$ =  [h$\nu$/k]/[exp(h$\nu$/kT$_{\rm B}$)-1]; 
 and the added `$\star$' aims at reminding that. See radio astronomy textbooks (e.g.\,Kraus 1986, and also Lequeux 2005) for complete technical definitions.}.
 If T$_{{\rm exc}}$ varies within the cloud, the line intensity is determined by the excitation temperature of the outer layers; in particular CO lines often form at LTE and reflect the kinetic temperature of the boundary layers of clouds. 

\subsubsection{Interstellar masers.} In extreme cases, the divergence from LTE may even lead to cases of `population inversion' where the population n$_{\rm u}$/g$_{\rm u}$ of the upper level is larger than the lower one n$_{\rm l}$/g$_{\rm l}$. The excitation temperature T$_{{\rm exc}}$ is then negative, as well as the line optical depth, leading to an exponential amplification exp($\vert$$\tau_{{\rm ul}}$$\vert$) of the line intensity. The resulting very powerful maser lines have been observed since the beginning of molecular radio astronomy in a few transitions of OH ($\lambda$\,=\,18\,cm with four hyperfine lines) and H$_2$O ($\lambda$\,=\,1.35\,cm) in Galactic massive star forming regions (see e.g.\,\,Lo 2005 for a list of general references about interstellar masers). The highly non linear amplification favours emission in the directions where the velocity configuration leads to the largest amplification, with very small solid angles $\Omega_s$ for the emission regions, which are measurable by VLBI with milli-arcsec resolution. From the large observed line flux densities, S$_\nu$, one infers extremely high brightness temperatures T$_{\rm b}$ = S$_\nu$/$\Omega_s$$\times$$\lambda^2$/2k, ranging up to 10$^{14}$\,K. Much more luminous OH and H$_2$O maser emission (`mega-masers') was discovered in the nuclear regions of external galaxies with line luminosities $\sim$10$^2$-10$^4$\,L$_\odot$, $\geq$\,10$^6$ times more luminous than typical Galactic masers (see Lo 2005 and Sections 4 \& 10).
	
		Conditions leading to such powerful Galactic and extragalactic maser emission are extremely complex. Population inversion is a necessary condition, together with velocity coherence along the line of sight allowing a significant gain. The pumping mechanisms achieving population inversion are not always fully understood. They result from complex combinations of collisional and IR-line radiative effects whose net result favours the excitation of the upper level of the maser transition, either through cascades from upper levels or because spontaneous IR emission from the lower maser level is faster than from the upper one. It is certainly not just a chance that the two main molecules with strong interstellar maser emission, OH and H$_2$O, are abundant and have a complex rotational level structure. For extragalactic mega-masers, it is generally agreed that the intense  mid-infrared radiation in extreme starburst regions plays a major role in OH pumping, while the energy required to achieve the collisional pumping of H$_2$O in the central parsec of AGN comes from the AGN (see Lo 2005 and Section 11).

\subsection{Uniqueness of basic processes of interstellar chemistry} 

\subsubsection{Known interstellar molecules.} Immediately after the beginning of the harvest of discoveries of interstellar molecules by millimetre radio astronomy, it was realized that the interstellar chemistry is extraordinary and unique, through the number of detected exotic molecules and their peculiar abundances which were derived. There are now about 150 interstellar molecules identified in the local ISM of the Milky Way (see updated list in http://astrochemistry.net/, see also http://www.cv.nrao.edu/$\sim$awootten/allmols.html, and http://physics.nist.gov/cgi-bin/micro/table5/start.pl by F.J. Lovas for NIST recommended frequencies for observed interstellar molecular microwave transitions). Formed mainly from H, C, O \& N atoms, they  can also include S, Si, metals, etc. They belong to two main classes:  expected common stable species with up to $\sim$10 atoms; and a large number of exotic unstable species, very uncommon and rare because of their instability in normal laboratory conditions, including radicals (such as CH$_{\rm n}$ with n=2 to 6), ions (such as HCO$^+$ and N$_2$H$^+$), long polyyne chains (HC$_{{\rm 2n+1}}$N with n=1 to 3), small cycles (such as C$_3$H$_2$), or isomers (such as HNC). Most of the latter are extremely unstable and hardly observed in the laboratory, except as very transitory intermediate reaction products. A number of them were not even known before their discovery in interstellar space. Clearly, the existence of such exotic compounds is directly linked to the incredibly low interstellar densities, which allow them to survive for 10$^5$-10$^6$\,yr and to be observable provided they are efficiently formed. They are obviously completely out of equilibrium with respect to very low interstellar temperatures. Three-body processes are practically excluded in the interstellar gas phase. Grain processes had long been proposed for molecule synthesis, but they cannot account for most of such unstable species. On the other hand, the latter are logically explained by reaction chains initiated by energetic particles such as UV photons and cosmic rays. Only exothermal reactions with practically no activation barrier are generally possible, implying radical and mostly ion reactions. However, the creation of the initial chemical bonds remains difficult without a third body. It practically needs dust grain reactions, at least for the formation of H$_2$. The interpretation of interstellar chemistry, combining gas phase and grain processes, has been excellently reviewed a number of times in the last three decades (see e.g.\,\,Solomon \& Klemperer 1972, Herbst \& Klemperer 1973, Watson 1976, Tielens 2005, Lis, Blake \& Herbst 2005, and references therein). The basic features have not much changed during this period, with however the addition of deeper insights on various particular points, including new information from observations, laboratory and theory. Current views may be summarized as follows.

\subsubsection{Gas phase chemistry.} Although the importance of grain processes is not to be underestimated, the main success in detailed modelling has come from gas phase reactions. They are simpler and pretty well understood, especially for charged species. The most direct possible analogy with other out-of-equilibrium chemistry we are used to in dilute gases, is probably  photochemistry in the upper earth atmosphere. {\it Ultraviolet radiation} from massive stars is ubiquitous in the interstellar medium, except in molecular clouds. It is certainly even stronger in many other, younger, galaxies than the Milky Way, with more UV photons generated from star formation, or less shielding by dust. Shielding from UV radiation is crucial for the survival of interstellar molecules, and there are practically no UV photons deep in molecular clouds. However, photochemistry is essential in all regions more or less permeated by interstellar UV radiation, either ambient or coming from a local source: boundary regions of molecular medium; translucent, mostly atomic, clouds; outer circumstellar shells, etc. The most spectacular photochemistry occurs in the {\it photodissociation regions (PDRs)}, and such photo-processes are among the best understood ones in interstellar chemistry. Their main effect is the destruction of molecules, explaining the low abundances of molecules in the diffuse interstellar medium, and even their quasi-absence in high-redshift optical absorption systems (see Section 7.1). However, the radicals and ions they generate may rival with, and even supersede, cosmic rays for initiating reaction chains,  forming various species. The physics and the chemistry of PDRs are entirely dominated by the UV field which is many orders of magnitude larger than the standard interstellar field. In addition to the enhancement of photochemistry itself, an important feature of PDRs is the high temperature which allows opening new reaction channels, such as C$^+$, O or OH with H$_2$, overcoming small activation energies (Tielens \& Hollenbach 1985, Tielens 2005, Lequeux 2005, and references therein). Note that the UV radiation also achieves the excitation and dominates the life cycle of PAHs, especially in PDRs (see Section 8).
	
	As the interior of molecular clouds is deprived from UV photons, it is agreed that the chemistry leading to the formation of the abundant unstable species is mainly initiated by {\it cosmic rays} which easily propagate inside (X-rays may play a similar role in some cases, especially AGN, see Section 11). The energy reservoir in cosmic rays is quite significant, and indeed comparable to the other forms of interstellar energies which are all roughly in  equipartition, the gas thermal energy in particular. With typical energies $\sim$\,0.1-10\,GeV, they dominate the heating and ionization of molecular clouds and, from the ionization of H$_2$ and He, they trigger their gas phase chemistry with molecular ions. Ion reactions are particularly efficient because of the long range r$^{-4}$ charge-induced dipole interaction which leads to very large pseudo-capture cross-sections and most often to the absence of activation energy for exothermal reactions. Many ion reactions thus proceed with very fast rates, close to the universal Langevin pseudo-capture rate, $\sim$\,2\,x\,10$^{-9}$ cm$^{-3}$s$^{-1}$ (e.g.\ Su \& Bowers 1979). Basic modelling (Herbst \& Klemperer 1973, Watson 1973, 1976) shows that they well explain many key properties of interstellar chemistry including the presence and the central role of molecular ions such as H$_3$$^+$ and HCO$^+$, the high abundance of CO, the presence of key radicals such as OH and CH, and isomers such as HNC/HCN, and the high abundance of deuterated species.                                                   
	
	  Standard models of gas phase interstellar chemistry also include (see publicly available codes from e.g. http://www.physics.ohio-state.edu/~eric/research.html, and the UMIST reaction rate database http://www.udfa.net/): radical reactions without activation barrier whose rates are generally more uncertain than those of ion reactions; universal destruction of molecular ions by dissociative recombination; radiative association generally very efficient for very large molecules, but slow and uncertain for small species; etc. This yields large networks of reactions, with up to hundreds of species and thousands of reactions. However, many reaction rates can only be guessed. The large time constants involved, $\sim$\,10$^5$-10$^6$\,yr, make better to search for time dependent solutions than just the equilibrium state that may never be achieved. Despite success on various particular points, detailed quantitative predictions remain overall difficult for various reasons: uncertain rates for most important processes; difficult modelling of grain processes; lack of information about the physics, structure and evolution of the clouds; coupling between chemical and physical evolution (even able to lead to bistable equilibrium); etc. Therefore, the most notable advances in the last decades have rather addressed particular questions often triggered by observational results, such as the PAH chemistry, or various extensions especially to the warm gas: hot molecular cores, shocks, PDRs, AGN, turbulent heating, etc.
	  
\subsubsection{Grain processes}are fundamental because they are by far the main source of new chemical bonds and also because grains are one of the main repositories of heavy elements and a potential source for complex species including PAHs. However, despite significant advances, they remain the most difficult part of interstellar chemistry because of the complexity of grain structure and chemical evolution and of the interactions with the gas. Surface reactions are the only efficient mechanism {\it to form H$_2$}, and hence generate most interstellar molecular bonds by subsequent gas phase reactions, at a sufficient rate to compensate destruction processes. We have a basic logical scheme for this formation since the beginning of modern interstellar chemistry ( Hollenbach \& Salpeter  1971, Watson \& Salpeter 1972), which has little changed since (see e.g.\,\,Tielens 2005 for current views): available H atoms in the gas hit grains every $\sim$\,10$^6$\,x\,[300\,cm$^{-3}$/n] yr, have to stick on the surface through physisorption -- i.e.\ weak binding through Van der Waals forces -- with a high probability and easily migrate so that they almost surely find another H previously fixed on a chemisorption defect site, and react to form an H$_2$ molecule which is immediately ejected from the grain because its adsorption energy is too weak. Such a scheme is well supported by the general knowledge of surface physics and a few laboratory results with some materials. However, it is highly dependent on the unknown actual detailed surface state so that quantitative estimates remain out of reach and entirely rely indeed on the knowledge of the interstellar H$_2$ abundance and destruction rate.
	  
	  For {\it all other gaseous molecules} which could significantly form in grains, the rates are even qualitatively highly uncertain, because they depend on complex processes allowing the return to the gas of compounds much more tightly bound to the grains. However, there are a number of facts well established about chemical grain processes (see e.g.\,\,Tielens 2005 and references therein). Any gas species periodically hits a grain and may stick with a probability which exponentially depends on the temperature and the physisorption energy. All species, except H$_2$ and He, thus eventually stick on the coldest grains of dense molecular cores. Through subsequent surface reactions, grains therefore accrete mantles of compounds more or less volatile depending on their temperature, whose composition may be traced by infrared spectroscopy. Ice mantles, primary products of O + H chemistry are thus commonly found in dense clouds. Sensitive infrared spectroscopy reveals that ice is mixed with various impurities, expected products of accretion plus surface reactions in the H rich interstellar context (Tielens 2005 and references therein): CH$_3$OH, CO$_2$, NH$_3$, CH$_4$, CO, OCS, etc. In the cloud boundary layers such mantles may be exposed to UV  photochemistry leading to more stable mantles of organic polymers with radical defects. It is natural to think that such a rich chemistry may be at the origin of the synthesis of a number of interstellar molecules, especially those which are not easily formed by gas reactions. In particular, abundant deuterated species should be synthesized through the strong isotopic fractionation associated with the very low grain temperature. However, a main problem for the assessment of the importance of these processes remains the way that such mantles may be desorbed with injection of complex molecules into the gas. 
	  
	  It is well established that grain mantles are submitted to a harsh processing when they leave the interior of molecular clouds. The volatile mantles are first sublimated when their temperature rises and they are exposed to UV radiation. More stable polymer mantles and even refractory cores of silicate or amorphous carbon may eventually be destroyed by sputtering or grain shattering when they are periodically submitted to interstellar shocks or hot gas (Draine \& Salpeter 1979a,b). However, such processes are probably too violent to preserve most molecular bonds. Milder UV photodesorption is probably efficient for keeping grain surface clean of physically adsorbed molecules such as H$_2$O at the edge of molecular clouds. However, its efficiency is more uncertain for  heavier molecules more or less chemically bound to the grains. Intermediate desorption processes have also been proposed, such as cosmic ray driven mantle explosions (see e.g.\,\,L\'eger et al.\ 1985, Tielens 2005). This key question of grain desorption is currently addressed by several experimental and modelling studies with significant recent progress (e.g.\ see Collings et al.\ 2004, Garrod et al.\ 2007).
	  
	 Products of a warm chemistry, associated with grain desorption products, is observed in interstellar shocks and {\it hot cores} surrounding many protostars. The latter are warm ($>$\,100\,K) and dense ($>$\,10$^6$\,cm$^{-3}$) regions where abundant hydrogenated molecules such as H$_2$, NH$_3$, CH$_3$OH are found, similar to those of interstellar ices. Their origin from desorption of molecular mantles from heated grains is also supported by the very large abundance of deuterated species reflecting the important isotopic fractionation at the very low temperature of the grains. However, the multitude of more complex organic species also observed in the gas probably requires active gas reactions from the primary desorbed products (see e.g.\,\,Caselli et al.\ 1993, Tielens 2005). Similar, but more extreme processes, occur in {\it interstellar shocks} (see e.g.\,\,Hollenbach \& McKee 1979, 1989, McKee \& Hollenbach 1980, Flower et al.\ 1995, Flower \& Pineau des For{\^e}ts 1995). Post-shock temperatures higher than 1000\,K allow reactions with endothermal or activation energies such as the formation of OH and H$_2$O from O + H$_2$. The gas composition suggests not only the grain desorption of icy mantles, but also sputtering of silicates when abundant SiO is observed.

\subsubsection{Isotope fractionation} is the general result of the dependence of the zero-point vibration energies on the atomic masses at the very low interstellar temperatures. The energy difference for deuterium, D, {\it versus} hydrogen, H, $\Delta$E,  is very large, several hundreds K, i.e. much larger than kT in cold media, allowing enormous overabundance of deuterated compounds (Watson 1974, Roueff et al.\ 2000, 2005 and references therein). It is still significant for heavy elements such as $^{13}$C {\it versus} $^{12}$C, a few tens K, and thus for the determination of element isotope ratios such as $^{13}$C/$^{12}$C reflecting nucleosynthesis processes. However, the ratios of molecular isotope varieties are rarely just determined by the exponential factor exp(-$\Delta$E/kT) of thermal equilibrium. In particular for deuterated molecules such as HD/H$_2$, DCO$^+$/HCO$^+$, DCN/HCN, D$_2$CO/H$_2$CO, ND$_3$/NH$_3$, they depend in a complex way on the chemical processes and the thermal history of the molecular material. Their observational determination may thus provide precious information, such as the importance of cold grain chemistry in warm media. A recent breakthrough in this field is the discovery of poly-deuterated molecules (Lis et al.\,\,2002, van der Tak et al.\,\,2002), such as ND$_3$ (see references in Roueff et al.\,\,2005).
	
	The ubiquitous interstellar polycyclic aromatic hydrocarbons (PAHs) (Section 8) are in some respects intermediate between usual interstellar molecules and grains for their chemistry as well as for their physical properties (see e.g.\,\,Tielens 2005 and references therein, Omont 1986, Puget \& L\'eger 1989). Among significant processes are: 1) ionization processes: single and multiple ionization, Coulomb explosions, electron recombination and attachment; 2) photochemistry and other chemical reactions proceeding through unimolecular reactions in intermediate complexes: photo-dehydrogenation and loss of side groups, eventual photo-destruction, accretion of H and other gas atoms, ions and small molecules, etc.; 3) interactions with grains: accretion onto grains, PAH injection into the gas from grain desorption or shattering.
	
\subsection{Relation between interstellar and other cosmic molecules}

	Most cosmic media displaying significant amounts of molecules share some kind of similarity, especially for the element abundances, and to a lesser degree for the physical conditions such as temperature, density and UV radiation. It is thus not surprising that the other cosmic molecules present many similarities with interstellar molecules, such as the overwhelming importance of H$_2$, but also large differences. The latter are particularly obvious when there are large differences in density or chemical composition. 
	
	In the atmosphere of cool stars (red giants, supergiants and dwarfs) and brown dwarfs, the relatively high density leads to local  thermodynamical equilibrium (LTE) for the molecular abundances which are just determined by the laws of chemical equilibrium at temperatures of a few thousand Kelvin. With `normal' element abundances (C/O\,$<$\,1), one finds CO and H$_2$O as dominant molecules (after H$_2$), similarly to the ISM. However, the gas is much less rich in complex and C-bearing molecules. Another striking difference is the absence of dust (except in brown dwarfs and some M dwarfs: in fact below 200K dust plays a major role in the opacities of cool stars). The absence of dust makes metals free to efficiently form diatomic oxides, which, for a few of them, such as TiO, display spectacular spectral bands in the visible. In the special case of the atmospheres and circumstellar envelopes of carbon-rich giants (C/O\,$>$\,1), most oxygen is locked in CO; the dominant other molecules include then CN in the atmosphere and C$_2$H$_2$ and HCN in the colder circumstellar envelope.

	The stellar media which are the closest to the interstellar medium, are those of the extended `circumstellar envelopes' and `planetary nebulae' ejected at the end of the AGB red giant phase. The conditions in the cold, UV-shielded, circumstellar envelopes may be very similar to molecular clouds, but with a peculiar chemistry characterised by short time constants, clear-cut photochemistry and cases with C/O\,$>$\,1 (see e.g.\,\,Glassgold 1996). Planetary nebulae are more similar to 
	interstellar \HII~regions, with molecules and peculiar photodissociation regions in the youngest ones (Huggins et al.\ 2005). 
	
	The atmospheres of planets are close to LTE at low temperature, with abundant CH$_4$ and NH$_3$ in jovian planets, an absence of primitive H$_2$ and peculiar composition in telluric ones. Photochemistry in the upper layers may have some broad similarities with that in the ISM.
	
	However, the most obvious and fundamental connection is between interstellar molecules and comets. The molecules detected in comas are essentially a subset of those identified in the ISM. Their presence and abundance are very well explained from the action of photochemistry on parent molecules directly ejected from the comet nuclei (see e.g.\,\,Bockel{\'e}e-Morvan et al.\,\,2000). The latter are pretty similar to the most abundant interstellar molecules, and more precisely to those of the dust grain mantles.
	
	Indeed, comet nuclei have a direct relationship with interstellar dust and its UV-processed mantles accreted from interstellar gas. They are in some respects samples of the interstellar medium and their composition partially reflects that of both interstellar dust and gas (see e.g.\,\,Ehrenfreund \& Charnley 2000). The detailed analysis of comet samples will thus be fundamental, not only to understand the process of comet formation, but also as a unique information about properties and composition of the interstellar dust and even gas. In particular, they can cast some light on the existence and formation of molecules in the context of formation of the Solar System, and maybe in the general interstellar medium. A good example is that of amino-acids. Their detection is not yet confirmed by radio astronomy in the interstellar medium. Even if they are marginally detected in the future, the detectability of molecules with this type of complexity will remain at the limit allowed by the spectral confusion of the accumulation of all other weak interstellar lines. On the other hand, amino-acids have been  detected in  meteorites (see e.g.\,\,Cronin \& Pizzarello 1997, Ehrenfreund \& Charnley 2000). 
	
	Anyway, the rich variety of interstellar molecules just slightly less complex than amino-acids, detected by millimetre radio astronomy, shows that the special processes of interstellar chemistry may be very efficient in synthesizing the building blocks of pre-biotic molecules. The information one can expect from the combination of interstellar  molecules, comet nuclei, asteroids and meteorites, is at the heart of astro-biology and certainly  fundamental for eventually progressing in understanding the origin of life.
	
	\section{CO observations and gross structure of molecular gas in the Milky Way}
	
	\subsection{Introduction: atomic and molecular interstellar gas}

	Studying interstellar gas at galactic scales is essential to understand the structure and dynamics of galaxies, and the way they form stars. Except for the impeding factor of interstellar dust extinction, the bulk of the insterstellar medium is not easily accessible to optical astronomy because of its lack of emission at low temperature, the lack of adequate optical absorption bands and the blurring of absorption lines by large dust extinction. Its exploration has thus relied on longer wavelengths, essentially radio astronomy. First, since its discovery (Ewen \& Purcell 1951, Muller \& Oort 1951, Pawseyl 1951), the 21-cm line of atomic hydrogen, \HI, has shown the perfect tool for tracing the atomic gas, even in the mainly molecular medium. The ubiquity of \HI~which is generally the most massive component of the ISM in galaxies, the sensitivity of the 21-cm line, its lack of interstellar and atmospheric important absorption, together with relatively easy technical developments, have made \HI~21-cm studies one of the main drivers of the spectacular development of radio astronomy. Its multiple achievements are invaluable for tracing the interstellar gas and its various properties: amount and distribution in the Milky Way and various types of galaxies; kinematics tracing rotation and global masses including dark matter; details of structures and gravitational potentials including bars, spiral structure, warps; major gas complexes driving massive star formation; perturbed dynamics in galaxy interactions; exchanges with the intergalactic medium in inflows and outflows; magnetic field; etc (see e.g.\,\,Hartmann \& Burton 1994 and Kalberla et al.\,\,2005 and references therein for reviews and data about \HI~in the Milky Way, and Rogstad et al.\,\,1973, Scodeggio \& Gavazzi 1993 and Hoffman et al.\,\,1996 in galaxies). 

	Molecular gas is the next important component of the interstellar medium. As it traces the densest parts of the gas, its importance increases on average with significant concentrations of gas, especially in the disks of spiral galaxies such as the Milky Way. Immediately after the discovery of interstellar CO (Wilson et al.\,\,1970), it  became clear that molecular gas is ubiquitous in the Galactic disk, and it was rapidly identified in external galaxies (Rickard et al.\,\,1975). This has completely renewed our views about star formation in galaxies. The amount, distribution and properties of molecular gas are thus essential for understanding the present star formation and evolution of galaxies. The distribution of the molecular gas in the Milky Way, mostly within giant molecular clouds, has been established by CO surveys of the Galactic disk. It may serve as a benchmark to compare with molecular gas in other spiral galaxies which display a quite significant range of molecular material. Star formation is the key issue related to the molecular medium. Its global rate  eventually relies much on feeding by extra-galactic gas, and it may be greatly enhanced by galaxy interaction and merging, leading to violent starbursts (Section 5). The amount of molecular gas is generally smaller in galaxies other than spirals. It is also more or less correlated to the total amount of gas and star formation activity. The molecular gas has special properties in central regions of galaxies, where it may display high concentrations and strong starbursts, and may have a close relationship with bars and eventually the activity of the nucleus.
	
\subsection{Molecular gas in the Milky Way at galactic scale}

	While H$_2$ constitutes at least 99\% of the molecular gas, its lack of permanent dipole moment and the cold temperature, well below the excited energy levels, make it almost impossible to directly observe in most of the molecular medium (see Section 10). Therefore, the observation of the molecular gas essentially relies on other molecules and especially on CO (indeed most often the main isotope variety $^{12}$C$^{16}$O) which proves to be readily observable even in quite tenuous molecular gas. CO lines, and in particular the J=1--0 transition at 115\,GHz, are thus somewhat equivalent to the 21-cm line for the study of the molecular interstellar gas. CO is practically always the easiest molecule to detect in all molecular clouds of the Milky Way, as well as in any galaxy at any redshift, and it is systematically used to estimate masses in molecular gas. 
	
	We have a good view of the overall amount of CO, the remarkable statistical properties of molecular clouds and their distribution in the Milky Way, from the various Galactic $^{12}$CO and $^{13}$CO surveys (see e.g.\,\,Dame et al.\,\,1987, 2001,  Sanders, Solomon \& Scoville 1984, Clemens et al.\,\,1988,\ and other early references in Combes 1991, and papers in Clemens et al.\,\,2004 for recent surveys including BU-FCRAO and NANTEN [e.g.\ Tachihara et al.\ 2002]). 
	
	Most of the molecular mass ($\approx$\,90\%) appears to be in massive structures, distributed in clumps, the giant molecular clouds (GMC) with diameter $\sim$\,50 pc, masses $\sim$\,10$^{5-6}\,$\,M$_\odot$ and densities $\sim$\,10$^2$\,cm$^{-3}$ (see e.g.\,\,Scoville \& Solomon 1974, Williams et al.\,\,2000, Evans 1999 and references therein). A striking property is the relations between the internal linewidth $\sigma$ of the clouds, their size R and their mass M (Larson 1981), namely $\sigma$\,$\propto$\,R$^{\alpha_s}$ and M\,$\propto$\,R$^{\alpha_m}$, with $\alpha_s$ close to 0.5 and $\alpha_m$ close to two. They imply that the GMCs appear to have a constant average surface density and are in some way close to virial equilibrium. However, Larson's relations are indeed valid over six orders of magnitude, covering a much broader range than GMCs, and they are reminiscent of the Kolmogorov scaling law and suggestive of a fractal medium. Similar properties are revealed in the power spectrum of widespread \HI~emission, and at many, maybe most, scales, cloud boundaries are likely to be dynamic and transient.  It is today well agreed that the ISM and its structure are dominated at all scales by a complex variety of turbulent processes combining self-gravity, stellar pressures and magnetic fields. Understanding the turbulent ISM is constantly progressing from observations and simulations (see e.g.\,references in Williams et al.\,\,2000, Falgarone et al.\,\,2005a, and Elmegreen \& Scalo 2004 whose conclusions give a good idea of the great difficulties which have to be overcome to properly understand interstellar turbulence).  
	
	Because of the large isotope abundance ratio $^{12}$CO/$^{13}$CO  $\approx$\,\,60--70, but only an average factor $\sim$\,5 for the line ratio, $^{12}$CO and $^{13}$CO surveys are complementary to map the Galactic molecular gas. As $^{12}$CO lines are completely saturated in the densest parts of molecular clouds, observations of $^{13}$CO are better able to trace such regions and to provide detailed cloud structures  in surveys such as BU-FCRAO (Jackson et al.\,\,2006) and AST/RO (Martin et al.\,\,2004), although in many cases $^{13}$CO proves to be still optically thick and C$^{18}$O should be better. However, the higher optical thickness of the $^{12}$CO(1--0) line and the low gas density needed to excite this line are better matched to trace the large-scale envelopes containing most of the mass of the clouds. Therefore, the most comprehensive molecular survey of the Milky Way, practically complete for Galactic CO in GMCs, was carried out in this line during more than  two decades by the two 1.2\,m telescopes of the CFA group (Dame et al.\,\,2001). This survey provides information on individual molecular clouds in most regions and displays the main structural features of the molecular Galaxy. As known since the early surveys (e.g.\,\,Burton \& Gordon 1978 and Sanders et al.\,\,1984 and references therein), most of the molecular mass is concentrated in the region of Galactic radius between 3 and 7\,kpc, known as the `5\,kpc Molecular Ring' (Scoville \& Solomon 1975). The origin of such a distribution, which is indeed exceptional among other galaxies (Section 4.1), might be related to the Galactic bar (Combes 1991) and its relation to the Galactic inner bulge (Blitz \& Spergel 1991). Except for a few GMCs at high Galactic latitude, most of the H$_2$ mass is concentrated within a couple of degrees from the Galactic plane. The average H$_2$ column density decreases by a factor $\sim$\,30 from $|$$b$$|$ = 5$^\circ$ to 30$^\circ$ (Dame et al.\,\,2001).

	The $^{12}$CO lines are generally optically thick, and so care must be placed in interpreting the quantities derived. However, it has been empirically proven that the velocity-integrated CO(1--0) intensity, I$_{{\rm CO}}$ = $\int$T$^*$$_{\rm B}$dv, is not only the most sensitive qualitative tracer of the molecular gas, but also a good quantitative tracer, providing the column density N$_{{\rm H2}}$ of H$_2$ to a factor of a few. The conversion `$X$-factor'
\begin{equation}
	 {\rm X = N_{{\rm H}_2}/I_{CO}\ \  cm^{-2}\,K^{-1}\,km^{-1}\,s} 
\end{equation}
%%\noindent
may be determined for CO emission in the solar neighbourhood by different ways: correlation with diffuse gamma-ray emission, with far-infrared plus 21-cm surveys, optical or X-ray extinction (Combes 1991, Dame et al.\,\,2001, Lequeux 2005). Although $X$ varies by a factor of a few, in particular with the galactic latitude $b$, it is remarkable that this variation is not larger. Indeed, it shows little systematic variation from the value of 
\begin{equation}
		{\rm X = (1.8\pm 0.3)\,\times\,10^{20} cm^{-2}\,K^{-1}\,km^{-1}\,s}
\end{equation}
 for large clouds out of the Galactic plane ($|$$b$$|$ $>$ 5$^\circ$) (Dame et al.\,\,2001). See also the very good concordance with other derivations of X from $\gamma$ rays for local clouds 1.74$\pm$0.03 (Grenier et al.\,\,2005), or Galactic averages 1.9$\pm$0.2 (Strong et al.\,\,2004) and 
\begin{equation}
	 {\rm X = (1.56\pm 0.05)\,\times\,10^{20} cm^{-2}\,K^{-1}\,km^{-1}\,s}
\end{equation}
(Hunter et al.\,\,1997) which is probably the best average Milky Way estimate from $\gamma$ ray emission. 

	It is difficult to analyse in detail the origin of such a tight correlation between the intensity of $^{12}$CO emission and the mass of molecular gas since the CO line intensity results from a complex combination of the CO abundance, its excitation, and radiative transfer with random cloud size and velocity distribution (see e.g.\ Scoville \& Solomon 1974). However, a  qualitative insight may be gained by stressing that (see e.g.\,\,Solomon et al.\,\,1987, Maloney \& Black 1988): 1) for $^{12}$CO Galactic and extragalactic surveys, telescope beams encompass an ensemble of quasi-virialised molecular clumps obeying the general scaling relations; 2) most of their mass is located out of their cores in regions optically thin in the CO(1--0) line, and there is at most one clump on a line of sight emitting at given radial velocity; such a low filling factor avoids saturation effects when one increases the clump density; 3) for abundances close to solar, the photodissociation boundary of CO is close to that of H$_2$, and a nearly constant fraction of the cooling emission occurs in the CO(1--0) line, so that for each clump the CO emission is proportional to the H$_2$ mass if the heating by cosmic rays remains standard.
	
	The total derived mass of H$_2$ has the same uncertainty as the X-factor. For the Milky Way, it could be $\sim$\,1.0$\times$10$^9$\,M$_\odot$, excluding He, (Blitz 1996), which is significantly smaller than the \HI~ mass, $\sim$\,5$\times$10$^9$\,M$_\odot$ (Wouterloot et al.\,\,1990). 

	While most of the molecular mass in the Milky Way and other galaxies is accounted for by molecular clouds and especially GMCs, it is important to remind that H$_2$ and other molecules are also found in a number of other Galactic environments in amounts much smaller, but quite significant for the variety and the importance of the media implied, and their wide spread nature at Galactic scales. They include atmospheres of cold stars, giants  and dwarfs, and brown dwarfs; AGB envelopes and planetary nebulae; planetary atmospheres; supernova remnants; various stellar outflows, extragalactic outflows; etc. The most relevant case for a general description of the interstellar medium is that of the molecules of the diffuse `atomic' gas, i.e. of all the \HI~regions. Despite the fact that molecules are rapidly photodissociated by the ambient UV radiation, reformation of H$_2$ on grains maintains a certain abundance of H$_2$, which initiates the formation of small amounts of other simple molecules through  cosmic ray reactions and photochemistry. As the diffuse gas is not UV optically thick, absorption spectroscopy of strong UV lines is the main tool for studying first H$_2$ and a few other molecules such as CO and OH, provided there are lines of sight with strong background UV sources such as hot stars or quasars. Let us also recall the importance of molecules of photodissociation regions at the boundary of the molecular and atomic media, and of the very large aromatic molecules (PAHs) which pervade the photodissociation regions and the whole diffuse medium (Section 8).

	It has been proposed that in addition to the large mass of H$_2$ in regular molecular clouds traced by CO millimetre emission, an even larger mass of H$_2$ might be hidden in the shape of extremely cold H$_2$ in the outer Galactic disk (Pfenniger, Combes \& Martinet 1994, Pfenniger \& Combes 1994, Wardle \& Walker 1999, Pfenniger 2004, Combes 2006, Bell et al.\ 2006). It has even been considered that such cold H$_2$ could explain part or the totality of the Galactic dark matter implied by the Galactic rotation curve, especially in the outer disk. There are interesting arguments in favour of such an assumption. They  include: an apparent excess of very cold dust with respect of the amount of \HI~traced by the 21\,cm line and H$_2$ traced by CO emission (e.g.\ Cambresy et al.\,\,2001 ); a similar excess in $\gamma$ ray emission with respect to \HI~+ H$_2$ (Dixon et al.\,\,1998, Grenier et al.\,\,2005)[but the lacking gas needed to match cold dust or $\gamma$ rays could as well be in the shape of additional cold \HI~gas difficult to measure with precision]; 
	%%the similarity of the distribution of \HI~and dark matter in some dwarf galaxies (Broells et al.\,\,1992);
 the interest of a large reservoir of gas to explain the evolution of spiral galaxies; etc. However, direct evidence of such very cold H$_2$ is still lacking. Its wide distribution is ruled out by the lack of significant detection by absorption of H$_2$ UV lines, despite the number of lines of sight studied, in particular by FUSE. The only possibility would be that H$_2$ is condensed in very small globules of mass and core radius maybe as small as the Earth (Pfenniger 2004), or more diffuse gaseous H$_2$ with $\sim$\,10$^{-3}$M$_\odot$ and $\sim$\,10 AU (Rafikov \& Draine 2001). Such globules should be very difficult to detect because of their very small filling factor ($\sim$1\%) and their opacity to UV radiation. 
 %%Lensing of background objects in particular stars of nearby galaxies should be almost the unique way to reveal their presence. 
 The lack of absorption evidence in the search for microlensing events in the direction of the Large Magellanic Cloud constrains the properties of the proposed globules (Alcock et al.\,\,2000, Lasserre et al.\,\,2000) constrains the properties of the proposed globules, but does not completely rule them out (Rafikov \& Draine 2001). Clumps of ionized and \HI~interstellar gas have been observed in radioastronomy in `Extreme Scattering Events' in front of radio quasars, and in VLBI observations of the 21\,cm absorption line. They have the right size and could be ionized and atomic envelopes of such molecular globules immersed in the interstellar UV radiation; but the implied individual atomic or ionized gas masses have no common measure with the much higher masses invoked in self-gravitating H$_2$ globules. 

	To summarise, apart from elusive cold H$_2$ globules, we have a very good view of the properties and distribution of molecular gas in the Milky Way from CO surveys. The bulk of this gas is distributed in GMCs along the thin Galactic disk, where most of young stars form. Despite the complex structure of GMCs, the integrated intensity of the CO(1-0) line provides the most widely used quantitative estimate of the amount of H$_2$ through the `H$_2$/CO conversion factor'. However, the physical origin of this tight empirical relation turns out to be complex. Besides, complementary $^{13}$CO surveys are needed to trace the densest parts of molecular clouds.

%\end{document}

	\section{CO observations and gross structure of molecular gas in other galaxies}

\subsection{Molecular gas in spiral galaxies from CO surveys of local galaxies}

% Fig. 1
   \begin{figure*}
    \centering %%\includegraphics[width=14.cm, angle=270]{M31_abb1.eps}
\caption{({\it reproduced from Fig.\ 1 of Nieten et al.\ 2006)} Distribution of CO emission in M\,31 ({\it Andromeda}). {\bf (a)} ({\it top}) The velocity-integrated intensity distribution of the
$^{12}$CO(1--0) spectrum, observed with the IRAM 30-m telescope. The X and Y coordinates are taken along the major and minor axis, respectively. The
dashed line marks the border of the area surveyed which is about one
degree squared. 
{\bf (b)} ({\it bottom}) The velocity field as traced by the CO
emission. }
 \end{figure*}
		
		Because H$_2$ is so difficult to detect in the normal interstellar medium (Section 10), most of our knowledge about the global molecular gas of other galaxies than the Milky Way also relies on CO observations, with a similar problematics to precisely infer the amount of H$_2$. In parallel with mapping molecular gas in the Milky Way, since the first detection of extragalactic CO (Rickard et al.1975), enormous efforts have been devoted in the last three decades to perform the inventory of molecular gas in galaxies, through systematic millimetre observations of CO in a large, diversified sample of local galaxies. This enterprise has first addressed the main reservoir of molecular gas, large spiral disk galaxies with their various classes, including that of the Milky Way. Single dish CO observations, with relatively low angular resolution, were generally first performed to derive global properties of the molecular gas, its total amount, proportion with respect to \HI  , relation with metallicity, morphology, star formation, etc. Detailed studies of fine details have then been undertaken, mostly with interferometers. In parallel, the studies have been extended to other classes and peculiar galaxies where the importance of molecular gas is more marginal.
		
		The results are impressive and provide a remarkably complete view of molecular gas in (local) galaxies. The emerging picture is perhaps more complex than initially thought. Molecular gas is extremely intricated with atomic gas, and the most important quantity for the evolution of galaxies and even in some way star formation, is the total amount of gas \HI~+ H$_2$. Molecular gas is intricately woven with the atomic gas. Because of the constant exchanges between \HI~and H$_2$, the gas should be considered globally with its dynamics, transport properties including exchanges with the intergalactic medium and the central regions, and condensation processes leading to star formation and its feedback action on the gas. We will defer most problems related to star formation, including starburst galaxies, to Section 5.
		 
		The total number of local (D\,$\la$\,100\,Mpc) galaxies where CO data  has been detected, approaches 1000,  whith more than half being `normal' spirals Sa--Sb (see, e.g.\,\,the global analysis by Casoli 1998, the catalogue for non interacting galaxies compiled by Bettoni et al.\,\,2003, and more recent observations including Boselli et al.\,\,2002, Sauty et al.\,\,2003, Hafok \& Stutsky 2003, Yao et al.\,\,2003, Garland et al.\,\,2005, Leroy et al.\,\,2005). These data were mostly obtained in dedicated surveys of the $^{12}$CO(1--0) line with 12-15\,m radiotelescopes whose intermediate size is well fitted to such programmes: the most organized, initial long-term effort was performed with the FCRAO 14\,m telescope and achieved more than one third of all these detections (Young et al.\,\,1995, 1986, 1989). Other important contributions came from various groups (see references in Bettoni et al.\,\,2003 and recent results quoted above), using mainly the NRAO 12\,m dish and SEST 15\,m, as well as other telescopes such as BTL, KOSMA, JCMT, etc., and the larger dishes of Onsala 20\,m, IRAM 30\,m and Nobeyama 45\,m.
		
		Most of these observations, carried out with a limited angular resolution (e.g.\,\,45'' with 14\,m FCRAO, $\sim$\,2\,kpc at 10\,Mpc), were single pointings aimed at a global CO detection from the central regions in objects with small angular extension, or often a few pointings along the major axis in order to trace the large-scale radial distribution of molecular gas (e.g.\,\,Young et al.\,\,1995). However, several tens of detailed maps have been carried out with the much better angular resolution of the largest single dishes (e.g.\,\,Nishiyama \& Nakai 2001) or especially interferometers such as the BIMA survey of 44 nearby galaxies (Regan et al.\,\,2001), and the  high-resolution
(\,0.5''--1'') NUGA survey at IRAM of 12 low luminosity AGN (Garc{\'{\i}}a-Burillo et al.\,2003). Comprehensive high resolution maps were performed for most of the members of the Local Group:  M\,31 with FCRAO (Loinard et al.\,\,1999) and IRAM-30m  (Neininger et al.\ 1998, 2001, Nieten et al.\,\,2006; see Fig.\ 1); M\,33 (Engargiola et al.\,\,2003), dwarf spheroidals (Blitz \& Robishaw 2000) with BIMA; and the Magellanic Clouds with SEST and NANTEN (see below). See also the review by Blitz et al. (2007) of the properties of GMCs in the Local Group., and CO maps of M\,51 by Scoville \& Young (1983) and Schuster et al.\ (2007).

		Typical massive spirals are the standard habitat of molecules in the Universe. Such a large sample of several hundreds provide a very comprehensive view of their molecular medium with its complexity. Various analyses of the properties of molecular gas in spiral galaxies have been published in the last two decades (e.g.\,\,Young \& Knezek 1989, Young \& Scoville 1991, Bregman et al.\,\,1992, Braine et al.\ 1993, Latter et al.\ 1996, Casoli et al.\,\,1998, Boselli et al.\,\,2002, Wong \& Blitz 2002, and references therein). The first striking observation is the enormous range of variation of the molecular content of galaxies, even within the relatively homogeneous class of massive spirals Sa-Sc (e.g.\ Bettoni et al.\,\,2003). The ratio I$_{{\rm CO}}$ /M(HI), approximately proportional to M(H$_2$)/M(HI), spans two orders of magnitude for a given value of any galactic scaling parameter such as the blue or far-infrared luminosities L$_{\rm B}$ and L$_{{\rm FIR}}$. The average values of I$_{{\rm CO}}$ /M(HI), however, clearly correlate with various quantities such as the metallicity, the dynamical mass, L$_{\rm B}$ or L$_{{\rm FIR}}$.

%Fig. 2
    \begin{figure*}
    \centering
\caption{{\bf (a)} ({\it left, reproduced from Fig.\ 2 of Boselli et al.\ 2002)} The relationship for a template sample of nearby galaxies between the X conversion factor from CO line intensity to
H$_2$ column density (Eq.\ 1) and the metallicity index 12 + log(O/H). 
{\bf (b)} ({\it right, reproduced from Bettoni et al.\ 2003}) The mean molecular to atomic gas content ratio as a function of galaxy Hubble type t. The open symbols are derived from the catalogue of Bettoni et al.\ (2003), while full symbols
represent ratios published by Casoli et al.\ (1998). 
}
 \end{figure*}
 
%\end{document} 
 		
		The difficulty of precisely estimating the conversion factor X = N$_{{\rm H2}}$/I$_{{\rm CO}}$ adds a major source of uncertainty in the values of M(H$_2$)/M(\HI). Its calibration for the various types of galaxies remains a very complex and uncertain process. There is presently no firm basis for a precise physical calibration because of the complexity of the interstellar medium, although better understanding the physics may provide the best constraints for its value. Its empirical calibration relies on the detailed information on the local ISM of the Milky Way where the gamma rays are the fundamental calibrator (Section 3.2), and on less accurate methods applied to a sample of various types of well studied galaxies, generally nearby (see e.g.\,\,Boselli et al.\,\,2002). The most useful of these methods for estimating the mass of the molecular gas seems the virial equilibrium of giant molecular clouds (Young and Scoville 1991), or the determination of the mass of cold dust from millimetre/submillimetre observations assuming a metallicity-dependent gas-to-dust ratio (Gu\'elin et al.\ 1995). However, even in nearby galaxies such determinations of X remain very uncertain. It should also be reminded that, inside a given object, X may change by a factor $\sim$10 from the diffuse medium to the core of the GMCs (Polk et al.\,\,1988). Despite such difficulties, the most recent analysis by Boselli et al.\,\,(2002) has confirmed that the most massive spiral neighbours of the Milky Way have comparable X factors, but with non negligible variations up to a factor $\sim$2 of the best average MW estimate from $\gamma$ ray emission (Eq.3), 1.56\,$\pm$\,0.05 10$^{20}$\,cm$^{-2}$\,(K$\times$km$\times$s$^{-1}$)$^{-1}$ (Hunter et al.\,\,1997). Many former studies use a previous MW calibration of X, 2.38\,10$^{20}$; they should thus be recalibrated by at least a factor $\sim$1.5, maybe more, as estimated for the average of X with a larger sample by Boselli et al.\,\,(2002). Consequently, it seems now well established that the global amount of molecular gas is on average significantly smaller than the atomic one, with average values of M(H$_2$)/M(\HI) $\approx$ 0.2--0.4 for large spirals depending on the sample (see Fig.\ 2b and last discussions by Casoli et al.\,\,1998, Bettoni et al.\,\,2003 and Boselli et al.\,\,2002), which is similar to the Milky Way value $\approx$ 0.2 (e.g.\,\,Blitz 1996).
		
	It is also well established that the molecular gas radial distributions in most spirals galaxies are centrally peaked with exponential profiles, and thus markedly different from what is seen in \HI~(see e.g.\,\,Young 2000). Indeed, the Milky Way is one of the rare exceptions with its  CO depression interior to the molecular ring. For the vertical distribution, the molecular gas is more concentrated in a thin disk than \HI, as seen from the CO emission of edge-on galaxies. However, the arm-interarm contrast is generally not very strongly marked (but see Loinard et al.\,\,1999 for M\,31).
	
	An easily accessible quantity, the isotopic line ratio I($^{12}$CO)/I($^{13}$CO), may give some indication about the value of X and its variation across a galaxy. However, this ratio is sensitive to variation in temperature and column density, as well as isotope fractionation and especially isotope-selective photo-dissociation. Mapping surveys of $^{12}$CO and $^{13}$CO emission in nearby galaxies have been used to trace the properties of their molecular gas and the variation of X, especially using the 4x4 receiver array of FCRAO by Paglione et al.\,\,(2001 and references therein). They have concluded that similar physical processes may affect the value of X and I($^{12}$CO)/I($^{13}$CO), and that X might decrease by factors of 2--5 from disks to nuclei.
	
	Stellar bars drive gas into the circumnuclear region of galaxies. CO studies of the molecular gas in bars are important to investigate their structure and dynamics, and their influence on star formation in circumnuclear regions. Such CO observations of the  bar and the inner region (about the central kiloparsec) have been carried out in a few tens of  barred galaxies (e.g.\,\,Kenney et al.\ 1992, Regan et al.\ 2002, Lee et al.\ 2006, Jogee et al.\ 2005). It is found that the mean nuclear molecular gas
surface density of barred spirals is significantly higher than that of unbarred spirals, explaining in part why powerful starbursts reside there.

\subsection{Molecular gas in other types of galaxies}
%%(starbursts deferred to Sec. 4)

	Even within the relatively homogeneous class of massive spirals, there are very significant variations of the relative amount of molecular material and the ability of CO to trace H$_2$ through the X-factor, between different galaxies and within a given galaxy.  The relationships between the H$_2$/CO conversion factor X and various galactic parameters -- UV radiation field, metallicity,  blue and near-IR luminosities -- have been analysed by Boselli et al.\ (2002) and displayed in their figure 2. As expected, there is a strong dependence of X on these parameters, e.g.\ the CO abundance, and hence X$^{-1}$, decrease with the UV strength and increase with the metallicity  (see e.g.\ Fig.\ 2a for the metallicity). 
%%	Since they may be mostly explained by variations in gas amount, metallicity and physical conditions, especially UV intensity,
It is thus not surprising that the range of the variations of X is much increased when one encompasses the full variety of galaxies. 
Other galaxies are generally less favourable than spirals for the survival of  molecules in general and CO in particular, their molecular gas is therefore less spectacular, and CO emission even relatively weaker. Even the latest-type spirals are in general not strong CO emitters  (B\"oker et al.\,\,2003, Young \& Knezek 1989). Recent models of galaxy evolution, incorporating the formation of H$_2$ out of HI gas, have also explored the possibility of a diffuse H$_2$ gas phase outside star-forming regions (e.g. Pelupessy et al.\ 2006). The mass of such diffuse warm H$_2$ should be significantly underestimated by CO observations in metal-poor regions (Papadopoulos et al.\ 2002).
	
	The Magellanic Clouds are particularly interesting because of their proximity which allows us to check the molecular gas with much detail about conditions of star formation, UV intensity and metallicity very different from the Milky Way. A first full coverage in the $^{12}$CO(J\,=\,1–-0) emission line at 115\,GHz, was performed at low angular resolution with the Columbia 1.2\,m telescope (Cohen et al.\,\,1988). Then both LMC and SMC were the object of systematic survey studies of $^{12}$CO and $^{13}$CO  with the SEST 15\,m telescope (Rubio et al.\,\,1991, Israel 1997, Israel et al.\,\,1993, 2003, and references therein), and with NANTEN (Mizuno et al.\,\,2001a,b, Yamaguchi et al.\,\,2001) (see also studies of higher-J lines of CO  with AST/RO, Bolatto et al.\,\,2005). At the distance of the Clouds, the SEST beam size, 13\,pc, is smaller than the typical size, $\sim$20\,pc, of their giant molecular clouds, itself significantly smaller than GMCs in the Milky way, $\sim$50\,pc. Dwarf galaxies such as the Magellanic Clouds are generally poor CO emitters, so that most SEST studies were limited to regions with significant CO emission, mainly traced by far-infrared IRAS emission. CO was detected at strengths significantly smaller than those expected from Galactic sources at Magellanic Cloud distances, typically three times weaker in the LMC and an order of magnitude lower in the SMC (Israel et al.\,\,1993). The ratio X\,=\,N$_{{\rm H}_2}$/I$_{{\rm CO}}$ is also 2--3 times larger than in the Milky Way. Similarly, the emission of CO associated with well developed \HII~regions remains quite modest, and the lack of diffuse CO emission there (e.g., Lequeux et al.\,\,1994) suggests that these molecular clouds are generally part of photo-dissociation regions (PDRs). 
	 
	Large samples of Magellanic type dwarf galaxies and irregulars have been observed: for instance, Albrecht et al.(2004) detected 41 galaxies with the IRAM 30m-telescope, and Leroy et al.\ (2005) detected 28 with BIMA (see also Hoffman et al.\ 2003). Leroy et al.\  found that the CO luminosity is most strongly correlated with the K-band and the far-infrared luminosities. There are also strong correlations with the radio continuum and B-band luminosities and linear diameter. Conversely, they found that far-IR dust temperature is a poor predictor of CO emission within the dwarfs alone, although a good predictor of normalized CO content among a larger sample of galaxies.

	Various studies of low surface brightness galaxies (LSBs), (e.g.\,\,Schombert et al.1990, O'Neil \& Schinnerer 2004, Matthews et al.\,\,2005) have demonstrated that despite their typical low metallicities and low mean gas surface densities, some LSB galaxies contain a molecular medium that is traced by CO. M$_{{\rm H2}}$ and M$_{{\rm H2}}$/M$_{{\rm HI}}$ values fall within the ranges typically found for high surface brightness objects, albeit at the low end of the distribution,

	CO was detected in several tens of elliptical galaxies (e.g.\,\,Wiklind et al.\,\,1995, Knapp \& Rupen 1996, Sofue \& Wakamatsu 1993, and references therein). It was found that the CO-to-dust abundance ratio in elliptical galaxies is approximately the same as that for spirals and for local molecular clouds. The molecular gas masses range from 2$\times$10$^6$ to 10$^9$\,M$_\odot$, and appear to be unrelated to the underlying stellar population. This suggests an external origin of the gas.  Low excitation temperatures for CO transitions in galaxies with cold dust could lead to an underestimate of the molecular gas mass by a factor of 5. The average M$_{{\rm H2}}$/M$_{{\rm HI}}$ ratio for the elliptical galaxies is 2-5 times lower than for normal spiral galaxies.

	Millimetre CO emission has been detected in the cooling flows of a dozen central massive elliptical galaxies of clusters (e.g.\,\,Edge \& Frayer 2003, Wilman et al.\,\,2006, Salom{\'e} \& Combes 2003, 2005, 2006 and references therein). It is important to understand the exact nature of their complex structures (bubbles, cavities, cold fronts) unveiled by X-ray data, which  may contain huge optical nebula. It seems now established that cooling flows entertain some fueling of the AGN activity which reheats the intra-cluster gas. CO was also detected in the context of galaxy collisions in the tidal debris of violent galaxy-galaxy interactions (Braine et al.\ 2000), and particularly in the Stephan's Quintet group of galaxies (Lisenfeld et al.\ 2004). The cold gas is probably a mixture of gas falling down on the central galaxy and of uplifted gas dragged out by a rising bubble in the intracluster medium. Its peculiar morphology and kinematics argue for the picture of an intermittent cooling flow scenario where the central AGN plays an important role.
	
	%%%%		
		To sum up, the picture of the molecular medium coming out of CO observations of many hundreds of galaxies is complex, reflecting the variety of the history, mass, luminosity and  metallicity of the host galaxies. However, as expected, definite correlations emerge between the last parameters and the amount of molecular material as well as with the `X conversion factor', N$_{{\rm H2}}$/I$_{{\rm CO}}$. In spiral disks, the average fraction of the interstellar gas in molecules does not much depend on the precise morphological type of the galaxy and is comparable to the Milky Way value, $\sim$\,0.2. But it significantly decreases in earlier and later types, i.e.\ ellipticals and irregulars, enhancing the dearth of molecular gas there where the total amount of gas is low. The increase of the X factor itself with decreasing metallicity renders more difficult the detection of CO in irregular dwarfs. However, the proximity of the Magellanic Clouds provide us with very sensitive benchmarks of the CO distribution in such cases. The detection of CO in a number of ellipticals shows that it can be an interesting tracer of gas when present, as well as in cooling flows of central massive ellipticals of clusters.	
	
\section{Molecules as tracers of star formation at galactic scales}

\subsection{Introduction. Star formation rate}

	Star formation is a combination of complex processes of the interstellar medium, eventually culminating in fragmentation and collapse of stellar size clumps (e.g. McKee \& Ostriker 2007). Most of the steps imply densities where the interstellar gas is necessarily molecular. 
One may thus expect strong correlations between the amount of molecular gas and the efficiency of star formation at all scales. Indeed, in external galaxies only massive star formation is detectable from the UV energy it generates. The star formation rate is therefore directly characterised by the UV luminosity, or by the induced H$\alpha$ or far-infrared luminosities. The derived star formation rate (SFR, in units of M$_\odot$/yr) relies on usual assumptions about the stellar initial mass function (IMF). The star formation efficiency is usually defined as the star formation rate per unit mass of interstellar gas. We will stress in Section 5.2 that there are also good reasons to define it as star formation rate per unit mass of {\it molecular} gas. The relevant scale for massive star formation is indeed that of giant molecular clouds which have the right mass to eventually form one or several clusters of massive stars. In the Milky Way, it is well proved that most star formation takes place in GMCs. In estimating SFRs, dust extinction should be carefully taken into account for correcting the UV or H$\alpha$ luminosities, or estimating the fraction of the UV energy which is processed into the far-infrared. When this fraction is large, the far-infrared luminosity, L$_{{\rm FIR}}$ is a better indicator of the star formation rate. GMCs are also the right scale for discussing the various feedback processes associated with massive star formation, either positive ones propagating star formation by compressing the interstellar gas by stellar winds or supernovae blast waves; or negative ones by cloud destruction.

\subsection{Star formation and molecular clouds in non-starburst galaxies}

	Since the first extragalactic CO surveys, it was noticed that there is a strong correlation between the CO intensity I$_{{\rm CO}}$	and both the far-infrared and H$\alpha$ luminosities. However, making this correlation quantitative with SFR raises several difficulties even for non-starburst galaxies including the most important case of normal spirals. First, a precise value of SFR is difficult to estimate from both L$_{{\rm FIR}}$  and L$_{{\rm H}\alpha}$, although it can be approximately and consistently calibrated on both, yielding SFR roughly proportional to I$_{{\rm CO}}$  for normal spirals (see e.g.\,\,Young 2000, Gao \& Solomon 2004b,  
and references therein). The use of L$_{{\rm FIR}}$  has the advantage to be very well fitted to the extension to starburst galaxies (see Section 5.3). But for normal spirals, it is not easy to estimate the fraction of the UV radiation emitted by young stars which is absorbed by dust and reemitted in the far-infrared. L$_{{\rm H}\alpha}$  is thus often preferred for normal spirals; however, it must be corrected for extinction, either by a uniform average factor (e.g.\,\,Kennicutt 1998a) or, better, individually for each galaxy (e.g.\,\,Boselli et al.\ 2002).
	
	For galaxies with low CO emission (Section 4.2), it is not surprising that the correlation between SFR and I$_{{\rm CO}}$  is very poor since CO no longer well reflects the H$_2$  mass. However, if one properly determines M$_{{\rm H2}}$  with the right X-factor, the correlation between SFR and M$_{{\rm H2}}$  remains at the same level of accuracy as for normal spirals, i.e. with the same average value of the star formation efficiency SFE = SFR/M$_{{\rm H2}}$, with a similarly large dispersion.
	
	It has been shown that there is a similar correlation between SFR and the total amount of gas, M$_{{\rm H2}}$  + M$_{{\rm HI}}$, with the obvious explanation that the average value of M$_{{\rm H2}}$ /M$_{{\rm HI}}$ is roughly constant for spirals (Boselli et al.\ 2002). Kennicutt (1998a,b) suggested that the total amount of gas surface density is indeed the most fundamental factor for determining SFR. This is certainly plausible for the average star formation rate on cosmological time scales. But the present star formation rate is much more likely related to the amount of molecular gas, and even the amount of dense molecular gas (see Section 5.3), than to M$_{{\rm HI}}$. The definition of the star formation efficiency with respect to M$_{{\rm H2}}$, SFE = SFR/M$_{{\rm H2}}$, could thus be  preferable (Boselli et al.\ 2002, Wong \& Blitz 2002).
	
	In the most accurate determinations of SFE, the value of SFE is thus comprised between $\sim$10$^{-8}$ and $\sim$10$^{-10}$\,yr$^{-1}$, irrespectively of the morphological type, for most non-starburst galaxies (see e.g.\,\,Fig.\,9 of Boselli at al.\ 2002). Such short timescales, between $\sim$10$^8$  and 2$\times$10$^9$\,yr, for the consumption of the current molecular gas by star formation, means that it should be renewed at similar rates from the \HI~reservoirs, first Galactic and eventually extragalactic.

\subsection{Molecular gas and starbursts in luminous and ultra-luminous infrared galaxies (LIRGs and ULIRGs)}
 
	The advent of far-infrared astronomy, mostly with IRAS, has revealed the existence of galaxies with L$_{{\rm FIR}}$  one or two orders of magnitude larger than for normal galaxies: luminous infrared galaxies (LIRGs) with L$_{{\rm FIR}}$\,$>$\,10$^{11}$ L$_\odot$  and ultra-luminous ones (ULIRGs) with L$_{{\rm FIR}}$\,$>$\,10$^{12}$ L$_\odot$ (see e.g.\,\,Sanders \& Mirabel 1996, Lonsdale et al.\ 2006). These high luminosity galaxies are directly powered by gigantic starbursts mostly dust enshrouded, with star formation rates of several ten or hundred M$_\odot$/yr, which may be directly inferred from L$_{{\rm FIR}}$ if the general relation, SFR\,$\approx$\,2x10$^{-10}$\,L$_{{\rm FIR}}$ (Kennicutt et al.\  1998a), applies. From their perturbed morphology, especially for the most luminous ones (L$_{{\rm FIR}}$  $>$ 10$^{12}$L$_\odot$), it is clear that these starbursts are often triggered by strong interactions with a close companion, eventually leading to a complete or partial merging. Their luminosities are dominated by dust heating within molecular clouds of circumnuclear starbursts. However, some nuclear activity at a relatively low level is also present in many of them. Both starburst and AGN activities are fueled by the presence of huge amounts of molecular gas which has been driven into the merger nucleus. Even if a large number has been identified by IRAS at z\,$\la$\,0.1--0.3, LIRGs and especially ULIRGs are relatively rare in the local universe, but they are orders of magnitude more numerous at high redshift (see Section 9). High-z ULIRGs may represent important steps in the formation of elliptical galaxies, and also in the growth of their massive black holes, and thus in the genesis of quasars.
	
	Up to redshifts of $\sim$0.1 for LIRGs and $\sim$0.3 for ULIRGs, they are well within the range of sensitivity of the best present facilities for comprehensive studies of the most prominent molecules, CO, HCN and OH masers. In addition, interferometric studies in the radio continuum (e.g.\ Turner \& Ho 1994) may provide high angular resolution diagnosis of the structure of the starburst and hence of the molecular medium through the general extraordinary FIR-radio correlation (Condon 1992) between the star formation power generated in the starburst and the synchrotron radio luminosity of its supernovae.

	It is clear that the neutral gas of the central CO-emitting region is almost entirely molecular. However, it is known that using the Milky Way value for the molecular gas mass to CO intensity ratio, X\,=\,N$_{{\rm H2}}$/I$_{{\rm CO}}$, overestimates the gas mass in ULIRGs since it may yield a molecular gas mass comparable to and in some cases greater than the dynamical mass of the CO-emitting region (Sanders et al.\ 1986, Sanders et al.\ 1991, Scoville et al.\ 1991, Downes et al.\ 1993, Solomon et al.\ 1997, Solomon \& Vanden Bout 2005). The reason is probably that the structure and the temperature of the molecular gas in the centers of ULIRGs is different from the individual virialized clouds of disks of normal galaxies. Extensive high-resolution mapping of CO emission from ULIRGs shows that the molecular gas is in
rotating disks or rings. Kinematic models (Downes \& Solomon 1998) in which most of the CO flux comes from a moderate density warm intercloud medium, have yielded conversion factors for deriving the mass of molecular mass from CO emission approximately {\it 4-5 times lower} than standard values for the Milky Way; namely X = N$_{{\rm H2}}$/I$_{{\rm CO}}$ $\approx$ 0.4\,$\times$\,10$^{20}$ cm$^{-2}$\,(K\,km\,s$^{-1}$)$^{-1}$ 
%%and $\alpha$ = $M_{\rm gas}/L^\prime_{\rm CO}$ $\approx$ 0.8\,M$_\odot$\,(K\,km\,s$^{-1}$\,pc$^{2})^{-1}$. 
Such a value for X is often used by observers, even for high-z ULIRGs where such a calibration is more uncertain (Section 9).

	There is a correlation in LIRGs and ULIRGs, as well as in other galaxies, between {\it the CO luminosity} L$_{{\rm CO}}$ and the far-infrared luminosity L$_{{\rm FIR}}$ which traces the star formation rate (see e.g.\,\,Sanders \& Mirabel 1996, Kewley et al.\ 2002). However, such a relation is not linear, and the ratio between L$_{{\rm CO}}$ and L$_{{\rm FIR}}$ decreases with increasing L$_{{\rm FIR}}$ (Sanders \& Mirabel, 1996, Solomon et al.\ 1997, Gao \& Solomon 2004b). The reason is probably that CO mainly traces the low density gas of giant molecular clouds (see Section 3 \& 4), but not their active star forming hot cores. On the other hand, HCN {\it is a much better tracer} of the dense regions and thus of star formation. This has been shown by Gao \& Solomon (2004a) who found a tight correlation between the HCN luminosity L$_{{\rm HCN}}$ and L$_{{\rm FIR}}$ in a sample of 65 normal spiral and starburst galaxies. The correlation remains almost linear over a factor of 10$^3$ in luminosity from normal galaxies to LIRGs and ULIRGs. Wu et al.\ (2005) have recently shown that the correlation between L$_{{\rm HCN}}$ and L$_{{\rm FIR}}$ continues up to the much smaller scale of Galactic dense cores, and they argue that it could be explained if the basic unit of star formation in all galaxies is a dense core similar to Galactic ones. A large CO, HCN multi-transition survey of 30 LIRGs is nearing completion with JCMT and the IRAM 30-m telescopes (Papadopoulos et al.\ 2007), and the properties of the dense molecular gas have been studied in a sample of 17 nearby LIRGs and ULIRGs through observations of HCO$^+$, HCN, CN, HNC and CS (Gracia-Carpio, J. et al.\ 2007).
	
\subsection{OH mega-masers.} Since luminous infrared galaxies are those where the molecular medium is the most enhanced, it is not surprising that nearby LIRGs such as M\,82 provided the first extragalactic detections of many molecules (Section 6 and Table 2). This is true not only for molecules such as HCO$^+$ or CS which, together with HCN, are well known tracers of dense regions, but also for widespread molecules such as OH. Absorption 18\,cm lines of OH were detected very early in M\,82 (Weliachew 1971). Later, OH emission lines were detected (Nguyen-Q-Rieu et al.\ 1976), with maser amplification of the background radio continuum and intensities 10 times stronger than bright Galactic OH masers. However, it was a surprise to discover OH maser emission with  many orders of magnitude higher ( $\sim$10$^8$ times that of typical OH Galactic masers) first in the ULIRG Arp\,220 (IC\,4553) (Baan et al.\ 1982) and then in many luminous and ultra-luminous infrared galaxies (see detailed recent review on such OH `mega-masers' by Lo 2005, to which we refer, avoiding detailed developments on this important topic). OH mega-masers taking place in powerful starburst galaxies present a strong correlation between L$_{{\rm OH}}$ and L$_{{\rm FIR}}$, with L$_{{\rm OH}}$\,$>$\,10$^4$\,L$_\odot$ associated with L$_{{\rm FIR}}$\,$>$\,10$^{12}$\,L$_\odot$. A quadratic dependence between L$_{{\rm OH}}$ and L$_{{\rm FIR}}$ was even  believed for a while, L$_{{\rm OH}}$ $\propto$ L$_{{\rm FIR}}^2$; but it is now proved from the combined analysis of 95 OH mega-masers that the relation is rather L$_{{\rm OH}}$ $\propto$ L$_{{\rm FIR}}^{1.2\pm 0.1}$ (Darling \& Giovanelli 2002). The main pumping mechanism is thought to be mid-infrared pumping by OH rotational lines. However, as for other interstellar OH masers, the whole process of mega-maser emission is very complex, as indicated in particular by the non understood absence of mega-maser detection in a large fraction (80\%) of infrared luminous galaxies. The possibility of performing high angular resolution VLBI studies of OH mega-masers provides important clues about their origin as well as the structure and physics of the starburst regions where they take place (see e.g.\,\,Lo 2005). It seems that OH mega-masers could mostly trace compact extreme starburst regions where the conjunction of very strong infrared and radio emission may create favourable conditions for mega-maser emission. Combined high resolution studies of OH and CO in nearby ULIRGs, such as Arp 220, are consistent with such ideas; however, further elucidation of the physics involved is required. 

\section{Molecular abundances in various local galactic environments}

%5.1 
\subsection{Summary of standard molecular abundances in various regions of the Milky Way, and models of interstellar chemistry}

	The broad features of the relative abundances of molecular species in various standard extragalactic environments share common properties which have many similarities with those observed with much more details in the Milky Way. The latter have been comprehensively studied for decades both by millimetre observations and theoretical chemistry modelling. As discussed in Section 2.4, a few archetypes have emerged for the general patterns of molecular abundances, depending on the physical conditions which directly affect the processes of molecular formation and destruction. Indeed, despite the interest of fine studies of the peculiarities of individual Galactic sources, allowed by their relatively close distances, waiting for ALMA (Turner 2007), only very broad classes of interstellar chemistry are really relevant for comparison with the coarse and global determinations of extragalactic abundances. For this purpose, two main patterns emerge for the molecular abundances of Milky Way interstellar sources. They correspond to cold, quiescent molecular clouds, and to hotter, massive-star forming regions, respectively. 
	
	In both cases, CO is by far the most abundant, easily observed molecule, with an abundance $\sim$10$^{-4}$  with respect to H$_2$, corresponding to a significant fraction of total carbon and oxygen. All other molecules are less abundant by at least a factor 100 (OH, H$_2$O), and rather $\sim$10$^4$-10$^6$ for most of them.

	{\it In very cold}, dense, dark molecular condensations, with high density but no massive star formation, such as TMC-1 or L134N, a large set of complex molecules, especially carbon-rich, are observed (e.g.\,\,Ohishi et al.\,\,1992). They are clearly built up by low temperature gas phase reactions. However, to account for the  observed abundances one needs to include a time dependent chemistry, and/or the long-term effects of grain accretion producing very large depletions and increasing the C/O abundance ratio (see e.g.\,\, the comprehensive review of chemical modelling by van Dishoeck 1998, and Wakelam et al.\,\,2006 for more recent references). Such detailed studies of small dark clouds are only possible at small distances in the Milky Way and are not directly relevant for extragalactic comparisons. However, it has been shown (see e.g.\,\,Terzieva \& Herbst 1998, Turner 2000) that a similar chemistry is at work even with the lower extinction (A$_{\rm v}$\,$\sim$\,1-5) of {\it translucent clouds}, where several tens of such species are broadly observed. The 38 species reported in Table 1 of Turner (2000), and partially displayed in Table 2, may be classified in four groups of similar size: basic simple interstellar molecules (OH, NH$_3$, H$_2$CO, HCO$^+$, N2H$^+$, CN, HCN, HNC); hydrocarbon radicals, from C$_2$H through C$_6$H which are the backbone of cold gas phase chemistry and form a highly homogeneous group (to which one may add c-C$_3$H$_2$ and HC$_{\rm n}$N); heterogeneous complex species such as CH$_3$OH, CH$_3$COH, CH$_3$CN, which have the complexity of hydrocarbons but contain O and N as well as C; a number of sulfur-bearing species more depending on the degree of grain desorption (plus SiO whose abundance remains very weak). Similar models may even be extended to diffuse clouds (A$_{\rm v}$\,$\sim$\,1), where, in addition to optically detected species (CH$^+$, CH, CN ...), the most detailed information comes from millimetre line absorption studies (Liszt \& Lucas 1999, Liszt et al.\,\,2005 and references therein) in which basic species of the first group above, plus CS, SO, C$_2$H, c-C$_3$H$_2$, are commonly detected. A number of the observed abundances (Table 2) are reasonably accounted for by models with standard lower depletion of heavy elements for the diffuse medium (Morton 1975). The addition of the effect of a few turbulence driven, weakly endothermic reactions (e.g.\ Spaans 1995, 1996, Turner 2000) may improve the agreement; but several serious problems remain (see e.g.\ Lequeux 2005 Sec.\ 9.4.1). Anyway, one may expect to find similar conditions in observations of cold extragalactic medium, either in beam averaged emission of cold giant molecular clouds, or rather {\it in millimetre line absorption} (Section 7.2).
 
%\clearpage
\begin{table}[ht]                                      
\caption{Comparison of  molecular abundances (with respect to H$_2$) in selected Galactic sources [dark cloud: TMC-1, photodissociation region: Orion Bar, hot core: Sgr\,B2(N), typical nuclear bulge cloud: Sgr\,B2(OH)], the starburst galaxies NGC\,253 and M\,82 (all from Tables 7 \& 9 of Mart\'in et al.\,\,2006a); LMC/N159 (Johansson et al.\,\,1994); and Galactic translucent clouds (Turner 2000). In addition to the molecules detected in NGC\,253 displayed here, molecules detected in external galaxies include (see Table 3 of Mart\'in et al.\,\,2006a and Muller et al. 2006): 
H$_2$, HD, CO, OH, CH, CH$^+$, CN, CO$^+$, H$_2$O, C$_2$H, HCO, HOC$^+$, N$_2$H$^+$, C$_2$H$_5$OH, DCN, DCO$^+$, $^{13}$CO, C$^{18}$O, $^{13}$CS, H$^{13}$CN, HC$^{15}$N, HC$^{18}$O$^+$, HC$^{17}$O$^+$, H$^{13}$CO$^+$, HN$^{13}$C, H$^{15}$NC, C$^{34}$S, H$_2$$^{34}$S; H$_3$$^+$, C$_2$H$_2$, CO$_2$.}
\begin{center}
\scriptsize
\begin{tabular}{l c c c c c c c c}
%\tableline %\tableline
Molecule       &NGC253& M82 &LMC  &  Sgr       &  Sgr      &  TMC-1   &    Orion & Translucent \\
               &     &  &N159  &  B2(N)   &    B2(OH)&             &     Bar  &    cloud  \\
%\tableline                                                                                              
HN$^{13}$C     & $-10.6 $ & $-9.5 $ &$...$&    $  -11.0 $ &  $     ... $&    $  ...$ &    $... $ &    $...$  \\ 
H$^{13}$CO$^+$ & $-10.4 $ & $-9.9 $ &$...$&    $  -11.4 $ &  $     ... $&    $     ...$ &    $   -10.3$ &    $...$  \\   
SiO            & $-9.9 $ & $<-9.9 $ &$...$&    $  -10.7 $ &  $     ... $&  $ < -11.6$ &      $   -10.3$ &    $ -10.0  $ \\   
NH$_2$CN       & $-9.7 $ & $... $ &$...$&    $  -10.1 $ &  $   -10.0 $&  $     ...$ &      $    ... $ &    $...$  \\   
C$_2$S         & $-9.7 $ & $... $ &$...$&    $   ...  $ &  $    -9.6 $&  $    -8.1$ &      $    ... $ &    $...$  \\   
CH$_3$CN       & $-9.5 $ & $-9.7 $ &$...$&    $  -6.7  $ &  $    -9.4 $&  $    -9.0$ &      $ < -10.3$ &    $ <9.0  $ \\   
c-C$_3$H       & $-9.5 $ & $... $ &$...$&    $ -10.5  $ &  $ < -10.9 $&  $    -9.3$ &      $    ... $ &    $ -8.0  $ \\   
HOCO$^+$       & $-9.4 $ & $... $ &$...$&    $  -10.5 $ &  $    -9.7 $&  $     ...$ &      $    ... $ &    $ -9.0  $ \\   
C$^{34}$S      & $-9.4 $ & $-9.3 $ &$...$&    $  -10.2 $ &  $     ... $&  $     ...$ &      $   -9.0 $ &    $...$  \\   
c-C$_3$H$_2$   & $-9.3 $ & $-8.1 $ &$...$&    $  -10.5 $ &  $    -9.8 $&  $    -8.0$ &      $   -9.7 $ &    $ -7.4  $ \\   
HC$_3$N        & $-9.2 $ & $-8.7 $ &$...$&    $  -7.5  $ &  $    -9.0 $&  $    -8.2$ &      $    ... $ &    $ -9.3  $ \\   
NS             & $-9.2 $ & $... $ &$...$&    $  -7.0  $ &  $     ... $&  $    -9.1$ &      $    ... $ &    $...$  \\   
H$_2$CS        & $-9.2 $ & $... $ &$...$&    $  -6.8  $ &  $    -8.7 $&  $    -8.5$ &      $    ... $ &    $ -7.6  $ \\   
SO$_2$         & $-9.1 $ & $... $ &$...$&    $  -6.6  $ &  $    -8.7 $&  $  < -9.0$ &      $   -9.9 $ &    $ -8.2  $ \\   
CH$_2$NH       & $-9.1 $ & $... $ &$...$&    $  -7.0  $ &  $    -9.2 $&  $     ...$ &      $    ... $ &    $ -7.8  $ \\   
H$_2$S         & $-9.1 $ & $... $ &$...$&    $  -9.9  $ &  $     ... $&  $  < -9.3$ &      $   -8.2 $ &    $ -7.6  $ \\      
HNC            & $-9.0 $ & $-8.8 $ &$ -10.2 $&    $   ...  $ &  $     ... $&  $    -7.7$ &     $   -9.0 $ &    $ -8.6  $ \\   
SO             & $-8.9 $ & $<-8.5 $ &$ -8.6  $&    $  -6.9  $ &  $    -8.7 $&  $    -8.3$ &     $   -8.0 $ &    $ -7.5  $ \\   
HCO$^+$        & $-8.8 $ & $-8.4 $ &$ -9.7  $&    $   ...  $ &  $     ... $&  $    -8.1$ &      $   -8.5 $ &    $ -8.7  $ \\   
HNCO           & $-8.8 $ & $<-8.8 $ &$...$&    $  -9.2  $ &  $    -8.4 $&  $    -9.7$ &      $< -10.8 $ &    $...$  \\   
H$_2$CO        & $-8.6 $ & $-8.2 $ &$ -9.3  $&    $  -9.3  $ &  $    -8.6 $&  $    -7.7$ &     $   -8.2 $ &   $ -8.2   $ \\   
OCS            & $-8.4 $ & $-7.9 $ &$...$&    $  -8.6  $ &  $    -8.3 $&  $    -8.7$ &      $    ... $ &    $ -9.2  $ \\   
HCN            & $-8.3 $ & $-8.4 $ &$ -9.7  $&    $   ...  $ &  $     ... $&  $    -7.7$ &      $   -8.3 $ &    $ -7.4  $ \\   
CS             & $-8.2 $ & $-8.2 $ &$ -9.4  $&    $   ...  $ &  $     ... $&  $    -8.0$ &      $   -7.6 $ &    $ -8.0  $ \\   
CH$_3$CCH      & $-8.3 $ & $-7.7 $ &$...$&    $  -8.4  $ &  $    -8.8 $&  $    -8.2$ &      $    ... $ &    $...$  \\   
CH$_3$OH       & $-7.9 $ & $<-8.3 $ &$...$&    $  -5.8  $ &  $    -7.3 $&  $    -8.7$ &     $   -9.0 $ &    $...$  \\   
C$_2$H         & $-7.7 $ & $-7.6 $ &$...$&    $  -9.7  $ &  $     ... $&  $    -7.1$ &      $   -8.7 $ &    $...$  \\   
NO             & $-7.2 $ & $... $ &$...$&    $  -6.0  $ &  $     ... $&  $  < -7.5$ &      $   -8.6 $ &    $...$  \\   
NH$_3$          & $-7.2 $ & $ ... $ &$...$&    $   ... $ &  $     ... $&  $   ... $ &      $    ... $ &   $ -7.7  $ \\   
%\tableline                                                                                                      
\end{tabular}                
\end{center}                     
\scriptsize
\end{table}

	{\it Warmer molecular regions} are associated with massive-star formation. In addition to warmer kinetic temperature favouring weakly endothermic reactions, one may then find enhanced photochemistry, and strong shocks whose main effect is probably efficient grain desorption not only of weakly bound species such as CO, CO$_2$, NH$_3$, ices and organic material, but also eventually refractory elements such as silicon. Archetypes of such regions are the so called {\it `hot molecular cores'}, whose best examples are Orion KL and SgrB2(N). Their chemistry has been studied in detail, both from observations and modelling (see e.g.\,\,van Dishoeck \& Blake 1998, Walmsley 1996, Ikeda et al.\,\,2001, Wakelam et al.\,\,2005 and papers in the proceedings of IAU Symposium 231 on astrochemistry, Lis, Blake \& Herbst 2005). Temperatures are above 100\,K, densities above 10$^7$cm$^{-3}$ and column densities in the range 10$^{24}$-10$^{25}$cm$^{-2}$. Hydrogenated species such as H$_2$O, NH$_3$, CH$_3$OH, CH$_3$CN, C$_2$H$_5$CN, CH$_3$CO and a variety of other complex organic molecules, have abundances at least an order of magnitude larger than in cold dark clouds (Table 2). Enormous enhancements of SO, SO$_2$, H$_2$S and SiO, as well as deuterated species, are also found. Detailed models reasonably well account for the observed abundances (see van Dishoeck \& Blake 1998 and references therein). The chemistry is driven by the evaporation of icy grain mantles containing a mixture of H$_2$O, CO, CH$_3$OH, NH$_3$, HCN, etc. Among the observed species, one may distinguish: very stable precollapse molecules, such as CO and C$_2$H$_2$, which were frozen on the grains and released into the gas-phase largely unaltered; molecules made in grain-surface reactions and released in the gas (H$_2$O, NH$_3$, H$_2$S, CH$_3$OH, etc.); and those which are produced by rapid gas-phase reactions between evaporated molecules.	In addition to mm-line surveys of many tens of species and isotope varieties, mid-IR infrared observations may probe species such as C$_2$H$_2$, HCN, OCS, NH$_3$, etc. Because of its proximity, the Orion-IRC2/KL region is by far the best studied. Within this region one may find local differences in chemistry, e.g.\,\,between complex O- and N-bearing organics. Important lifetime effects are found in models, e.g.\,\,for SO and SO$_2$, which may be used as clocks to date the age of various sources. 
	
	Strictly speaking, hot cores are very small ($\le$\,0.1\,pc), very dense and hot. On larger scales, more relevant for starbursts in galaxies, the chemistry of active star formation regions is still dominated by the influence of massive stars through photodissociation and various kinds of shocks. It partakes some properties of the hot core chemistry such as relatively high temperature and density, and short time constants. However, it may significantly differ depending of the relative contribution of UV photodissociation and of grain desorption by shocks. 
	
	Detailed physical and chemical properties of {\it photodisssociation regions} (PDRs) have been discussed by Tielens and Hollenbach (1985), Hollenbach \& Tielens (1997, 1999), Jansen et al.\,\,(1995), Tielens (2005). One fundamental feature is the rapid spatial variations of these properties and in particular of molecular abundances. Some species are specific of PDRs, such as ions (CO$^+$, SO$^+$, HOC$^+$) and in a less degree products of photo-chemistry such as c-C$_3$H$_2$ (Fuente et al.\,\,2005). On the other hand, complex molecules, such as CH$_3$OH, CH$_3$CN, etc., are underabundant in PDRs (such as Orion Bar, Table 2) compared to hot cores or Galactic Center clouds, because of rapid destruction by photodissociation.
	
	The special concentration of molecular clouds within 200\,pc of the {\it Galactic Center} 
 (e.g.\ Morris \& Serabyn 1996) is the medium in the Milky Way that is closest to extragalactic molecular sources. Molecules typical of grain desorption chemistry, such as C$_2$H$_5$OH and SiO (Minh et al.\,\,1992), are remarkably widespread, and indeed most of the various clouds exhibit abundances of complex molecules reminiscent of hot cores (Requena-Torres et al.\,\,2006 and references therein). The abundance ratios of various complex organic molecules relative to CH$_3$OH are roughly constant, suggesting that all complex molecules are produced by a similar chemistry initiated  by shock desorption of grains. However, the overall abundances may vary by orders of magnitude. They may be very high, even higher than in hot cores, and the abundance of CH$_3$OH may reach 10$^{-6}$. However, the prototype source, SgrB2(OH) in the envelope of the prominent star forming complex SgrB2, displays abundances smaller by one or two orders of magnitude than in the hot core Sgr(N) for molecules such as CH$_3$OH, CH$_3$CN, HC$_3$N, SO, SO$_2$ and SiO (Requena-Torres et al.\,\,2006, as shown in Table 2). Note that the observed abundances are dramatically smaller in the few regions very close to the Galactic Center submitted to intense photodissociation from the UV radiation of starburst clusters of massive stars.
	
%%%%
As a conclusion, in view of interpreting observed extragalactic abundances, Galactic translucent clouds seem a good reference for abundances derived from molecular {\it absorption lines} through the standard {\it cold} ISM (Section 7.2). On the other hand, typical observable extragalactic {\it emission lines} come from {\it starburst} regions and must obey some kind of warm chemistry similar to hot cores, shocks or PDRs. In the average over the telescope beam in external galaxies, one may indeed expect rather some kind of combination of these related, but distinct, types of chemistry, since they are known to coexist at very short distances in Galactic sources such as Orion or Sgr(B).	

%	6.2 
\subsection{Observed abundance variations in local galaxies: I. the Magellanic Clouds}

	  Because of the distances, extragalactic studies of molecular emission are severely limited compared to the details which are achieved in studies of the molecular gas in the Milky Way, such as those dealing with fine structure or the abundances of rare species. Even in our closest neighbours, the typical sensitivity for small sources is reduced in the Magellanic Clouds by almost two orders of magnitude with respect to the  Galactic Center, and by four orders of magnitude in the other galaxies of the Local Group such as Andromeda (M\,31). Outside of widespread CO (and to a less extend $^{13}$CO), practically all extragalactic molecular studies are limited to the strongest emitting sources, generally associated with active well localized massive-star formation regions, except in starburst galaxies where they are more widespread.

    Therefore, there is practically no detailed studies of a large set of molecular abundances except in a few sources of the nearby Magellanic Clouds, and in local starburst infrared galaxies or nuclear starbursts, where  star formation activity  fills most of the telescope beam. 
    
    The Magellanic Clouds have the special interest to provide tests of interstellar chemistry with a much lower metallicity and enhanced UV radiation compared to the Milky Way, a situation which could be similar in  many high redshift galaxies. One thus expects a large decrease in the overall abundances of molecules, and first of H$_2$ whose abundance is further reduced by the lower rate of formation on rarer dust grains. A large programme with the space telescope FUSE has detected H$_2$ UV absoption lines along $\sim$50  interstellar lines of sight in LMC and SMC (Tumlinson et al.\,\,2002). The amount of H$_2$ is on average an order of magnitude smaller than along lines of sight of the Galactic disk over a similar range of reddening. These results imply that the diffuse H$_2$ mass is only about 0.5\% and 2\% of the \HI~mass derived from 21 cm emission measurements in SMC and LMC, respectively. The high UV radiation enhances the excitation of upper rotational levels. Far-UV lines of CO and HD have also been measured in a few lines of sight (Bluhm and de Boer 2001, Andr\'e et al.\,\,2004). Recent VLT/UVES optical observations in a dozen of lines of sight have compared the abundances of CH, CH$^+$ and CN with Galactic ones (Welty et al.\,\,2006). The CH/H$_2$ ratio is comparable or smaller than the values found for Galactic diffuse clouds. The observed relationships between the column density of CH and those of CH$^+$ and CN show the same trends as in the local Galaxy. The authors discuss in detail the extension of chemical models for diffuse clouds to the smaller metallicities and higher UV radiation field of these galaxies. A significant fraction of the CH and CH$^+$ observed may arise in photon-dominated regions which should be more extensive than in our Galaxy.
    
   Millimetre studies of CO in the Magellanic Clouds are discussed in Section 4. Millimetre detections of a number of other molecular species ($^{13}$CO, CS, SO, CCH, HCO$^+$, HCN, HNC, C$_2$H, CN, H$_2$CO, and C$_3$H$_2$) achieved near two peaks in the CO emission of the LMC and SMC, have provided some information on molecular abundances, isotopic ratios, and cloud structure (Johansson et al.\,\,1994; Chin et al.\,\,1997, 1998). The molecular abundances are about an order of magnitude lower (and even more for CN) than the corresponding values found for Orion KL and TMC-1 (Table 2). However, molecular studies of the Magellanic Clouds have suffered from the low number of large millimetre facilities in the Southern Hemisphere. This situation is now changing with new facilities.  ATCA has already  produced first 3\,mm results at high angular resolution (Wong et al.\,\,2006). APEX (12\,m) has replaced SEST, allowing the extension of 
molecular studies to the submm range with a better receiver equipment (see also AST/RO, Bolatto et al.\,\,2005); the ASTE 10\,m and Mopra-22m telescopes have undertaken extensive survey works of the LMC, SMC and Bridge (Hughes et al.\ in prep., Muller et al.\ in prep.; see also the extensive contributions from the 4\,m NANTEN telescope for CO observations of the LMC and SMC). 
ALMA will allow a full development of interstellar chemistry in the Magellanic Clouds.

\subsection{Observed abundance variations in local galaxies: II. Nearby starbursts and other galaxies}
    In contrast to the Magellanic Clouds, the diversity of the molecular content is better documented in a few galaxies outside of the Local Group despite distances almost 100 times larger. This is due to the fact that these galaxies are powerful starbursts and are observable with the most sensitive millimetre facilities such as the IRAM-30m telescope. However, the number of different molecules presently detected, $\sim$40 (plus isotopic varieties, Table 3), remains far behind the total number, $\sim$150, known in local sources of the Milky Way (see updated list in http://astrochemistry.net/, see also http://www.cv.nrao.edu/~awootten/allmols.html). As for these archetype Galactic sources, the best success has been achieved by a systematic spectroscopic survey of a broad frequency range (most of the 2\,mm atmospheric window) that Mart\'in et al.\,\,(2006a) have recently achieved on the nuclear region of the starburst galaxy NGC\,253. This work extends and synthesizes the results of 20 years of continuous similar studies by members of this team (see e.g.\,\,Henkel et al.\,\,1991\, Garc\'ia-Burillo et al.\,\,2006a). More than 100 spectral features are identified in the 2\,mm band alone with the IRAM 30m telescope, corresponding to transitions from 25 different molecular species by Mart\'in et al.\,\,(2006a, see Table 2). The derived abundances show striking similarities with those observed in the molecular clouds of the Galactic Center, such as SgrB(OH), which are believed to be dominated by low-velocity shocks. A comparison of the chemical composition of the nuclear environment of NGC\,253 with other well observed nuclear starbursts of nearby galaxies (Mart\'in et al.\,\,2006a Fig.\,7) demonstrates the chemical similarity of galaxies such as IC\,342 and NGC\,4945 to NGC\,253. On the other hand, the chemistry of NGC\,253 appears clearly different from that of M\,82 which is another archetype of a nearby nuclear starburst galaxy. M\,82 abundances of molecules like SiO, CH$_3$OH, HNCO, CH$_3$CN and NH$_3$ are systematically low in comparison to NGC\,253 (Table 2), while species like HCO  (Garc\'ia-Burillo et al.\,\,2002)   and C$_3$H$_2$ (Mauersberger et al.\,\,1991) are overabundant. This suggests that photo-dissociation dominates the heating and the chemistry in most of the nuclear region of M\,82, which has much more many \HII~~regions than NGC\,253. The detection of 
widespread emission of HCO, HOC$^+$ and CO$^+$ in this galaxy disk reveals that the 
nucleus of this prototypical starburst is a giant ($\sim$ 650 pc) 
extragalactic PDR (Garc\'ia-Burillo et al. 2002, Fuente et al. 2005, 2006). The detection of a ~500 pc molecular gas chimney in SiO 
indicates the occurrence of  large-scale shocks in the disk-halo interface of 
this galaxy (Garc\'ia-Burillo et al. 2002). 
However, the detection of abundant and high excitation CH$_3$OH (Mart\'in et al.\,\,2006b) suggests its injection from dust grains and the existence of dense warm cores, shielded from the UV radiation and similar to the molecular clouds in other starbursts. Such examples show that we have already clues to understand the most active extragalactic chemistry in various starbursts. Similar less powerful Galactic sources, such as Galactic Center clouds, PDR and shock regions, may help to disentangle the complexity of extragalactic sources in various stages of evolution, including merger and AGN effects.

	A few molecules which have strong lines and are widely distributed, have been studied in greater detail than the bulk of the molecules reported in Table 2. Besides CO and OH (Sections 3 \& 4), HCN, CS and HCO$^+$ have a particular interest because of their abundance, their simple linear structure and their large dipole moments ($\sim$20-50 times larger than CO). As discussed in Section 5 for HCN, they are thus ideal tracers of high-density star-forming clouds. HCN is by far the most used for this purpose (see the systematic work of Gao and Solomon 2004a and Section 5.3). HCN has now been detected in the center of about 60 CO-bright galaxies, including mostly luminous and ultra-luminous infrared galaxies, and also a number of the nearest normal spiral galaxies, and mapped in the disk of some of them (see Gao \& Solomon 2004a for a complete list of references). CS has also been detected in some of those galaxies (see references in Gao \& Solomon 2004a), as well as HCO$^+$  (Nguyen-Q-Rieu et al.\ 1992, Brouillet et al.\,\,2005, Muller et al.\,\,2005).

	However, the number of detected molecules remains very limited in `normal', non starburst nearby galaxies. Even in M\,31, where large efforts have been devoted to intensive studies of the distribution of CO (Section 4 and Fig.\,1), only a few molecules such as HCN and HCO$^+$ (Brouillet et al.\,\,2005), could be detected. Even absorption line studies which have the advantage of a sensitivity independent of distance, are practically limited to the special cases of OH and H$_2$CO lines, or to the exceptional nearby radio source Cen\,A (Wiklind \& Combes 1997a) and a few exceptional high-z systems (Section 7.2).

  A few symmetrical molecules lacking permanent electric dipole cannot be detected through millimetre rotation lines, but only in the infrared. Besides H$_2$ (Section 10), the extragalactic detections are still limited to infrared absorption lines of a few prominent molecules in front of the  strong continuum of a very few starburst galaxies: H$_3$$^+$ in IRAS 08572+3915 NW with UKIRT (Geballe et al.\ 2006), CO$_2$ and C$_2$H$_2$ (together with CO, HCN, ice and silicates) in deeply obscured ULIRG nuclei with {\it Spitzer} (Spoon et al.\ 2004, 2005, Armus et al.\ 2006, Lahuis et al.\ 2007).

%	5.4 
\subsection{Abundance ratios of isotopic varieties, and inferences for chemical evolution of galaxies}

  The study of interstellar CNO isotope ratios is important for tracing the chemical evolution of galaxies and their nucleosynthesis. The widely separated molecular lines of different isotopic varieties (isotopologues) provide a unique way of measuring isotope ratios even in regions heavily obscured. However, great care must be taken in dealing with radiative transfer, line formation and isotopic fractionation (see Section 2 and e.g.\ Wilson \& Matteucci 1992). In the Milky Way significant variations of the CNO isotopic ratios are known in different environments, reflecting their various nucleosynthesis history (see e.g.\ the review by Wilson \& Rood 1994 and references therein, and examples of isotopic ratios in Table 3, reproduced from Muller et al.\ 2006, for the local ISM, the Solar System, the Galactic Center and a C-rich AGB star). 

  In external galaxies, since the first detection of $^{13}$CO by Encrenaz et al\. 1979, only a few isotopic abundance ratios have been reported. These studies were reviewed in particular by Henkel \& Mauersberger (1993) and Muller et al.\ (2006). The best studied cases are the LMC and a few prominent local starbursts such as NGC\,253, NGC\,4945 and M\,82 (Table 3). The trends observed in the isotopic ratios are explained by the various levels of nuclear processing. The ratios in NGC\,253 and NGC\,4945 appear  characteristic of a starburst environment in which massive stars dominate the isotopic composition of the surrounding interstellar medium with isotopes such as $^{16}$O and $^{18}$O. In contrast, the ISM in the LMC has been much enriched by the products of low-mass AGB stars such as $^{17}$O. The various Galactic components are intermediate. The isotopic ratios measured from absorption lines in the z\,=\,0.89 spiral galaxy PKS\,1830-211 (see Section 7.2) show the same trends as in starbursts, with very little contribution from low-mass stars, but less enrichment from massive stars than in starburst galaxies.

  The interstellar D/H ratio is an indicator of the degree of destruction in stars of the primordial deuterium produced in the Big Bang nucleosynthesis. However, the determination of the D/H ratio from observation of deuterated molecules is difficult because the abundance of deuterated molecules is greatly enhanced by fractionation in the cold and dense gas (Section 2.4). Modelling of deuterium chemistry (see e.g.\ Roberts \& Millar 2000) predicts D/H values as large as 0.01--0.1 in molecules in very cold clouds.  Such high ratios which have been observed in various Galactic sources, have been found in the LMC where DCN and DCO$^+$ were detected by Chin et al.\ (1996) (see also Heikkil{\"a} et al.,\,1997). This result is consistent with a D/H ratio of about 1.5x10$^{-5}$, as  observed in the Galaxy. However, this limit is not very constraining because of the uncertainty on the large fractionation enhancement. Upper limits for the DCN abundance were found for other galaxies (see e.g.\ Mauersberger et al.\,\,1995,  Muller et al.\,\,2006).

\begin{table*}[ht] 
\caption{Comparison of the C, N, O and S isotopic ratios in various Galactic and extragalactic environments (reproduced from Table 7 of Muller et al.\,\,2006) 
} 
\label{tab:ratio}
\begin{center} \begin{tabular}{lcccccc}
\hline \hline  
                     & $^{12}$C / $^{13}$C & $^{14}$N / $^{15}$N & $^{16}$O / $^{18}$O   & $^{18}$O / $^{17}$O & $^{32}$S / $^{34}$S \\ 
%\hline
\hline
Solar System (a)     & 89                  & 270                       & 490                   & 5.5                 & 22                 \\  
Local ISM (a)        & 59 $\pm$ 2          & 237 $^{+27}_{-21}$        & 672 $\pm$ 110         & 3.65 $\pm$ 0.15       & 19 $\pm$ 8          \\ 
Galactic Center (a)  & 25 $\pm$ 5          & 900 $\pm$ 200             & 250 $\pm$ 30          & 3.5 $\pm$ 0.2       & 18 $\pm$ 5          \\ 
IRC+10216(C-rich AGB)(a)        & 45 $\pm$ 3          & $>$ 4400                  & 1260 $^{+315}_{-240}$ & 0.7 $\pm$ 0.2       & 21.8 $\pm$ 2.6      \\ 
\hline
LMC (b)              & 62 $\pm$ 5          & 114 $\pm$ 14              & $>$ 2000              & 1.8 $\pm$ 0.4       & 18 $\pm$ 6      \\    
NGC\,253 (c)          & 40 $\pm$ 10         & --                        & 200 $\pm$ 50          & 6.5 $\pm$ 1         & 8 $\pm$ 2           \\ 
NGC\,4945 (d)         & 50 $\pm$ 10         & 105 $\pm$ 25              & 195 $\pm$ 45          & 6.4 $\pm$ 0.3       & 13.5 $\pm$ 2.5      \\ 
PKS\,1830-211(z=0.89)(e) &27 $\pm$ 2 &130 $^{+20}_{-15}$ &52 $\pm$ 4 &12 $^{+3}_{-2}$ &10 $\pm$ 1\\ 
\hline  
\end{tabular} \end{center}
\mbox{\,} %\vskip -.8cm
%$^\dagger$ Derived from a double ratio assuming $^{12}$C/$^{13}$C = 27 $\pm$ 2. \\
References: 
a) See references in Table 7 of Muller et al.\,\,(2006) for Galactic sources; 
%Anders \& Grevesse \cite{and89};
%b) Lucas \& Liszt \cite{luc98}, except for $^{18}$O/$^{17}$O taken from Penzias %\cite{pen81};
%c) Wilson \& Matteucci \cite{wil92}, Wilson \& Rood \cite{wil94} and references therein;
%d) Kahane et al.\ \cite{kah88}, \cite{kah92} and Cernicharo et al.\ \cite{cer00};
b) Chin (1999);
c) Henkel \& Mauersberger (1993), Harrison et al.\,\,(1999) and Mart\'in et al.\,\,(2005);
d) Wang et al.\,\,(2004);
e) Muller et al.\,\,(2006).
\end{table*}

\section{Trends in molecular abundances from absorption lines at high redshift}

	The dimming of the flux received from molecular emission with distance renders the measurement of molecular abundances at redshifts larger than a few tenths increasingly difficult with current sensitivities. Until now, only CO, HCN, HCO$^+$, HNC and CN (plus OH and H$_2$O masers and fine structure lines of \CI~and C$^+$) have been detected at z\,$>$\,0.1; and the few detections of HCN and HCO$^+$ do not reveal much differences from local ULIRGs (see section 9).
	%%, but see discussions about HCO$^+$ in high-z QSOs).
	 In contrast to emission lines, the independence of absorption line strength with distance makes absorption line studies an ideal tool to study the ordinary, widespread molecular gas at high redshift. However, it requires a strong background continuum source just aligned with the molecular gas. Such a coincidence is currently known in a few high-z lines of sight in two main domains: UV lines of molecular hydrogen in diffuse clouds in front of high-z quasars; millimetre and radio lines in denser clouds in front of high-z radio galaxies. The number of such rare cases is expected to considerably grow in the future with large surveys and increased sensitivities of large facilities.

\subsection{H$_2$ UV absorption lines in Damped Lyman-$\alpha$ systems of quasars}

	Damped Ly-$\alpha$ (DLA) absorption systems seen in quasar spectra correspond to relatively  large neutral hydrogen column densities: N(\HI)\,$\ga$ 2$\times$10$^{20}$\,cm$^{-2}$. There are various arguments supporting the case that DLA systems arise from galactic or proto-galactic disks, occur very close (within 10-15 kpc) to the center of typical L$^\star$ galaxies (see Schechter 1976 for the definition of standard L$^\star$ galaxies), and are the main neutral gas reservoir for star formation at high redshift (see e.g.\,\,Wolfe et al.\,\,2005, Srianand et al.\,\,2005a). DLA absorption lines are thus an essential tool to study the diffuse interstellar gas at high redshift and its evolution with respect to the redshift and type of galaxies. Although there is a major difficulty for directly detecting the emission of the corresponding dim high-z galaxies just in front of very bright QSOs, the (atomic) absorption lines by themselves may provide a rich diagnostic of their interstellar medium (e.g.\,\,Wolfe et al.\,\,2005).
	
	H$_2$ UV lines of the strong Lyman and Werner bands are known as ubiquitous along lines of sight of the Galactic disk and a large proportion of those in the Magellanic Clouds (Section 6.2). The number of different H$_2$ lines detectable along a line of sight allows a diagnostic of the column density and the abundance of H$_2$, as well as its rotational excitation, and hence of the ratio ortho- to para-H$_2$ and the kinetic temperature, the dust abundance and the intensity of the UV radiation field in the DLA molecular gas (see e.g.\,\,Savage et al.\,\,1977, Tumlinson et al.\,\,2002, Tielens 2005, Srianand et al.\,\,2005a,b). 
%%	However, detailed diagnostics may be difficult because of the posible complexity of some radiative transfer and chemical processes. 
Until a few years ago, the number of confirmed H$_2$ detections in DLAs remained extremely low after the first detection by Levshakov \& Varshalovich (1985) (see references in Ledoux et al.\,\,2003 and Curran et al.\ 2004). 
%% and Cui et al.\,\,2005
The major problem is the need to detect very weak absorptions in H$_2$ lines and especially to disentangle them from ubiquitous \HI~multi-line absorption in the Lyman-$\alpha$ forest. A major advance has been achieved by the systematic programme of Ledoux, Petitjean and Srianand and collaborators using ultra-sensitive high-resolution spectroscopy with UVES/VLT (Ledoux et al.\,\,2003, Srianand et al.\,\,2005a, Petitjean et al 2006). Reaching a detection limit of typically N(H$_2$) = 2$\times$10$^{14}$\,cm$^{-2}$, they have brought the total number of current detections of H$_2$ in DLAs and sub--DLAs to 14, all at z\,$>$\,1.8, including one system at z=4.2 (Ledoux et al.\,\,2006). H$_2$ is detected in about 20\% of the DLAs, and the H$_2$ column density is always small, mainly in the range 10$^{16}$-10$^{18}$\,cm$^{-2}$, with very low H$_2$/\HI~abundance ratio, $\sim$10$^{-3}$. The detection probability is practically independent of the total \HI~column density, but it has a very good correlation with the dust abundance. The latter is well traced by the degree of depletion of heavy elements in the gas and strongly depends on the metallicity, so that the probability of detecting H$_2$ exceeds 50\% in DLAs with metallicity larger than 0.1 solar (Ledoux et al.\,\,2003, Petitjean et al.\,\,2006, Noterdaeme et al.\ 2007).
	
	Such properties are similar to those found by H$_2$ studies in the LMC and SMC (Tumlinson et al.\,\,2002)
%%	, where the amount of H$_2$ is nevertheless smaller,
 and in the Galactic halo 
%% where the H$_2$ column densities, 10$^{14}$-10$^{17}$ cm$^{-2}$, are more comparable to those detected in DLAs 
(see Richter 2006 and references therein). Models of physical conditions in DLAs inferred from these H$_2$ detections (Srianand et al.\,\,2005a,b, Hirashita \& Ferrara 2005), imply T$_{\rm K}$  = 100-300\,K, n$_{\rm H}$ = 10-200 cm$^{-3}$, and internal UV radiation 1-100 times larger, and dust-to-gas ratio 10-100 times smaller than the respective standard Galactic disk values. Such high value of the UV intensity, derived from the excitation of the high-J rotational levels, is in agreement with the excitation of \CII~derived from the \CII$^\star$  absorption line which is detected in all the components where H$_2$ absorption lines are seen (Wolfe et al.\ 2003, Srianand et al.\,\,2005a). It is consistent with such H$_2$-detected DLAs being mainly located in the outskirts of high-z Lyman-Break star forming galaxies. 	
	
	Ultraviolet lines of deuterated molecular hydrogen, HD, have been identified in one absorption system at z=2.34 by Varshalovich et al.\,\,(2001). However, this is the only detection reported (Petitjean et al.\,\,2003). CO has also been detected in only one system (Petitjean priv.\  com.\,), and other heavy-atom molecules have never been detected in UV. This is fully consistent with the low molecular contents of DLA clouds and their low metallicity. 

	A very important product of such accurate measurements of H$_2$ lines at high redshift is the ability to check possible variation with time of the proton-electron mass ratio $\mu$\,=\,m$_p$/m$_e$. A recent reanalysis of very good quality H$_2$ spectral lines observed in the sight-lines of two quasars, based on highly accurate laboratory measurements of Lyman bands of H$_2$ and an updated representation of the H$_2$ level structure, may yield a fractional change in the mass ratio of $\Delta \mu$/$\mu$ $\sim$ (2\,$\pm$0.6)x\,10$^{-5}$, indicating that $\mu$ could have decreased in the past 12 Gyr (Ivanchik et al.\ 2005, Reinhold et al.\,\,2006).

	The rapid development of gamma-ray burst (GRB) observations is opening new prospects for studying high-z absorption systems and H$_2$ in particular. The first tentative evidence for H$_2$ molecules in a GRB absorber at z\,=\,4.05 has been recently found by Fynbo et al.\,\,(2006). Very sensitive studies of absorption systems on sight-lines of QSOs and GRBs are one of the major goals of the new generation of extremely large telescopes presently studied (Section 12.3.1).

	\subsection{Millimetre and radio absorption lines in lensing galaxies and radio-sources}
	
	As explained, absorption millimetre molecular lines are easier to detect than emission ones, especially at large distances, since for their detection the sensitivity depends on the intensity of the background continuum source and not on the distance. This is well known with the detection of 21\,cm \HI~absorption in a number of high-z DLAs (see e.g.\ Gupta et al.\ 2007 and references therein). While millimetre emission lines mostly probe the dense, warm gas, absorption line measurements are mainly sensitive to cold, diffuse gas where the rotational excitation is concentrated in the lowest energy levels. They are thus complementary to emission line studies to probe molecular abundances and physical conditions in the diffuse molecular interstellar medium not directly implied in star formation, especially at high redshift. They are indeed a very sensitive probe of cold gas at high redshift, able to detect individual clouds of a few solar masses (instead of 10$^{10}$\,M$_\odot$ for emission). However, they are limited to the molecular clouds which by chance are along the lines of sight of strong background continuum sources. This is a severe limitation outside of the Milky Way. Indeed, such extragalactic molecular line absorptions are detected up to now either directly in the host galaxy itself of a few radio continuum sources, or in a very few strong-lensing galaxies in front of strong radio sources. Such studies are still marginal for local galaxies (see Evans et al.\,\,2005, Liszt \& Lucas 2004, and references therein) except in nearby (4\,Mpc) Centaurus A where the detections include CO, $^{13}$CO, H$_2$CO, C$_3$H$_2$, HCO$^+$, HCN, HNC and CS, with abundances compatible with Galactic values (Wiklind \& Combes 1997a and references therein). However, despite their small number, the four systems known and comprehensively studied at z\,=\,0.25-0.89 (see e.g.\,\,Wiklind \& Combes 1994, 1995, 1996a,b, 1997a,b, 1998, 1999, 2005; Combes \& Wiklind 1997a,b, 1998, 1999; Henkel et al.\,\,2005, Muller et al. 2006, Combes 2007, Muller et al. 2007, and references therein) are very important for the direct information on the abundances and the properties of the molecular interstellar gas at high redshift.

	In these high-z systems, a total of at least 15 different molecules have been detected, in a total of more than 30 different transitions. These include CO, HCO$^+$, HCN, HNC, CS, CN, C$_2$H, OH, H$_2$O, N2H$^+$, NH$_3$, H$_2$CO, C$_3$H$_2$, HC$_3$N. Many isotopic varieties were also detected especially by Muller et al.\ (2006), including $^{13}$CO, C$^{18}$O, H$^{13}$CN, HC$^{18}$O$^+$, HC$^{17}$O$^+$, H$^{13}$CO$^+$, HN$^{13}$C, HC$^{15}$N, H$^{15}$NC, C$^{34}$S, H$_2$$^{34}$S. Two of the four known absorption systems, PKS 1413+357 and B3 1504+377, are situated within the host galaxy itself of the `background' continuum radio source. The two other absorption systems occur in intervening galaxies acting as strong gravitational lens to the background continuum source: B0218+357 and PKS 1830-211, which are among the most strongly lensed radio galaxies, and were thus among the first objects selected for such studies. These two lines of sight have a very large extinction, A$_{\rm v}$\,$\sim$\,10-100, and thus a very high molecular fraction. Several isotopic species are detected there, showing that the main lines are saturated. Nevertheless, the absorption lines do not reach the zero level, indicating that the coverage of the continuum source by obscuring molecular gas is only partial, but this gas is optically thick, as verified by mm-wave interferometry (Menten \& Reid 1996, Wiklind \& Combes 1998, Frye et al.\,\,1997). Such absorption data are consistent with the presence of a diffuse gas component, dominating the observed opacity, and a dense component, accounting for most of the mass (Wiklind \& Combes 1997b). It is clear  that without knowledge of the small scale structure of the absorbing molecular gas, one can only derive lower limits to the column density. However, within such limitations,  Wiklind \& Combes (1997a) have compared the column densities of simple molecules (HCO$^+$, HCN, HNC and CS) in these high-z systems with absorption measurements in Galactic low and high density gas and in Cen A. Despite the presence of a considerable scatter, the most striking impression is the remarkable correlation, over more than three orders of magnitude, in column density. The high-z systems do not show any peculiarites compared to the local values, except maybe for the ratio HNC/HCN, suggesting that the molecular ISM and its chemistry display similar conditions at earlier epochs as it does in the present one. 
	
	Among the other important information brought by observations of high-z molecular  absorption lines, let us note: i) the high sensitivity to the detection of the fundamental sub-mm transition of ortho-water (Combes \& Wiklind 1997a, Wiklind \& Combes 2005), which should lead to important studies with Herschel and ALMA; ii) the detection of NH$_3$ absorption lines (Henkel et al.\ 2005); iii) measurements of {\it isotopic ratios} (Section 6.4, Table 3; Muller et al.\,\,2006). The last  work is particularly remarkable, with the high sensitivity of the IRAM interferometer making it possible to measure reliably the C, N, O and S isotopic abundance ratios in HCO$^+$, HCN, HNC, CS and H$_2$S.  These  ratios are much more reliable than those which result from the observation of weak and broad emission lines, showing the power of the observation of absorption lines; iv) the number of significant limits for undetected molecules, including particularly important ones such as O$_2$ and LiH (Combes et al.\,\,1997, Combes \& Wiklind 1998); v) the surprising absence of detection of molecules with heavy atoms in the sight-line of another radio galaxy, PMN J0134-0931, where the four 18\,cm lines of OH are detected (Kanekar et al.\,\,2005), while there is a good correlation between OH and HCO$^+$ in the other four absorption systems (see also Curran et al.\ 2007). 
	
	The narrow widths of the lines, 1-30\,km/s, imply depths of parsec scale for the sampled clouds. This situation is favourable to detect {\it time variations} in the absorption profile, if there exists knots in the radio source moving close to the velocity of light. Variations on time-scales of a month then correspond to structures of $>$\,10$^3$ AU. Wiklind \& Combes (1997b) have reported possible time variation affecting the relative ratio of two absorbing components. High-z absorption may thus be used to probe very small scale structures in the molecular gas. In gravitationaly lensed systems, such a variability might also be used to measure the time delay between lensed components. Monitorings were carried out for this purpose, with reasonable results for the value of the Hubble constant, in spite of possible micro-lensing events in one of the lines of sight (Wiklind \& Combes 2001). 
	%(Combes \& Wiklind 1999). 
	In the diffuse gas the rotational excitation temperature should be close to the cosmic background temperature T$_{bg}$. Multi-line observations of absorption systems could thus provide a direct measurement of the variation of T$_{bg}$ with redshift. However, line saturation effects make accurate determinations difficult (Combes \& Wiklind 1999).

	High-z radio molecular absorption lines also offer an alternative method than using UV atomic and molecular lines for probing possible {\it time variation of the fundamental constants}. The 18\,cm OH lines are probably the best for this purpose because of the strong dependence on both the fine structure constant $\alpha$ and the mass ratio $\mu$\,=\,m$_p$/m$_e$, and the possibility to cross-check the four OH 18\,cm lines and the \HI~21\,cm line. Kanekar et al.\,\,(2005) have detected the four 18\,cm OH lines from the z\,=\,0.765 gravitational lens toward PMN J0134-0931. Their measurements have a 2-$\sigma$ sensitivity of [$\Delta \alpha /\alpha ]\,<\,6.7\times 10^{-6}$ or [$\Delta \mu /\mu\,]<\,1.4\times 10^{-5}$ to fractional changes in $\alpha$ and $\mu$ over a period of $\sim$6.5 Gyr. These are among the most sensitive constraints on changes in $\mu$ (see Reinhold et al.\,\,2006), and complement the measurements using atomic lines on the $\alpha$ variation (e.g.\,\,Murphy et al.\ 2003, Chand et al.\,\,2006 and references therein). 

	It is surprising that only four high-z millimetre and radio absorption systems are known (e.g.\,\,Wiklind \& Combes 1999), plus one with only OH (Kanekar et al.\,\,2005), despite the considerable efforts to discover additional ones (e.g.\,\,Xanthopoulos et al.\,\,2001, Murphy et al.\,\,2003). This probably reflects various difficulties (see e.g.\ Curran et al.\ 2006): the need for the alignment of a molecular cloud with a strong millimetre continuum source, just by chance along the line of sight, or even in the host galaxy of the continuum source; the fact that, while the chance of such an alignment is much enhanced in lensing galaxies, most strong lenses are achieved by massive elliptical galaxies without much gas; the absence of tight correlation with 21\,cm \HI~absorption; the large extinctions of such molecular clouds which prevent using optical data for finding them; the rarity of exceptionally bright high-z lensed millimetre continuum sources which are required with the present sensitivity of millimetre facilities. Such reasons probably also mostly explain why the presently known systems are limited to z\,$<$\,1, although the lower average metallicity at higher z may also play a defavorable role for molecular abundances, including H$_2$. However, one may expect much progress in this field, especially at high redshift, with the gain in sensitivity in the continuum of one or two orders of magnitude of ALMA enabling a larger number of sources to be observed.

	\section{Molecules, dust and PAHs (large aromatic molecules)}  

%6.1 
\subsection{Introduction. Interplay between dust and molecules in galaxies}

	The main topic of this review is small molecules in the interstellar gas of galaxies. However, the overall molecular material, excluding H$_2$, is distributed with comparable masses, between the gas, dust mantles and in intermediate-size aromatic particles, mostly Polycyclic Aromatic Hydrocarbons (PAHs). The interstellar gas, dust and PAHs belong indeed to the same world, the ecosystem  of the interstellar medium, with constant exchanges and interplay. A detailed discussion of dust properties and its molecular compounds in galaxies is outside our scope. However, we have stressed the importance of the exchanges with dust for the gas chemical composition through accretion, surface reactions and desorption. PAHs may play a significant role in the chemistry of small carbon-based molecules such as C$_2$H or C$_3$H$_2$. They cannot be absent from this general panorama of molecules in galaxies, but they will be discussed briefly because their problematics is quite different from small molecules, somewhat midway between molecules and dust.
	
	 Dust plays a cornerstone role in the construction of molecular material in galaxies through H$_2$ formation. Through subsequent reactions, the latter is at the origin of most of molecular bonds in the gas phase. We have seen the resulting paramount importance of the various metallicity of galaxies for their molecular abundances in H$_2$, CO and other molecules. Efficient dust desorption is a more direct major factor which shapes many molecular abundances in star-forming media such as starburst galaxies. Dust accretion is a universal process which depletes molecular abundances in the cold dense gas. It is probable that all these processes depend not only on the dust abundance, but also on its actual composition which depends on the various galactic metallicity, environment and evolution. However, we still lack detailed evidence for such effects.
	 
	 Anyway, dust chemistry is intimately linked to that of the molecular gas, and it remains a central issue for understanding the various galaxies and their evolution. Dust is essential in many key processes in galaxies and the information we can get about their various regions (Section 2). Its abundance, resulting from the complex cycle of formation-growth-erosion-destruction, is thus crucial in various environments of metallicity, density and UV intensity. While most dust processes are still poorly known in the local interstellar medium, it is a real challenge to address them in the extreme conditions of the world of galaxies, such as galaxy formation and merging, or the vicinity of AGN. 
	 
	The biggest PAHs and their small complexes partially belong to the world of dust, and may indeed encompass the so called `Very Small Grains' (D\'esert et al.1990). More generally, there are certainly continuous exchanges between PAHs and carbonaceous dust grains, through accretion of PAHs, and release of PAHs by shattering.in grain-grain collisions.  Despite the stability of the aromatic hexagonal cycle, small aromatic molecules such as benzene (C$_6$H$_6$) are hardly detected in the local interstellar medium (Cernicharo et al.\,\,2001), and not at all in external galaxies, while the only widespread small cycle molecule is cyclic c-C$_3$H$_2$. On the other hand, PAHs with $\sim$\,50-100 atoms or more play a major role in the interstellar medium of the Milky Way and most galaxies. They are very different from the bona fide small interstellar molecules with $\la$10 atoms discussed in this review. Their size and number of vibration modes make their physics quite different from the latter. In addition, if their general nature and even size range are well established, not a single specific PAH molecule has ever been identified in the interstellar medium. It is neither known whether observed PAHs contain a significant amount of other atoms than C and H, such as nitrogen. However, this review would not be complete without some mention of PAHs as essential compounds of the interstellar medium of galaxies, so close to regular molecules by their nature and chemical exchanges. We will now briefly consider the state of our knowledge about PAHs in galaxies, especially from the results of ISO and {\it Spitzer}.
	%, with a special emphasis on their importance for the diagnosis of the interstellar medium and their possible impact on the chemistry of small carbonaceous molecules. 

%8.2 
\subsection{Mid-infrared emission of PAHs in galaxies}

	Mid-infrared spectra of most galaxies, as a variety of Galactic sources including \HII~regions, post-AGB stars, planetary nebulae, young stellar objects (YSOs), are dominated by a family of emission features at 3.3, 6.2, 7.7, 8.6, 11.2 and 12.7 $\mu$m (Fig.\,3), identified as aromatic infrared bands (AIB's) of C-H and C-C vibration modes (Duley \& Williams 1981). They are generally attributed to Polycyclic Aromatic Hydrocarbon (PAH) molecules (L\'eger \& Puget 1984, Allamandola et al.\,\,1985, Puget \& L\'eger 1989, Allamandola et al.\,\,1989), although the exact molecular identification of the carriers remains unknown. The intensity of these features suggests that PAHs are the most abundant complex polyatomic molecules in the interstellar medium, accounting perhaps for as much as 10-20\% of all carbon in a galaxy like ours. The physics of this infrared emission, induced by fluorescence after UV absorption, has been discussed in many places [in addition to the references above, see e.g.\,\,Sellgren (1984), Omont (1986), Peeters et al.\,\,(2002, 2004),  Rapacioli et al.\,\,(2005), Tielens (2005) and references therein]. This emission is generally attributed to small aromatic hydrocarbon species, transiently heated by the absorption of a single far-UV photon, because the band ratios, reflecting the very high "temperatures" of the emitters in cold environments, are practically independent of the UV intensity. This requires the emitting species to be typically of the order of 10-30\,\r{A} ($\sim$50--1000 carbon atoms, Sellgren, 1984, Draine \& Li 2006). PAHs are expected to exist in various proportion of the different ionization states: neutral, singly positively (cations) or negatively charged (anions) (see e.g.\,\,Bakes et al.\,\,2001a,b; Tielens 2005). In strong UV environments they have a good chance to be positively ionized and more or less dehydrogenated. Indeed, significant variations of the spectral system of AIBs have been observed in photodissociation regions, attributed to changes in the proportion of cations {\it vs} neutral PAHs. There is also evidence for colder components in the observed spectra (Rapacioli et al.\,\,2005). The latter probably correspond to larger aromatic particles with several hundred carbon atoms, which could be a part of the commonly called `very small grains' (VSG, D\'esert et al.\,\,1990). It has been proposed (Rapacioli et al.\,\,2005) that such particles with cold AIBs could be, at least partially, PAHs clusters, and that they could be photoevaporated into free-flying PAHs in strong UV radiation fields.

	%Fig3 Antennae
   \begin{figure*}
    \centering %%\includegraphics[width=7.cm, angle=0]{H2_antennae_map.ps}
\caption{{\it (reproduced from Figs.\ 1 \& 2 of Haas et al.\ 2005)}. ISO mid-infrared observations of the "Antennae" galaxy pair, which is at an early stage of galaxy collision: PAH aromatic bands and  exceptional H$_2$ line emission. ({\it left}) Optical three colour image of the Antennae with contours of the continuum-subtracted
H$_2$ S(3) line emission superimposed. The H$_2$ emission 
extends between the two nuclei of NGC\,4038/4039. 
({\it right}) Mid-infrared spectrum of the Antennae, obtained with the
ISOCAM-CVF mode. It has been derived from a 2'\,x\,2' region,
encompassing both nuclei and the entire overlap region. The prominent
emission features are those of polycyclic aromatic hydro-carbons
(PAHs) and Ne lines. Three H$_2$ lines are marked. While the H$_2$ S(5)
and H$_2$ S(2) lines are blended with [ArII] $\lambda$\,=\,6.99\,$\mu$m and [NeII] $\lambda$\,=
12.8\,$\mu$m, respectively, the H$_2$ S(3) line can unambiguously be detected
above the continuum (dotted). 
 }
 \end{figure*}
 
	Beyond serving as simple PAH indicators, the observation of mid-infrared (MIR) AIBs in galaxies can be used as redshift indicators, and as tracers of elemental and chemical evolution, and of environmental conditions. After early detections of AIBs in galaxies from the ground (Gillett et al.\,\,1975, and e.g.\,\,Aitken \& Roche 1985 and references therein), comprehensive surveys of MIR emission of local galaxies are among the most important programmes of the space observatories ISO (Kessler et al.\,\,1996) and {\it Spitzer}  (Werner et al.\,\,1984 (see e.g.\ Fig.\,3). The demonstration of the universality of almost pure AIB spectra in the MIR range for `normal' star-forming disk galaxies in the local universe is one of the major results from ISO (Mattila et al.\,\,1999, Laurent et al.\,\,2000, Lu et al.\,\,2003 and references therein). In particular, the atlas of MIR spectra of 45 disk galaxies by Lu et al.\,\,(2003) shows that most of them are completely dominated by AIBs and strikingly similar to each other and to the predominant type of MIR pattern in the interstellar medium of our own Galaxy. The combined luminosity of the AIBs in the region 5.8–-11.3\,$\mu$m is typically $\sim$10-20\% of the FIR luminosity in such spiral galaxies. However, there is a trend to decrease this ratio L$_{AIB}$/L$_{{\rm FIR}}$ in IR starburst galaxies, LIRGs and ULIRGs. This is interpretated as the result of the destruction of the PAHs by the hard UV radiation of \HII~regions, maybe complemented by shocks or reabsorption of mir-IR radiation. Such a trend is still enhanced in the central regions of AGN where the AIBs  disappear with respect to the strong emission of the hot dust of the AGN torus, so that it is a key element for discriminating AGN-dominated from starburst-dominated galaxies (Laurent et al.\,\,2000). In nearby
Seyferts, spatially resolved mid-infrared spectroscopy with ISO suggested that the absence or suppression of AIB emission is due to the fact that the dust is predominantly heated by processes related to the central AGN
(e.g.\,\,in NGC\,1068: Le Floc'h et al.\,\,2001, Circinus: Moorwood, 1999, NGC\,4151: Sturm et al.\,\,1999, Mrk\,279: Santos-Lle´o et al.\,\,2001), with the additional possibility that the PAHs are destroyed by the AGN X-rays (Voit 1992).
	
	Lu et al.\,\,(2003) have also confirmed that the weak NIR excess continuum, which has a color temperature of $\sim 10^3$\,K (Sellgren 1984), is well correlated with AIB emission, confirming  that they are produced by similar mechanisms and similar (or the same) material.\,\,But the precise origin of the NIR excess is still unknown. 

	The space observatory {\it Spitzer} has already produced a wealth of results about AIBs in local galaxies. Dedicated large programmes, such as the Legacy Project SINGS (Kennicutt et al.\,\,2003, Smith et al. 2007),  IRS-GTO (Houck et al.\,\,2005), MIPS-GTO (Gordon et al.\,\,2006), are studying MIR AIBs in a large sample of galaxies (Draine et al.\ 2007, often with good angular and spectral resolution, allowing their use for the diagnostic of the interstellar medium. For instance, it is confirmed that their emission has an extension comparable to the diffuse interstellar medium (see e.g.\,\,Engelbracht et al.\,\,2006 for M\,82), and that, in starburst regions, the bulk of AIB emission is not associated with spectacular superclusters of massive star formation, but with the diffuse medium (van der Werf \& Snijders 2006). The complex behaviour of PAH emission in the context of AGN is also further documented (e.g. Smith et al. 2007, Dale et al. 2007, Lutz et al.\ 2007). Another important result is the demonstration of the spectacular dependence on metallicity of the 8$\mu$m to 24$\mu$m flux density ratio in starforming galaxies, observed in systematic studies by Engelbracht et al.\,\,(2005). This ratio, related to that of PAHs to VSGs, drops by almost a factor 10 when the metallicity decreases from 0.5 to 0.3 solar metallicity. More generally, the cause of the low abundance of PAHs in low metallicity galaxies (see also Hogg et al. 2005, Wu et al. 2005, Smith et al. 2007, Jackson et al. 2006) is probably linked to that of the lack of dust and of all molecules.

	One major breakthrough of {\it Spitzer} is to provide MIR {\it spectra} of AIBs, and thus direct evidence of PAHs, in starburst galaxies {\it up to z\,$\sim$\,3} (Houck et al.\,\,2005, Yan et al.\,\,2005, 2007, Lutz et al.\ 2005a,b,  Weedman et al.\,\,2006a,b, Desai et al.\,\,2006, Armus et al.\ 2006, Teplitz et al.\ 2007, Sajina et al.\ 2007). The majority of the objects where weak AIBs are visible, are AGN-dominated. However, several of the reported MIR spectra are almost pure AIBs and the sources are thus clearly starburst-dominated, with best spectral templates similar to regular or warmer ULIRGs. Most of the z\,$\sim$\,2 objects ($\ga$40) where AIBs have been reported, have an infrared luminosity L$_{IR}$ $\ga$ 10$^{13}$\,L$_\odot$, and are thus of the class of hyper-luminous infrared galaxies. Some of them, mostly starburst-dominated, had  been detected at mm or submm wavelength (Lutz 2005b, Lonsdale et al.\ in preparation). They are thus sites of extreme star formation and represent a key phase in the formation of massive galaxies. Such results demonstrate the potential of using MIR AIBs to probe optically faint and infrared luminous high-z populations, and in particular to directly determine their redshifts.
 	
\subsection{Diffuse Interstellar Bands (DIBs) in galaxies}

	We end this section about aromatic molecules in galaxies with mentioning probable, albeit unproved, cousins of PAHs, the unidentified carriers of the `diffuse interstellar bands' (DIB). The latter are a series of broad absorption lines distributed between 4000\,\r{A} and 13000\,\r{A}, ubiquitously observed along sight-lines in the Milky Way which probe mostly the diffuse interstellar medium (e.g.\,\,Herbig 1995). Although several hundred DIBs are now known, %(e.g.\,\,Cox et al.\,\,2005)
 and the most conspicuous ones have been observed since the dawn of interstellar spectroscopy (e.g.\,\,Merrill et al.\,\,1934), the identification of their carriers has resisted intensive efforts during more than half a century. They are very likely gas phase carbonaceous compounds, maybe associated with metals. Possible candidates include, but are not limited to, PAHs, fullerenes, carbon nanotubes and carbon chains (e.g.\,\,Zhou et al.\,\,2006 and references therein). Their eventual identification and understanding their physics and chemistry might provide us with a new tool for the diagnosis of the conditions in the interstellar medium.
	
		The extension of comprehensive studies of DIBs to external galaxies is now possible with recent advances in instrumentation and telescope capabilities. They offer unique possibilities for studying very different conditions  from those observed in the Milky Way, with respect to metallicity, UV intensity, gas-to-dust ratio, dust extinction properties such as the 2175\,\r{A} `bump' in dust extinction, etc. Most significant works include: i) Extensive observations of sight-lines of the Magellanic Clouds (e.g.\,\,Ehrenfreund et al.\,\,2002, Cox et al.\,\,2005; Welty et al.\,\,2006). On average, the  DIBs are weaker by factors of almost 10 (LMC) and about 20 (SMC), compared to those typically observed in Galactic sight-lines with similar N(\HI), presumably due to the lower metallicities and stronger radiation fields in the Magellanic Clouds (Welty et al.\ 2006). ii) Observations of several local galaxies (e.g.\,\,Sollerman et al.\,\,2005), and especially intense starburst galaxies by Heckman \& Lehnert (2000) 
%(see also Gallagher \& Smith 1999) 
who found that the DIBs are there remarkably similar to those in our Galaxy, while both the UV strength and the gas density are much larger than in the diffuse interstellar medium of the Milky Way. iii) The detection of several DIBs in a Damped Lyman-$\alpha$ system at cosmological distance with z$_{abs}$ = 0.524 (York et al.\,\,2006).
%Junkkarinen et al.\,\,2004, 
		
\subsection{Molecular infrared spectral features in extragalactic dust}

	The molecular compounds embedded in dust grains, especially in ices, are not directly within the scope of this review of free flying molecules in galaxies. They must nevertheless be mentioned  because of their relation with gaseous molecules, both as direct result of their accretion onto grains, and, more importantly, as sources of complex gaseous species in desorption of grain mantles (Section 2.4, and e.g.\,\,Tielens 2005). Verma et al.\,\,(2005) have reviewed the various ISO results about the first extragalactic detection of absorption features due to ices (H$_2$O, CH$_4$ and XCN) present in cold molecular components of starbursts, Seyferts and predominantly ULIRGs. For instance, in a heterogeneous sample of 103 active galaxies with high signal-to-noise mid-infrared ISO/SWS spectra, approximately 20\% display absorption features attributed to ices (Spoon et al.\,\,2002). While ice features are mostly weak or absent in the spectra of starbursts and Seyferts, they are strong in ULIRGs. The absorption dominated spectrum bears strong similarities to the spectra of embedded protostars (see e.g.\,\,Keane et al.\,\,2001, Tielens 2005), and the depth of absorption features implies deeply embedded sources as observed with {\it Spitzer} (Fig.\ 4, Spoon et al.\ 2006, 2007), and even in the AGN (see Section 11.2).
	
%Fig4 Ulirgs
   \begin{figure}
    \centering %%\includegraphics[width=7.cm, angle=0]{H2_antennae_map.ps}
\caption{{\it (reproduced from Figs.\ 2 of Fosbury et al.\ 2007 and Spoon et al.\ 2006)}. Spitzer-IRS spectra of various classes of ULIRGs, from (featureless) AGN dominated (top), to deeply embedded (bottom). In addition to prominent features from the PAH family at 6.2, 7.7, 8.6 11.3 and 12.7\,$\mu$m, and silicate absorption at 10 and 18\,$\mu$m, weaker absorption becomes apparent in some sources: water ice at 5.7\,$\mu$m, hydrocarbons at 6.85 and 7.2\,$\mu$m, and even warm CO gas at 4.6\,$\mu$m when the redshift is large enough to bring 
the IRS spectral coverage down to rest frame 4\,$\mu$m (Spoon et al. 2006, Spoon et al. 2007)}
 \end{figure}

%\end{document}

\section{Millimetre emission of molecules at very high redshift}  
%(prospects deferred to Sec. 10)

%%8.1 
\subsection{Star formation and ULIRGs at high redshift} 

	The importance of the molecular medium in the history of the Universe is directly related to that of star formation. The gross features of the global star formation history are now reasonably well established. They are  inferred not only from the analysis of the age of stellar populations in local galaxies, but, since about a decade, directly and more accurately from the observation and census of high-z star forming galaxies. The global star formation was much more important in the past, e.g.\,\,by about a factor of 10 more than presently at z\,$\sim$\,1 (see e.g.\,\,Le Floch et al.\ 2005 and references therein). The majority of the stellar mass of the Universe was thus formed in progenitors of spiral galaxies between z\,=\,1.5 and 0.5. Another important fraction is formed at higher redshift, including most of the stellar mass of elliptical galaxies. The exploration of the properties of the prominent molecular medium at these climax epochs of star formation in the Universe is fundamental to understand in detail the evolution of galaxies. This is mostly out of reach with current millimetre capabilities whose current limit of CO detection in a typical LIRG such as M\,82 is z\,$\sim$\,0.2 (however, see Combes et al.\ 2007). It is one of the main drivers of the ALMA project which should be able to detect LIRGs at very high redshift (Section 12). 
	
	As discussed in Section 5, the far-infrared (FIR) luminosity is both the most sensitive tracer of the star formation rate in major starbursts, and it is closely related to their molecular emission. The detection of the redshifted FIR emission of galaxies in atmospheric windows close to 1\,mm, mainly 850\,$\mu$m and 1.2\,mm, presents the remarkable property of being practically independent of redshift in a very broad range, z\,$\sim$\,0.5-5 (Blain \& Longair 1993). For a fixed observing frequency, there is a strong increase of the luminosity with z at the corresponding restframe frequency because of the very steep emission spectrum, $\propto$\,$\nu _{rest}^{4.5}$ in the submm range. This almost exactly compensates for the luminosity distance factor D$_{L}^{-2}$. Such an extraordinary advantage of the mm/submm window explains the remarkable success of mm/submm surveys of high-z ULIRGs first with SCUBA/JCMT and then with MAMBO/IRAM-30m (Coppin et al.\,\,2006, Voss et al.\,\,2006, and references therein) which have detected hundreds of high-z galaxies  with L$_{{\rm FIR}}$ sensitivity of a few 10$^{12}$L$_\odot$ potentially up to z\,$\sim$\,6.
	
	The comoving density of `submillimetre galaxies' (SMGs) at z\,$\sim$\,2 is more than two orders of magnitude larger than locally (e.g.\,\,Hughes et al.\,\,1998, Greve et al.\,\,2004, Coppin et al.\ 2006). They contribute significantly to the submillimetre background (see e.g.\,\,Smail 2006 and references therein). Their redshift distribution peaks at z\,$\sim$\,2-3 (Chapman et al.\,\,2003, 2005; Pope et al.\,\,2006). The starburst origin of their far-infrared emission is confirmed from their radio emission, CO emission (next section) and the absence of strong X-ray emission in most of them (Alexander et al.\,\,2005a, 2005c, 2006 and references therein). With typical star formation rates of a few hundred M$_\odot$yr$^{-1}$, they contribute significantly to the global star formation rate at z\,$\ga$\,1 (Le Floch et al.\,\,2005); and they are among the main contributors to star formation at z\,$\sim$\,2-3. It is agreed that they are probably the progenitors of massive elliptical galaxies and that their strong starburst is a major episode of the star formation history of these galaxies. From the general correlation between molecular gas and FIR emission, one may infer that they should be the objects with the largest molecular masses in the Universe. This is indeed confirmed by the millimetre detection of CO in a number of the most luminous ones at z\,$\sim$\,1-3 (next section). 
		
%7.2 
\subsection{CO studies} 

	To date, millimetre CO emission has been detected in about 40 sources with redshift $>$\,1, most of them with z\,$>$\,2. Molecular gas emission at high redshift has been reviewed by Solomon \& Vanden Bout (2005) (see also Cox et al.\ 2005). Therefore, here we only summarise the situation of this major topic for the discussion of molecules in galaxies, 
	%%together with the prospects with ALMA; 
	refering the reader to this review for details and a complete list of references.

	After the early millimetre detection of dust and CO in exceptional high-z objects, mostly strongly lensed, (Brown \& Vanden Bout 1991, 1992, Solomon, Radford \& Downes 1992, Barvainis et al.\,\,1992, 1994, McMahon et al.\,\,1994), it was realized that CO could be detected with current equipment in the brightest high-z ULIRGs even without the help of gravitational amplification. After about ten years of major efforts with millimetre facilities, there are 36 CO detections with z between 1.06 and 6.42 reported in Table 1 of Solomon \& Vanden Bout (2005) (this list is constantly increasing with new detections, e.g.\ Kneib et al.\ 2005, Iono et al.\ 2006b, Willott et al.\ 2007, Smail et al.\ in prep.). The detectability of CO at such prodigious distances  may be explained in a way similar to the easy detection of dust emission (Section 9.1). At large redshifts, there is practically always at least one CO line in the 3\,mm atmospheric window which is the most sensitive for such detections. For large redshifts, this line corresponds to a high value of the rotation number J, with a rotational line luminosity which strongly increases with J and hence with z, and again almost compensates for the effect of the large distance. The gas is dense and warm enough in these starbursts so that the CO energy distribution peaks in line 4-3 or higher. 
	
	The list of CO detections quoted in Appendix 1 of Solomon \& Vanden Bout (2005) includes 14 SMGs (and one Lyman Break Galaxy) with z\,=\,1.06-3.41 (see e.g.\,\,Neri et al.\,\,2003, Greve et al.\,\,2005, Tacconi et al.\ 2006), 16 QSOs with z\,=\,1.42-6.42 (see e.g.\,\,Omont et al.\,\,1996a, Ohta et al.\,\,1996, Walter et al.\,\,2003, Bertoldi et al.\,\,2003) and five radio galaxies with z\,=\,2.39-5.20 (e.g.\,\,Papadopoulos et al.\,\,2000, De Breuck et al.\,\,2005, Klamer et al.\,\,2005). Note that all the 11 objects with z\,$>$\,3.5 are prominent AGN. About half of the sources are strongly amplified by gravitational lensing with magnification in the range $\sim$2-20 (it reaches even 45 in a new detection by Kneib et al.\ 2005 displayed in Fig.\,6). Hovever, it is also certain that another large fraction has no significant magnification. The first CO detection in more than 80\% of these sources was achieved with the IRAM-PdB Interferometer, and most of the remaining ones with the OVRO interferometer (see Table 1 of Cox et al.\ 2005).

	In practically all cases with z\,$>$\,2, at least one line was detected in the 3\,mm atmospheric window, corresponding to relatively high rotation number, mostly J\,=\,3-2, 4-3 and 5-4 (in a few cases up to J\,=\,7-6). In some cases, more than a single millimetre CO line was detected, with different J values (see e.g.\ Fig.\,5), including sometimes detections in the 2\,mm window with the IRAM 30m-telescope (e.g.\ Weiss et al.\ in prep.).  
Low-J lines, J\,=\,1-0 or 2-1, were detected in the cm range in ten cases, mostly with VLA (see e.g.\ Fig.\,5) (see also one direct detection at z\,=\,5.19 with ATCA by Klamer et al.\ 2005, and detections with the NRAO Green Bank Telescope (GBT) and the MPIfR Effelsberg 100 m telescope by Riechers et al.\ 2006c). The value of the ratio of the intensities may provide information about the CO rotational temperature, and, hence, the kinetic temperature and the H$_2$ density. However, an accurate modelling of CO line formation remains difficult in the absence of detailed information about the extension and the structure of CO emission, and the clumpiness of the molecular gas (see e.g.\,\,Combes, Maoli \& Omont 1999, Solomon \& Vanden Bout 2005). This limits  the corresponding diagnosis that it could provide for the emitting galaxy. However, systematic programmes such as the one recently carried out on SCUBA-MAMBO SMGs at IRAM-interferometer (Neri et al.\,\,2003, Greve et al.\,\,2005, Tacconi et al.\ 2006), are efficient in providing basic information about the properties of such starburst galaxies at high redshift. The emerging picture is that they share many features with local ULIRGs, and in particular the CO emission is generally concentrated within the central kpc or so. However, starbursts where CO is currently detectable, are more extreme than typical local ULIRGs, with larger FIR and CO luminosities, higher dust temperatures and larger H$_2$ masses. Very recent observations have shown that some SMGs with velocity spread $\sim$1000\,km/s, are unresolved with the $\sim$0.2'' resolution now available with the IRAM interferometer (Tacconi et al.\ in prep.). This indicates that the 1000\,km/s spread occurs within a radius less than 1\,kpc, and implies masses of 10$^{10.5}$\,M$_\odot$ or greater are enclosed in this volume.
Many cases of powerful SMGs display evidence of interaction with neighbouring sources. There are a number of cases of resolved, often double, CO sources, among SMGs (see Section 3.4.2 of Solomon \& Vanden Bout 2005, Genzel et al.\,\,2003 and Tacconi et al.\,\,2006), radio galaxies (Papadopoulos et al.\,\,2000, De Breuck et al.\,\,2005, Greve et al.\ 2006a, and references therein) and prominent QSOs, especially the spectacular double CO source at z=6.42 (Walter et al.\ 2004) (Section 9.4).
	
	Even detections of a single CO line may provide interesting information about basic properties of these high-z starburst galaxies. The general H$_2$/I$_{{\rm CO}}$ relation, properly scaled for such objects (e.g.\,\,Solomon \& Vanden Bout 2005), provides an estimate of the mass of the H$_2$ gas, which may give an indication about the future duration of the starburst. Assuming some standard spatial extension for such a nuclear starburst, similar to the few cases where the sources have been resolved, one may infer an estimate of its total dynamical mass from the velocity width of the CO line, but it depends on the unknown inclination angle. The dozen of SMGs detected in the large programme with the IRAM interferometer have CO luminosities in excess of 10$^{10}$\,K\,kms$^{-1}\,$pc$^2$, and thus  M$_{{\rm H2}}$ of the order of a few 10$^{10}$\,M$_\odot$, and dynamical masses in the range of 10$^{11}$\,M$_\odot$. This implies very massive systems dominated by baryons in their central regions.             
	
	Most of the sources where CO has been detected have FIR luminosities L$_{{\rm FIR}}$\,$\sim$\,10$^{13}$\,L$_\odot$. Compared to the correlation between CO and FIR luminosities in local ULIRGs (Section 5), there is a trend of larger values for the ratio of the FIR luminosity to the CO luminosity. It thus appears that the star formation rate per unit mass of molecular gas was higher in these massive high z starbursts than in the less luminous local ULIRGs.

%Fig5 J1148
    \begin{figure*}
    \centering     %%\includegraphics[width=9.cm, angle=0]{J1148r_mod.ps}
\caption{
Spectra of redshifted CO and \CII~lines of the QSO J1148+5251 (z$_{{\rm CO/CII}}$\,=\,6.42): CO(6--5) (93.2\,GHz) and CO(7--6) (108.7\,GHz) from IRAM interferometer (Bertoldi et al.\ 2003); CO(3--2 (46.6\,GHz) from VLA (Walter et al.\ 2003); and \CII~($^2$P$_{3/2}$--$^2$P$_{1/2}$) (256.2,GHz) from IRAM 30m--telescope (Maiolino et al.\ 2005)(Note that the spectrum of CO(6--5) is reproduced twice to allow an easier comparison with the other spectra of CO and \CII~lines).}
%%157.74\,$\mu$m
 \end{figure*}

%7.3
\subsection{ Molecules in the host galaxies of high-z AGN}

	The fact that more than half the high-z galaxies where CO has been detected, are AGN, mainly bright QSOs and powerful radio galaxies, is probably due to the combination of two reasons. First, there is a high probability of finding strong starbursts in the host galaxies of such AGN as proven by the many detections of dust submm emission: MAMBO-IRAM studies have shown that the probability to detect 1.2\,mm continuum with a flux density $\ga$ 2\,mJy is as high as 25-30\% around bright QSOs with z $\ga$ 2 (Omont et al.\,\,1996b, Carilli et al.\,\,2001, Omont et al.\,\,2001, 2003, Beelen 2004). The probability is even higher in prominent high-z radio galaxies observed with SCUBA-JCMT by Archibald et al.\,\,(2001). Furthermore, such objects are very good cases for searching CO emission for two main reasons:  there is a general good correlation between CO and FIR emission, and the FIR emission is strong in many of these AGN. It is also much easier to have a good determination of the redshift of the molecular gas of AGN, so that it is accurate enough to warrant that the CO line lies within the narrow bandwidth of  millimetre detectors used up to now (note that the bandwidth is significantly increased to more than 5000\,km/s with the current new generation of receivers, and soon to larger values as already 25000\,km/s at Mopra-22m). This explains in particular the absence of non-AGN sources among current CO detections with z\,$>$\,3.5. Practically no SMGs are known with such redshift, probably partly because they are rarer and mostly because their radio detection needed for redshift determination is too difficult (see e.g.\,\,Chapman et al.\,\,2003, 2005). 
	
	Compared to CO-detected SMGs without strong AGN (Section 9.2), the properties of the molecular emission of these AGN host galaxies are not fundamentally different. They have similar H$_2$ masses, a few 10$^{10}\,$M$_\odot$, total virial masses, 
	%$\sim$\, xx 10$^{10}\,$M$_\odot$, 
 and CO linewidths $\sim$\,200-800\,km/s, with QSOs rather at the lower end of this width range and radio galaxies at the upper end (e.g.\,\,Carilli \& Wang 2006). They also display a high fraction of interacting objects, especially for radio galaxies (Papadopoulos et al.\,\,2000, De Breuck et al.\,\,2005). There is nevertheless some indication that their temperature might be somewhat higher than regular SMGs without strong AGN, as suggested by the dust temperatures induced from 350\,$\mu$m emission (Bendford et al.\,\,1999, Beelen et al.\,\,2006). 

		This class of high-z CO AGN includes the brightest high-z AGN known, especially strongly lensed ones, and the QSO with the largest redshift known, discussed in Section 9.4. These AGN also include most of the few cases where other lines than CO (HCN, HCO$^+$, HNC, CN, \CI, C$^+$) have been detected up to now.

%9.4
\subsection{Other species and detailed studies through strong gravitational lensing}
%7.4.1
\subsubsection{Detectability of other millimetre lines}

	All other emission lines from the bulk of the interstellar gas of these high-z starbursts are significantly more difficult to detect than CO. There are already six species outside of CO currently detected - HCN, HCO$^+$, HNC and CN and the fine structure lines of \CI~and \CII~- all in a very few sources, from one to five (H$_2$O was also searched in several sources, without success up to now,  e.g.\ Riechers et al.\ 2006b, Wagg et al.\ 2006). Most of the detections have been achieved thanks to strong gravitational amplification in the few most exceptional lensed sources discussed in Section 9.4.2. Solomon \& Vanden Bout (2005) have devoted a detailed discussion to detections of HCN, \CI~and \CII~which were known at the time, as well as to very exceptional sources. We summarise this below with updates.

	{\bf HCN}. As discussed in Section 5, HCN emission is a very good tracer of dense gas and star formation, including major starbursts such as those of high-z Emission Line Galaxies. However, because of its large electric dipole, high-J rotational levels are difficult to excite, so that HCN detection is easier in practice in the J\,=\,1-0 line redshifted into the cm range (however see 3\,mm detections in Wagg et al.\ 2005 and Gu\'elin et al.\ 2007). There is thus no possibility of compensating the effect of high-z large distances by observing stronger higher-J lines as for CO. Waiting for Extended VLA (EVLA) (and SKA in the long term ), the detection of high-z HCN(1-0) is just at the limit of current facilities (VLA, GBT) for the strongest sources. Four detections have been reported (F10214, Cloverleaf, APM 08279, VCV J1409) together with a few upper limits (see Section 9.4.2; Solomon \& Vanden Bout 2005, Carilli et al.\,\,2005 and Greve et al.\ 2006b). In all cases, the ratios of the FIR and HCN luminosities are within the scatter of the relationship between HCN and far-IR emission for low-z star-forming galaxies, although they have a trend to be larger than the average value of this ratio at low redshift (Carilli et al.\ 2005).
	
  {\bf HCO$^+$} emission is a star formation indicator similar to HCN, tracing dense molecular gas (Section 5). It has been recently detected in the Cloverleaf and APM 08279+5255  (Riechers et al.\,\,2006a, Garc{\'{\i}}a-Burillo et al.\,\,2006b). HCO$^+$ and HCN have similar luminosities, and there is evidence that they come roughly from the same circumnuclear region. 
  
  {\bf HNC} and {\bf CN}. The J\,=\,5-4 line of HNC has been detected, and the N\,=\,4-3 line of CN tentatively detected in APM08279+5255 by Gu\'elin et al.\ (2007, see also Riechers et al.\ 2006d). Both lines intensities are about half that of HCN J\,=\,5-4, so that the [HNC]/[HCN] abundance ratio seems similar to its value in the cold Galactic clouds and much larger than in the hot molecular gas associated with Galactic \HII~regions. 

	{\bf [\CI]}. Atomic carbon may offer an interesting diagnosis of the molecular gas, especially relatively cold, with its two submm ground state lines and their simple excitation pattern (see e.g.\ Bayet et al.\ 2006 for a comparison of \CI~and CO in nearby galaxies). High redshifts often bring these lines in a much easier atmospheric frequency range than at z=0. The relatively small intensity ratio, $\sim$1.5--5, between adjacent CO lines and \CI~lines, makes the latter detectable in a significant fraction of the high-z sources where CO is detected. \CI~has been detected up to now in five high-z sources: F10214, Cloverleaf, SMM J14011, PSS J2322 and APM 08279 (including one detection of the higher excitation line $^3$P$_2$- $^3$P$_1$). The luminosity ratio between the \CI~and CO lines is a potential diagnostic of the gas excitation and carbon chemistry.

	{\bf [\CII]}. The $^2P_{3/2} \rightarrow {^2P_{1/2}}$ fine-structure line of $\rm C^+$ at 157.74\,$\rm \mu m$ (1900.54\,GHz) is known as the most powerful emission line of the interstellar gas of galaxies. It traces in particular photo-dissociation regions associated with star formation, and is thus an important potential tool for studying the corresponding molecular gas. In local galaxies, with far-infrared luminosities $\rm L_{FIR}\,\sim$\,10$^{10}$-10$^{11}\, L_\odot$, the ratio of the $\rm C^+$ luminosity to the far-infrared luminosity, L$_{{\rm [CII]}}/L_{{\rm FIR}}$, is typically a few 10$^{-3}$. However, for ULIRGs with $\rm L_{FIR} \ga 10^{12}\, L_\odot$, this ratio drops by about an order of magnitude (see e.g.\,\,Fig.\,2 of Maiolino et al.\,\,2005). The search for C$^+$ emission at high z has for long suffered from various handicaps: the absence of any `inverse K-correction' compensating the distance factor; the lack of sensitivity in the submm range where the line is redshifted at z\,$\la$\,6; the lack of known adequate sources at z\,$\ga$\,6.4 where the line enters the 1.2\,mm band of current sensitive equipment. Therefore, the repeated efforts to detect the \CII~line at high z remained unsuccessful for more than ten years until two sources were detected very recently: J1148+5251 at z\,=6.42 with the IRAM 30m-telescope (Maiolino et al.\,\,2005; see Fig.\,5) and with more details with the IRAM interferometer (Walter et al.\ in prep.); BR1202-0725N at z\,=4.69 with the SMA (Iono et al.\,\,2006a). The ratio L$_{{\rm [CII]}}/L_{{\rm FIR}}$ is $\sim$\,2--4\,10$^{-4}$, i.e.\,\,comparable to the value observed in the most luminous local ULIRGs such as Arp 220. These results are important because they confirm that, with the gain in sensitivity of at least two orders of magnitudes with ALMA, the \CII~line will be easily detectable in all high-z ULIRGs and even LIRGs in the redshift ranges corresponding to the atmospheric submm windows.

\subsubsection{Strongly lensed and other prominent sources}

	Among the $\sim$40 high-z sources where CO detection has been reported, some deserve a special mention either because they have extraordinarily large amplification by gravitational lensing allowing early detection and detailed studies, or they have the largest redshifts or luminosities, or they are representative of various special types. The most prominent are:
	
	{\bf IRAS FSC10214+4754} (z=2.286). It was discovered by Rowan-Robinson et al.\,\,(1991) as an extraordinarily bright high-redshift IR source among IRAS data. The detection of CO J=3--2 emission  (Brown \& Vanden Bout 1991; Solomon, Radford \& Downes 1992) was the first detection of molecular gas at high redshift. It was later found to be gravitationally lensed by a factor $\sim$10 (FIR and CO) to $\sim$50 (mid-IR). The correction for magnification reduces its properties to those typical of local ULIRGs. It is still the best studied ultraluminous infrared galaxy at high redshift. It contains both a dust-enshrouded quasar obscured by Compton-thick material, responsible for the mid-IR luminosity (Alexander et al.\,\,2005b), and a much larger molecular ring starburst responsible for a substantial fraction of the FIR luminosity. The high magnification has allowed the angular resolution of the CO emission and the detection of HCN and \CI.

%Fig6 Lens A2218
    \begin{figure*}
    \centering %%\includegraphics[width=15.cm, angle=0]{jpkneib-coul-012805.ps}
\caption{(({\it Reproduced from Figure\,1 of Kneib et al.\ 2005}). Observational results at $\lambda$\,$\sim$\,3\,mm from the IRAM interferometer of a faint
submillimetre galaxy, SMM J16359+6612 (SMM1), lying at $z=2.516$ behind the
core of the massive cluster A\,2218.  The foreground gravitational lens
produces three images with a total magnification of 45. ({\it Left}) The $^{12}$CO(3-2) map of SMM1
superposed on the optical {\it HST} image of A\,2218. Three millimetre images SMM1-A, B and C are clearly identified, all three well centered on their optical
counterparts. 
({\it Right}) This panel displays (from top to bottom) the velocity
profile of the three different submm images SMM1-A, B and C, and of the
sum of these three spectra.  A double-peak profile is clearly observed
for SMM1-B and -C, as well as in the sum of all three components
(bottom panel). The two components are separated by
$\sim$280\,km/s and agree well with  two features ($\alpha$ and $\beta$) also identified in optical spectroscopy.}
 \end{figure*}

	{\bf The Cloverleaf} (HH1413+1143, z=2.558). This broad absorption line QSO is a lensed object with spectacular four
bright image components. Barvainis et al.\,\,(1992) discovered strong FIR and submillimeter radiation similar to that of IRAS F10214, showing that bright optical high-z quasars may also be extremely FIR (and mid-IR) luminous. The CO lines are stronger than in any other high-z source (Barvainis et al.\,\,1994), owing to both powerful intrinsic line luminosities and magnification. As a result, this source is the best studied in detail both for the number of CO lines detected (Barvainis et al.\,\,1997, Weiss et al.\,\,2003), and for the angular resolution (Yun et al.\,\,1997, Alloin et al.\,\,1997, Kneib et al.\,\,1998, Venturini \& Solomon 2003, Weiss et al.\,\,2003 and references therein). It is also one of the few high-z sources where HCN, HCO$^+$ and \CI~lines have all been detected.

	{\bf APM 08279+5255} (z=3.911). This extremely bright broad absorption line quasar has both one of the highest magnifications ($\mu$\,$\sim\,$100 at optical-IR wavelengths but only $\sim$\,7 for CO emission), and one of the largest intrinsic FIR and CO luminosities. This has again allowed multi-line CO studies, including strong J\,=\,9-8 emission from $\sim$200\,K hot gas of sub-kiloparsec size (Downes et al.\,\,1999), and even the J=11-10 line
(Weiss et al.\,\,in preparation), as well as angular resolution of the central nuclear emission (Lewis et al.\,\,2002, Papadopoulos et al.\,\,2001). Both HCN J\,=\,5-4 and HCO$^+$ J\,=\,5-4 line emission from the dense molecular gas has been reported (Wagg et al.\,\,2005; Garc\'ia-Burillo et al.\,\,2006b), as well as \CI~(Wagg et al.\,\,2006), HNC and CN (Gu\'elin et al.\ 2007).

	{\bf BR 1202-0725} (z=4.69). This prominent quasar is one of the brightest sources at 1.2\,mm (McMahon et al.\,\,1994), and also the third one which was discovered in CO emission (Omont et al.\,\,1996a, Ohta et al.\,\,1996). Both CO and 1.2\,mm emission are split in two components, 4'' apart, of comparable strengths; each of them being among the strongest high-z sources, with apparent (if non lensed)  $\rm L_{FIR} \ga 10^{13}\, L_\odot$ (but modest combined mid-IR luminosity, Hines et al.\,\,2006). There is an enormous difference in optical/near-IR extinction between the QSO component and the northern source which is not optically detected (even in the near-IR K-band at its centre). Radio emission has also been detected (both components), as well as recently (Iono et al.\,\,2006a) X-ray (mostly QSO component) and \CII~line emission (northern component). The origin of the double structure, lens or two interacting prominent starburst galaxies, remains a puzzle; see Solomon \& Vanden Bout (2005), Carilli et al.\,\,(2002), Iono et al.\,\,(2006a), Sameshima (2006) for arguments pro and con each possibility. However, two galaxies seem more probable.

	{\bf SDSS J1148+5251} (z=6.42). This is the most distant quasar known to date (Fan et al.\,\,2003). Its multi-wavelength detection (see references in Solomon \& Vanden Bout 2005, Hines et al.\,\,2006,  Charmandaris et al.\,\,2004, Beelen et al.\,\,2006), including CO(3--2) with VLA (Walter et al.\,\,2003) which proves to be resolved (Walter et al.\,\,2004), and CO(6--5) and CO(7--6) with IRAM interferometer (Bertoldi et al 2003) (see Fig.\,5), are typical of other high-z strong starburst QSOs. This shows the presence of a giant starburst and large amount of molecular gas, with heavy elements, less than one billion years after the Big Bang. It is also the first high-z source where the \CII~line emission was detected (Maiolino et al.\,\,2005).

	Figure 6 presents the results of CO observations from the 
IRAM interferometer 
%(Fig.\,6a) 
of the triple image of another spectacular lensing case, 
the submillimetre galaxy {\bf SMM J16359+6612} lying at $z=2.516$ behind the
core of the massive cluster A\,2218, which produces a magnification of $\sim$45 of an intrinsically weak SMG (Kneib et al.\ 2005).  	
	%Other important sources individually discussed by Solomon \& Vanden Bout (2005) include VCV J1409 (z=2.583), PSS J2322+1944 (4.119), SMM J14011 (2.565), SMM J02399 (2.808), SMM 16359+6612 (2.517) and 4C 41.17 (3.796). 

\section{Infrared H$_2$ emission, tracer of warm molecular gas, shocks and photodissociation regions
}   
 
\subsection{Basic features and physics of H$_2$ emission, and Milky Way observations}

	In normal conditions, H$_2$, the overwhelming compound of the molecular interstellar medium, is not easy to directly detect (Shull \& Beckwith 1982, Combes \& Pineau des Forets 2000). UV absorption (see Section 7.1) is inefficient to probe deeper than the outskirts of molecular clouds with A$_{\rm v}$\,$\la$\,1 (however, see Keel 2006). Rovibration infrared lines are limited to weak quadrupole lines. They include: i)  vibration lines with $\Delta$v\,=\,1,2...\ -- S(J$_{\rm l}$), Q(J$_{\rm l}$), O(J$_{\rm l}$) with $\Delta$J\,=\,+2, 0 and -2, respectively -- with wavelengths in the good~2\,$\mu$m atmospheric window for the most important lines with $\Delta$v\,=\,1; ii) pure rotation lines S(J$_{\rm l}$) ($\Delta$J\,=\,2, which do not connect the rotation levels corresponding to different nuclear spin configuration (ortho-H$_2$ with I=1 and J odd and para-H$_2$ with I=0 and J even) which span the whole mid-IR range: S(0) 28.22\,$\mu$m, S(1) 17.03\,$\mu$m, S(2) 12.28\,$\mu$m, S(3) 9.66\,$\mu$m, S(4) 8.03\,$\mu$m, S(5) 6.91\,$\mu$m, S(6) 6.11\,$\mu$m, S(7) 5.51\,$\mu$m, etc. All these lines have high excitation energy, from 510\,K for S(0) to 7200\,K for S(7) pure rotation lines, and 5600\,K and 10000\,K for v\,=\,1-0 and v\,=\,2-1 lines respectively.  
	
	Such IR lines are practically useless in absorption (however, see Lacy et al.\,\,1994, Usuda \& Goto 2005) because of the extremely large values for the column density N$_{{\rm H2}}$ they require to be detectable: typical values of N$_{{\rm H2}}$ needed to achieve an optical depth  $\tau$\,=\,0.01 are a few 10$^{23}$\,cm$^{-3}$ (A$_{\rm v}$ $\sim$ a few 10$^2$) for v\,=\,1-0 2\,$\mu$m ro-vibration lines and a few 10$^{24}$\,cm$^{-3}$ (A$_{\rm v}$ $\sim$ a few 10$^3$) for pure rotation lines (see e.g.\,\,Shibai et al.\,\,2001).
		
	On the other hand, despite the high excitation energies, near-IR and mid-IR emission lines of H$_2$ have proved to be widespread and important diagnosis tools in all molecular media harbouring energetic processes, related to massive star formation, shocks or AGN, able to achieve upper level excitation (see e.g.\ the introduction of Roussel et al.\ 2007 for a review of previous references on extragalactic H$_2$ infrared lines). The physics of the different excitation processes is various; but the most important ones are well understood for a molecule as simple as H$_2$. This  allows very precise modelling when the astrophysical conditions are well defined. Thermal excitation is generally dominating for the pure-rotational lines (Section 10.3). An excitation temperature $\sim$200\,K is enough to ensure a substantial excitation of the S(0) 28.22$\mu$m line, $\sim$300-400\,K for S(1), but $\sim$1400\,K for an upper line such as S(7) (see e.g.\ Rosenthal et al.\ 2000, Higdon et al.\,\,2006a). Pure rotational lines of H$_2$ have been detected in various Galactic sources with ISO (see e.g.\ the Galactic Center by Lutz et al.\,\,1996, Orion by Rosenthal et al.\,\,2000) and {\it Spitzer}.
	
	The case of the higher energy required to excite vibration lines is more complex. Following the first detections of such lines thirty years ago, their excitation has been the object of elaborated modelling which allows a precise use of such lines for diagnostic of the conditions in the emitting medium. The two main excitation routes are UV fluorescence and collisions in the hot gas of shocks, although it has been proposed that rovibrational excitation in the process of H$_2$ formation may play a significant role, in particular in recombination on the grain surfaces (see e.g.\,\,references quoted in Dalgarno 2001). In many cases, the properties of the H$_2$ near-IR emission are compatible with thermal collisional excitation in a hot gas ($\sim$2000\,K) such as found in strong shocks like prominent Galactic sources (see e.g.\,\,in the close vicinity of the Galactic Center, Gatley et al.\,\,1984) or Orion, Beckwith 1981). In such shocked regions, the H$_2$ rovibration populations in the ground electronic state are usually well thermalised by collisions (Shull \& Beckwith 1982; Black \& van Dishoeck 1987 and references therein). A direct indication of dominent collisional shock excitation is a very small value for the ratio of similar v\,=\,2-1 and v\,=\,1-0 lines ($\sim$0.13 at 2000\,K).
	
	Alternatively, in the vicinity of strong radiation sources, such as photodissociation regions, ultraviolet pumping may dominate the vibrational excitation of H$_2$. There is then no energy limitation for populating the higher v levels. The accurate knowledge of the various ultraviolet and infrared transition probabilities allows a precise prediction of the resulting intensities of the emission infrared lines. As discussed e.g.\,\,by Dalgarno (2001) and references therein, particularly useful is the ratio of the S(1)(2-1) and S(1)(1-0) intensities. It is found to be close to the predicted value, 0.54, in various Galactic reflection nebulae including the best studied one NGC\,2023. UV excitation by photons from OB stars is also believed to dominate the excitation of the wide H$_2$ emission observed in the inner 400 pc region of the Galaxy, where the ratio of the H$_2$ to far-IR luminosity agrees with that in starburst galaxies (Pak et al.\,\,1996).
	 
	 The multiplicity of the rovibrational lines may span a broad range of excitation energy (e.g.\,\,from 500\,K to 17000\,K in Orion, Rosenthal et al.\ 2000). The accuracy of the models then allows various other diagnostics, in particular of their selective extinction. For high gas densities, collisions and fluorescence may compete, and the relative line intensities can be used to infer densities and radiation fields (Sternberg \& Dalgarno 1989). In addition, other heating mechanisms than shocks may be important, such as X-ray illumination where hard X-ray photons are capable of penetrating deeply into molecular clouds and heating large amounts of gas (e.g.\,\,Maloney et al.\ 1996). Even heating by strong UV in photodissociation regions may be the main excitation mechanism of the first pure-rotational lines.

	%%2 pages a revoir avant Opticon

\subsection{2\,$\mu$m H$_2$ emission in galaxies}

	The studies of the H$_2$ near-infrared lines in galaxies were initiated in the late 1970s by Gautier et al.\ (1976), at the begining of infrared spectroscopy. Since this time they have followed all the progress of the telescopes and instrumentation of near-infrared astronomy. This was already a mature field in the 1980s and 1990s. It remains at the forefront of major developments in progress such as adaptive optics studies, JWST and extremely large telescopes.
	
	As it could be inferred from the results in the Galaxy and especially in the Galactic Center region, detectable near-IR emission from external galaxies comes from large concentrations of hot molecular gas found in extended regions of strong photodissociation or shocks, especially in the central regions hosting AGN or major starbursts.
	
	Various H$_2$ rovibration transitions were observed in a number of objects in the Magellanic Clouds (see Israel \& Koorneef 1991, Pak et al.\,\,1998, and references therein). Consistent conclusions are that UV radiative excitation is the energetically dominant mechanism. 
	
	The 2\,$\mu$m lines of H$_2$ are relatively easy to detect in the central regions of nearby galaxies especially when they present some AGN or strong starburst. Many, more or less detailed, spectroscopic studies have thus been carried out. However, the analysis of the early results was often not obvious and controversial (see e.g.\ the reviews by Mouri 1994; Goldader et al.\,\,1997). Indeed, more comprehensive results very often reveal a mixture of various excitation mechanisms from UV, shocks and even AGN X-rays, at various spatial scales. The most fruitful observations have best succeeded in disentangling such a complexity by high angular resolution and velocity-resolved spectroscopy. 
	
	The most widespread extragalactic H$_2$ emission is related to starbursts. In the central regions of starbursts galaxies, such as the archetypes M\,82 or NGC\,253, the great majority of the 2\,$\mu$m line emission arises from energy states excited by ultraviolet fluorescence (Pak et al.\,\,2004). It is also found that the ratio of the H$_2$ v=1--0 S(1) line to FIR continuum luminosity is constant over a broad range of galaxy luminosities, as well as in normal late-type galaxies (including the Galactic center) as in nearby starburst galaxies, and especially in LIRGs (Goldader et al.\,\,1997, Pak et al.\,\,2004). This is consistent with a common origin of FIR and H$_2$ emission from the UV radiation from photodissociation regions (PDRs) illuminated by recently formed OB stars. The spatial distribution of the H$_2$ emission also correlates well with the submillimetre continuum emission and the CO emission (Pak et al.\,\,2004).
	
	Even a good part of the H$_2$ line emission showing evidence of thermalisation in starburst nuclei and especially ULIRGs (Davies et al.\,\,2003, Pak et al.\,\,2004) should come from the densest parts of PDRs (n$_{{\rm H2}}$\,$\ga$\,10$^4$\,cm$^{-3}$). While the v\,=\,1 levels are thermalised at $\sim$1000\,K, UV-pumped gas is needed to account for the higher levels. A similar conclusion applies for the extended H$_2$ emission within a few hundred parsec around many AGN (Davies et al.\,\,2006 and references therein).  However, Quillen et al.\,\,(1999) argued that shock excitation may also be dominant in a fraction of the Seyfert galaxies they studied (see also Rodr{\'{\i}}guez-Ardila et al.\,\,2005). Another case of shock dominated emission is the relatively nearby, double nucleus, late merging, luminous LIRG NGC 6240. We group the discussion of its extraordinarily strong H$_2$ rovibration lines, discovered by Joseph et al.\,\,(1984), with that of the equally strong rotation lines in Section 10.3.

	Modern instrumentation on 8-10\,m telescopes gives new opportunities for near-infrared H$_2$ studies of the central regions of starburst and AGN (see e.g.\ observations with the Integral Field Unit of the Gemini Near-Infrared Spectrograph reported by Riffel et al.\ 2006a). The availability of efficient adaptive optics instrumentation, such as integral-field spectroscopy with SINFONI at VLT, opens the possibility to directly study up to size scales ($\la$\,10\,pc) comparable to those on which models predict the molecular torus around AGN should exist (e.g.\,\,most recently Schartmann et al.\,\,2005), as shown by the first results (Davies et al.\,\,2006, Mueller-Sanchez et al.\,\,2006, Zuther et al.\,\,2006, Neumayer et al.\ 2007, Reunanen et al.\ 2007).

	Very extended near-IR H$_2$ emission have also been detected {\it up to 20\,kpc}, in the extended regions of the central galaxies of several tens of cooling-flow clusters (Donahue et al.\,\,2000, Edge et al.\,\,2002, Jaffe et al.\,\,2005, Johnstone et al.\ 2007). Warm H$_2$ ($\sim$1000-2500\,K) seems present wherever there is ionization in the cores of cooling flows, and in most cases it also coincides with CO emission. The relative strentghs of the P$\alpha$ line to the H$_2$ lines might indicate a source of UV excitation hotter than 10$^5$\,K, whose nature is still unknown.
		
	The prospects appear extremely rich for exploiting the ubiquitous emission of near-IR lines of H$_2$ in strong starbursts and shocks, with the expected new capabilities of infrared astronomy: both from the ground for local galaxies, with the developments in instrumentation, adaptive optics, extremely large telescopes and interferometry; and from space, especially at high redshift, with the jump in performance expected from JWST (and even SPICA), with respect to {\it Spitzer} and ISO.

\subsection{Mid-IR H$_2$ pure-rotational lines in galaxies}

	The emission of rotational lines of H$_2$ by galaxies pertakes many features with that of rovibrational lines. In particular, the main emitters are again the various types of starbursts and the AGN. However, there are two main differences: their excitation energies are a factor 5-10 smaller which makes their collisional excitation much easier, so that they probe the more widespread moderately warm molecular gas of only a few hundred Kelvin; they  are extremely difficult to observe from the ground so that practically all the results have come from the space missions ISO and {\it Spitzer}.
	
	The results of ISO on molecular hydrogen and warm molecular gas in galaxies have been recently reviewed by Verma et al.\,\,(2005) (see also Habart et al.\ 2005). In addition to the detections that they quote in the spectra of half-a-dozen active galaxies, the most comprehensive work is the analysis of a sample of 21 starburst and Seyfert galaxies presenting pure rotational lines from S(0) to S(7) by Rigopoulou et al.\,\,(2002). A multi-line analysis, including S(0), in this sample yields a temperature around $\sim$150\,K for the bulk of the warm gas, both for starbursts and Seyferts. The mass of this warm gas is about 10\% of the total mass of molecular gas probed by CO in starbursts, and a larger fraction in Seyferts. Such a temperature and mass of warm gas are compatible with PDR heating in starbursts, and with additional heating of a larger mass fraction by X-rays in Seyferts. However, low velocity shocks may also contribute.
	
	{\it Spitzer} has detected H$_2$ in much more galaxies. However, published results and their analysis are still very incomplete. Higdon et al.\,\,(2006a) have reported multi-line detections of a large sample of $\sim$60 ULIRGs. The results extend those of Rigopoulou et al.\ (2002). However, the lack of sensitivity for the S(0) line tends to favour the emission of smaller masses of warmer gas ($\sim$300\,K) in the S(1)-S(3) lines. Hotter gas is also revealed in a small fraction of the sample by the detection of the S(7) line.
	
	A major {\it Spitzer} result is the determination of the properties of warm H$_2$ in the central regions
of normal galaxies derived by Roussel et al.\ (2007) from measurements of rotational lines in the Legacy Program 
SINGS (Kennicutt et al. 2003). This study extends previous extragalactic surveys of emission lines
of H$_2$, to fainter and more common systems (L$_{{\rm FIR}}$ = 10$^7$ to 6x10$^{10}$ L$_\odot$) of all morphological
and nuclear types. It has securely detected the 17 $\mu$m S(1) transition in about 45 
galaxies, probing the range 100--1000\,K. The derived column densities amount to a significant fraction of column densities of the total molecular hydrogen, between 1\% and more than 30\%. The H$_2$ line intensities scale tightly with the emission
in the PAH bands which can be understood from a dominant origin in photodissociation
regions. However, many sources classified as AGN strongly depart from the rest of the sample, in having warmer
H$_2$, smaller mass fractions of warm gas, and an excess of power emitted in H$_2$ with respect to PAHs, favouring shock excitation.  In many star-forming sources, deviations from an
apparent ortho to para ratio of three are detected, consistent with the effects of pumping by
far-UV photons combined with incomplete ortho-para thermalization by collisions.

		Most of these ISO and {\it Spitzer} studies were focussed on the central regions of galaxies. However, see Valentijn \& van der Werf (1999); and among the most remarkable works are a few observations of extended regions in strongly interacting /merging galaxies. With ISO, Lutz et al.\,\,(2003) found very strong mid-IR H$_2$ lines, from S(0) to S(11), in the late merging, 'double active nucleus', system NGC 6240. This luminous LIRG was known to display complex, high-velocity H$_2$ 2$\mu$m emission extending over $\sim$5\,kpc and peaking between the two nuclei (van der Werf et al.\,\,1993, Tecza et al.\,\,2000). Shocks due to the turbulent central velocity field and the superwind created in a nuclear starburst are likely to dominate these extraordinary levels of emission. From a re-analysis of archival ISOCAM-CVF data of the early merger the Antennae, Haas et al.\,\,(2005) have found that the strongest H$_2$ emission is displaced from the regions of active star formation (Fig.\,3). This indicates that the bulk of excited H$_2$ gas is shocked by the collision itself in the region where the two galaxies overlap. 
	
	Even more exceptionally strong H$_2$ lines have been reported by Egami et al.\ (2006) in the infrared-luminous brightest galaxy of the cluster Zwicky 3146 (z=0.29). The line luminosities and inferred warm H$_2$ gas mass (~10$^{10}$\,M$_\odot$) are six times larger than those of NGC 6240. 
	Strong H$_2$ pure-rotational emission lines are also seen in cooling-flow clusters (Johnstone et al.\ 2007), and in some mid-IR weak radio galaxies, where these lines are excited by shocks induced by galaxy mergers or by AGN jets in the interstellar medium (Ogle et al. 2006). 
	One particularly spectacular result is the detection by Appleton et al.\ (2006) of a powerful high-velocity H$_2$ emission associated with an intergalactic shock wave in the Stephan's Quintet group of galaxies. The molecular emission extends over 24 kpc along the X-ray emitting shock-front, and seems to be  generated by the shock wave caused when a high-velocity intruder galaxy collides with filaments of gas in the galaxy group. 
	The S(1) and S(2) lines of warm molecular hydrogen ($\sim$400\,K) were also recently detected with {\it Spitzer} by Higdon et al.(2006b) in two tidal dwarfs galaxies formed in the tidal trails of NGC 5291. 
	
	Many other important results may be expected from {\it Spitzer}, including serendipiteous ones, thanks to the remarkable capabilities of the Infra-Red Spectrometer (IRS) and the large amount of time it has devoted to spectra of galaxies. 
	
%%Appleton

\subsection{Prospects for detecting H$_2$ in forming galaxies}
%with large far-IR telescopes in space}
% (SPICA, SAFIR project, etc).}

	  \subsubsection{Warm molecular gas at various redshifts}.
	  
	There is no doubt that the role of infrared H$_2$ lines to probe the warm molecular gas in various violent contexts in galaxies will further develop in the future with the expected increase of ground and space capabilities. The expectations for the near-IR rovibrational lines with ELTs, adaptive optics and JWST have been discussed in Section 10.2. JWST will bring a similar breakthrough for pure-rotational lines at moderate  redshifts, with the tremendous gain in capabilities expected for its mid-IR instrument MIRI with respect to {\it Spitzer} IRS (Section 12.3.1). Space missions more oriented toward far-IR, such as SPICA (Matsumoto 2005) and the SAFIR project (Benford et al.\,\,2004), or dedicated to H$_2$ lines such as the H2EX project (Falgarone et al.\ 2005b) will allow an extensive exploration of the bulk of the warm H$_2$ gas in the low-z universe and in more or less deep extensions at high z.

	  \subsubsection{Importance of H$_2$ lines in the physics of galaxy formation}. 
	  
	 The presence of molecules is thought to be essential for various steps occuring in the process of galaxy formation. However, outside the marginal cases of tidal dwarfs and other mergers discussed above, true cases of galaxies in formation, especially primordial ones at very high z, are rather out of reach of current observations of molecular lines. Their detection is an important goal for JWST and eventually ALMA. Similarly to star formation, molecules are thought to be essential for cooling the gas below $\sim$10$^4$\,K, which is a prerequesite for full gravitational collapse of proto-galaxies and most proto-stars. In primordial gas, before the formation of heavy elements in the first stars, the species available for achieving such a cooling from $\sim$10$^4$\,K to $\sim$10$^2$\,K, are very limited, only H$_2$ and HD indeed. However, in the absence of the extremely efficient formation of H$_2$ on grains, the common belief is that molecules may exist only as very small amounts, despite the fact that the gas thermodynamical conditions would be highly in favour of converting all hydrogen atoms into H$_2$. However, even small amounts of H$_2$ or HD may be enough for reaching the required cooling efficiency of condensations of primordial gas of various sizes from proto-galaxies to massive proto-stars. 
	 
	 The first question to address is then the expected amount of H$_2$ and HD in the post-Big Bang primordial gas issued from the recombination process, until it forms the first dark-matter and baryonic condensations leading to the first stars and galaxies. Despite the  limited observational information, this is certainly one of the part of the Universe where the chemistry should be the best understood because of its extremely simple element abundances, well determined physical parameters  and the absence of structure (until the first collapses of density peaks into dark-matter halos). Indeed, the exquisite measurements of the CMB features, such as the acoustic peaks (e.g.\ Spergel et al.\ 2006), confirm how amazingly well are understood the details of the physics of hydrogen recombination (see e.g.\,\,Seager et al.\,\,2000 for a detailed description of this physics).

	 The derivation of molecular abundances in this primordial gas is indeed a splendid academic problem which might have born fundamental implications for galaxy formation and even CMB anisotropies. It is interesting to realize that the calculations of the primordial H$_2$ abundance have extended over more than 40 years, practically since the discovery of the CMB. The main processes forming H$_2$ in such conditions were immediately identified, using H$_2$$^+$ as an intermediate product  (Saslaw \& Zipoy 1967)  or more importantly H$^-$  (Peebles \& Dicke 1968). Later refined works (e.g.\,\, Lepp \& Shull 1984, Puy et al.\,\,1993, Palla et al.\,\,1995, Galli \& Palla 1998, Stancil, Lepp \& Dalgarno 1996, 1998,  the review by Lepp, Stancil \& Dalgarno 2002, Puy \& Pfenniger 2005, N{\'u}{\~n}ez-L{\'o}pez et al.\ 2006) included other intermediates such as HeH$^+$ (also negligible with respect to H$^-$), calculations of deuterium and lithium chemistry, and updated cosmological parameters and reaction rates. Some of them brought substantial revisions to the important abundances of HD and LiH.  Even the recent work by Hirata \& Padmanabham (2006) have brought a non negligible decrease of the H$_2$ abundance from 2.6\,10$^{-6}$ to 6 \,10$^{-7}$, by taking into account the effects of the nonthermal background produced by cosmic hydrogen recombination.

	 However, 
	 %it could seem that all these improvements remain rather of academic interest, since 
it is now well agreed that the abundance of H$_2$ in the non-condensed  primordial gas, $\sim$\,10$^{-6}$ (e.g.\ Hirata \& Padmanabhan; as well as that of HD $\sim$\,10$^{-10}$ e.g.\ N{\'u}{\~n}ez-L{\'o}pez et al.\ 2006), is far too small for cooling of early halos.  
Indeed, while it is widely believed that H$_2$ (and HD) cooling is essential in the final  collapses forming the first stars and galaxies, it is recognized that the only H$_2$ and HD important for this are formed in the enhanced density gas of already partially collapsed halos (e.g.\ Tegmark et al.\ 1997). However, the basic chemical processes responsible for the formation of H$_2$ in this denser gas are mostly the same as in the non-condensed primordial gas, i.e.\ mainly the H$^-$ channel.  Similarly, the abundance of primordial LiH is now found to be so small (Lepp et al.\,\,2002) that its possible role in scattering CMB photons seems negligible, and is anyway much smaller than that of Li.
	 
	 Indeed, the question of the effect of H$_2$ and HD in galaxy formation is made more difficult because their abundance in the stages where they may be important, much depends on the highly complex and nonlinear processes involved in the formation of galaxies. In this context, eventually important molecular processes at galactic scales are strong shocks which should unavoidly form large amounts of molecules in the postshock dense cooling gas. Given the importance of H$_2$ emission in local galactic-size shocks, there is no doubt that H$_2$ (and HD) lines should be the main coolants of such shocks in primordial gas below $\sim$10$^4$\,K. The rovibrational lines of H$_2$ emitted in the strongest shocks should be detectable by JWST (e.g.\,\,Ciardi \& Ferrara 2001), and some of them 
%#
perhaps by SPICA as well as some pure-rotational lines 
%#
(Mizusawa et al.\,\,2005\, see also Omukai\& Kitayama 2003 for the more ambitious SAFIR project). As discussed above, {\it primordial} HD should also be negligible in galaxy formation. However, various studies (e.g.\,\,Uehara \& Inutsuka 2000, Flower \& Pineau des For\^ets 2001, Galli \& Palla 2002, N{\'u}{\~n}ez-L{\'o}pez et al.\ 2006, Johnson \& Bromm  2006, Greif et al.\ 2007) underline the substantial contribution of newly formed HD to gas cooling below $\sim$500\,K during the collapse of primordial clouds. The possibility of 
%#
marginally detecting HD rotational lines with ALMA, especially the first one, 1-0, with rest wavelength 112\,$\mu$m, was discussed in particular by 
%\#
%Kamaya \& Silk (2003) and 
N{\'u}{\~n}ez-L{\'o}pez et al.\ (2006)
%#
(see also Mizusawa et al.\ 2005). Despite the uncertainty about the abundance of HD, there is a 
%# TBC
marginal chance that this line may be detected with ALMA, in massive starburst galaxies, when their redshift is large enough (z\,$\ga$\,6.3) to move the line into a good atmospheric window.
	  
	  The above conclusions about H$_2$ lines may apply to various objects displaying strong shocks in the era of formation of the first stars and galaxies and reionization (e.g.\ Johnson \& Bromm  2006, Alvarez et al.\ 2006, Wise \& Abel 2007). More generally, modelling the behaviour of H$_2$ and its emission lines is an essential part of the current intense simulation  activity of this crucial era (see e.g.\ Yoshida et al.\ 2003, Ricotti \& Ostriker 2004, Susa \& Umemura 2004, Reed et al.\ 2005 and references therein). With the sensitivity of JWST and maybe of SPICA, H$_2$ lines will be an essential diagnostic tool for the most massive objects of this yet unexplored epoch. %However, with the formation of the first metals, it is sure that dust and other molecules rapidly provide important alternative cooling channels.

	 	 \section{Molecules and Active Galactic Nuclei (AGN)}

\subsection{Interplay between super-massive black holes and their host galaxy}

		The central super-massive black holes are an essential ingredient of galaxies despite their relatively small mass compared to the total galactic mass. Indeed, the total energy radiated during the AGN phases may be comparable to the total energy radiated by stars, and thus may deeply perturb the whole galaxy. It is now well established  that many -- and perhaps all -- massive luminous nearby galaxies contain central supermassive black holes at their centres with masses $\sim$\,10$^6$-10$^9$\,M$_\odot$, with tight relations between the black-hole mass and the spheroid which harbours it (e.g.\,\,Kormendy \& Richstone 1995, Magorrian et al.\,\,1998, Kormendy \& Gebhardt 2000,  Ferrarese et al.\ 2006). While the tightest correlation is with the central velocity dispersion, $\sigma$, of its host bulge, with M$_{{\rm BH}}$ $\propto$ $\sigma$$^4$ (e.g.\,\,Tremaine et al.\,\,2002, Ferrarese 2002), it also results that M$_{{\rm BH}}$ is approximately proportional to the spheroid mass, M$_{{\rm BH}}$ $\approx$ 2.5 10$^{-3}$ M$_{{\rm bulge}}$ (see e.g.\,\,Merritt \& Ferrarese 2001, McLure \& Dunlop 2004). Although the origin of the M$_{{\rm BH}}$-$\sigma$ relation is not yet fully established, various suggestions have been made  (see e.g.\,\,a list in Di Matteo et al.\,\,2003, 2005 and Begelman \& Nath 2005). Most of them, following e.g.\,\,Silk \& Rees (1998), have stressed the probable dominant influence of strong feedback on the black hole growth due to the action of quasar flows on the galactic gas reservoir.
		%either at large distances  or in the immediate nuclear environment. 
		
		It is therefore not surprising that there is some correlation between the phases when the galaxy and the black hole build most of their respective masses, and thus between AGN and starburst activities. An essential factor for explaining such a correlation is certainly the fact that both the black-hole accretion and the starburst are fed from the interstellar gas which must be strongly perturbed in both cases, either to be transported to the galactic centre, or to be compressed to initiate star formation. In both cases the gas must also eventually be dense. It is thus natural that the association of molecules with AGN is important in various respects and at different scales. We will distinguish three scales: i) on parsec scale, the accretion molecular disk and accompanying jets and outflows, and the much thicker molecular torus (Krolik \& Begelman 1988, Elitzur \& Shlosman 2006 for recent references); 
		%accounting for obscured AGN (`Type 2'); 
ii) on hundred parsec scale, the dense gas with complex structure, location of nuclear starbursts and essential for transporting the gas to the very centre; iii) the scale of a whole galaxy, or at least its bulge for spirals, where the relations with the black-hole mass hold. This large scale is the theatre where the AGN feedback eventually plays in dispersing the interstellar gas. This is also the case of galaxy mergers. They generate major molecular starbursts, both extended and nuclear, and affect the AGN by transporting large amounts of gas to the centre. Molecular clouds also contribute to the dissipation which eventually allows black-hole mergers (e.g.\,\,Makino \& Funato 2004 and references therein).
		
		We will not develop much further the peculiarities of the molecular medium of AGN host galaxies at such large galactic scales, because much of this have been discussed in other sections: about the general relation between starbursts and AGN in local LIRGs and ULIRGs (Section 5), in high-z ULIRGs (SMGs) (Section 9.2), in high-z bright QSOs and radio galaxies (Section 9.3); as well as about H$_2$ emission in AGN host galaxies and at very large distance in shocks triggered by AGN jets (Section 10).
		
		Let us stress nevertheless that the main effect of AGN on the molecular medium of galaxies at all scales is its eventual destruction in the final feedback, and when this effect is total, the AGN emission has also stopped. However, there are interesting cases to consider, especially for QSOs, when there is significant and even strong AGN emission without much molecular gas. This occurs especially at relatively low redshift, $\la$\,1. In the feedback model, it could take place just before complete feedback stops the accretion, or at the occasion of minor late mergings. Let us also recall the case of cooling flows in massive cluster galaxies which are powerful AGN with radio jets, strong resulting feedback, and often CO emission (Section 4.2). In this context the CO and cold dust emission of QSOs and radio galaxies is interesting for tracing the presence of starbursts. It has been seen that very strong CO and dust millimetre emission is frequent in z\,$\ga$\,2 QSOs and radio galaxies. However, it is quite different at z\,$\la$\,1. Both systematic surveys of CO in local (z\,$<$\,0.15) radio galaxies (Evans et al.\,\,2005), and 1.2\,mm dust emission in 3CR radio galaxies and quasars (Haas et al.\,\,2003) have shown that starburst ULIRGs in the host galaxy of powerful AGN are much rarer locally than at high z. Nevertheless, CO studies show that a strong starburst activity, typical of LIRGs and nuclear starbursts, is present in a large fraction of prominent local AGN, including: luminous QSOs (Palomar Green, PG; Scoville et al.\,\,2003; however, see Bertram et al.\,\,2006), hard X-ray selected Seyferts (Rigopoulou et al.\,\,1997), and galaxies hosting F-R I and compact radio jets (Evans et al.\,\,2005). See also the large fraction of PAH emission observed by Schweitzer et al.\,\,(2006) in a sample of 27 PG QSOs at z $\la$\,0.3. Even ULIRG starbursts are present in  local infrared ultraluminous QSOs (e.g.\,\,Hao et al.\,\,2005) with large FIR excess, and CO has been detected in a number of them (Solomon et al.\,\,1997, Evans et al.\,\,2001, 2006, etc.). Detailed interferometric studies of prominent objects such as 3C\,48 (Krips et al.\,\,2005a) and PDS\,456 (Yun et al.\,\,2004) confirm the existence of rich merger structures and give clues about the joint AGN-starburst evolution. However, it is shown by Ho (2005) that the evidence of molecular gas is not enough to infer the presence of a strong starburst, since the star formation efficiency in otherwise gas-rich host galaxies may be suppressed in the presence of strong AGN feedback.

%·	11.2 
\subsection{Molecules in the central regions and fueling the AGN}	
%%Problematics and models for AGN fueling

	The region in the vicinity of the `molecular torus' is fundamental for the diagnostic of AGN processes. It is complex as regards its structure, dynamics, radiative processes, outflows, turbulence, shocks, magnetic field, etc. Besides, this region has the interesting property to be strongly irradiated by the X-rays of the AGN and allows one to study their physical and chemical effects on the surrounding molecular medium (Maloney et al.\,\,1996). 
	
	As discussed in Section 10.2, H$_2$ near-infrared emission lines are widely observed in nuclear and circumnuclear regions of AGN (see e.g Riffel et al.\ 2006b for a near-infrared atlas of 47 AGN including lines of H$_2$, and Roussel et al.\ 2007 for H$_2$ rotational lines in nearby Seyferts and LINERs). The lines of H$_2$ are commonly used as diagnostic of the conditions in these regions. Comparing the ratios in several H$_2$ lines allows one to probe the physical conditions in warm molecular clouds which are irradiated by  ultraviolet (or X-ray) radiation or heated by shocks. With the ongoing increase in angular resolution they will become more important to probe the complex structure, dynamics and star formation of the immediate vicinity of the molecular torus, and their relation with feeding the black-hole accretion.
	
	The current sensitivity and angular resolution of millimetre interferometers is well adapted to map one or several CO lines in the central kpc around nearby AGN. The processes which removes the angular momentum of the gas, brings it to the centre and feeds the AGN are complex and not very well understood (e.g.\,\,Combes 2005). Several CO surveys have thus addressed the central structure and dynamics of the molecular gas in a number of nearby AGN (Jogee et al.\,\,2001, Garc\'ia-Burillo et al.\,\,2003 [NUGA], Helfer et al.\,\,2003). The most striking result is the large variety of circumnuclear disk morphologies found, especially in the most detailed survey, NUGA, of a dozen of low luminosity AGN with the IRAM interferometer. The results allow a detailed study of each individual object (e.g.\,\,Garc\'ia-Burillo et al.\,\,2005, 2006c and references therein, Combes et al.\,\,2004, Krips et al.\,\,2005b). They propose an interpretation of the various observed dynamical states as
the different epochs of the evolution cycle driven by bars: from the formation of a bar through gravitational instability of a cold disk, to destruction of the bar by the gas flow driven by the bar, and replenishment of the gas disk through accretion, the various stages of secular evolution concur to fuel the AGN and assemble bulges at
the same time. However, the variety observed is rather challenging and urges to refine current dynamical models  (Garc\'ia-Burillo et al.\,\,2004, 2005, 2006c). 
		
	The field of observations of molecules other than CO in the central regions of AGN is active at various wavelengths. However, understanding the chemistry remains difficult, especially the direct effect of the AGN on the various layers through X-ray irradiation and heating. For instance, Evans et al.\,\,(2006) have carried out a systematic observation of HCN in QSOs to  study their dense molecular gas and the role of star formation in their host galaxies. The observed enhanced ratio of infrared to HCN luminosities compared to cool IRAS galaxies may appear to be an indication that the AGN contributes significantly to heating the dust, but other interpretations are possible. The influence of buried AGNs at the cores of ULIRGs have been invoked for explaining the observed intensities of HCO$^+$ or HNC (Imanishi et al.\ 2006, Aalto et al.\ 2006, Yamada et al.\ 2007), but the case remains unclear.
	 
	%The high angular resolution of adaptive optics already allows mapping near-infrared molecular lines in the AGN torus (Davies et al.\ 2006). 
	The strong infrared continuum sources of AGN could look ideal for tracing absorption features such as ice, CO or other molecules, in deeply obscured sources. However, ISO and {\it Spitzer} infrared spectroscopic surveys of AGN (e.g.\,\,Spoon et al.\ 2002, 2005, 2007, Lutz et al.\ 2004) have found very few sources with absorption features. See nevertheless the case of NGC\,6240 observed by Armus et al.\ (2006), and the strong molecular features in the infrared L-band (3-4\,$\mu$m) and M-band (4-5\,$\mu$m) observed by Sani et al.\ (2007) with VLT-ISAAC spectroscopy. Aromatic infrared bands of PAHs are generally weak in AGN (Section 8), but they are present in many of them (e.g.\ Schweitzer et al.\ 2006). 
	
	The sensitivity and angular resolution of ALMA will be essential to disentangle the complex structures and use the diagnostic of other molecules than CO and H$_2$ in the central regions of AGN (e.g.\,\,Baker 2005). 
	%It will also allow the extension to starburst galaxies where CO has been detected in central starbursts (Sections 4 \& 8).

%·	10.3 
\subsection{H$_2$O mega-masers and AGN molecular disks}

	Similarly to OH (Section 5), extragalactic H$_2$O masers in the 6$_{1,6}$-5$_{2,3}$ transition ($\nu _0$\,=\,22.235\,GHz) were first searched in nearby galaxies on model of Galactic H$_2$O masers, mostly in nuclear starbursts such as M\,82, NGC\,253, NGC\,2146 (see references in Lo 2005; see also Brunthaler et al.\ 2006 for the Local Group). They showed up there with a luminosity of up to a few L$_\odot$, comparable to the most luminous Galactic masers and are similarly likely mostly associated with star formation. However, it was soon discovered (Dos Santos \& L\'epine 1979, Gardner \& Whiteoak 1982, Claussen et al.\,\,1984, Claussen \& Lo 1986) a class of much more luminous H$_2$O masers, now called mega-masers with luminosity of up to 10$^4$\,L$_\odot$ (such maximum luminosities may be intermittent, and overestimated because of anisotropic emission). They were soon proven to be associated with active nuclei. Such powerful H$_2$O mega-masers (see again the detailed recent review by Lo 2005, to which we refer for more detailed discussion and references) are certainly the molecular sources the most specifically associated with AGN. While OH mega-masers typically arise in LIRG extreme starburst regions that are distributed over a 100-pc scale (Section 5), H$_2$O mega-masers are found mainly within the central parsec of an AGN (although a few less powerful ones are also found in starburst galaxies). Either in the circumnuclear accretion disk, or in the gas associated with jets or outflows, they are uniquely tracing the dense molecular gas directly exposed to the power of the AGN. Either direct irradiation by penetrating X-rays, or (jet-)induced shocks produce the high temperatures (several hundred Kelvin) and the large H$_2$O abundance required for mega-maser emission. However, they are relatively rare; about 80 are known out of more than 1000 galaxies (including 450 AGN) searched following Lo (2005) and Kondratko et al.\ (2006a,b). With luminosities up to 10$^4$\,L$_\odot$, they are found mainly in Seyferts\,2 or LINERS (`Low-ionization nebular emission regions'). Such objects are probably the best for providing the long amplification paths and the protection of molecules which are required for large amplification. Among the known H$_2$O mega-masers, 50\% arise from Compton-thick and 85\% from heavily obscured ($>$\,10$^{23}$\,cm$^{-2}$) active galactic nuclei (Zhang et al.\,\,2006). Note that the detection has been recently extended to a single case of a Type 2 QSO at z\,=\,0.66 (Barvainis and Antonucci 2005) with a luminosity of $\sim$23000\,L$_\odot$.
	
	In NGC\,4258, mapping the H$_2$O mega-maser emission has provided the first direct evidence in an AGN for the existence of a thin Keplerian accretion disk with turbulence, as well as highly compelling evidence for the existence of a massive black hole. Such a splendid case study (see references in Lo 2005) has shown the unique power of H$_2$O mega-masers for probing molecular circumnuclear disks, their dynamics and their black-hole masses. The very high surface brightness of such maser emission allows VLBI observations to achieve milliarcsecond resolution, which corresponds to subpc resolution of the emitting molecular medium within 1\,pc of the nucleus. The NGC\,4258 mega-maser has also provided a geometric distance determination of extremely high precision. The current case of the use of H$_2$O masers to determine the mass of super-massive black holes is discussed in detail in Ferrarese \& Ford (2005). It is clear that with the advent of powerful new facilities such as EVLA (see e.g. Menten 2007) and especially SKA, studies of mega-masers will be useful high-resolution probes of AGN and will provide accurate determinations of black-hole masses and cosmological high-z distance determinations.
	
	The detection at IRAM of a new H$_2$O megamaser in the 3$_{1,3}$-2$_{2,0}$ line ($\nu _0$\,=\,183.310\,GHz) was recently reported in the local ULIRG Arp\,220 by Cernicharo et al.\ (2006). This result opens up the possibility of using the 183\,GHz H$_2$O line as an additional tool to explore the physical conditions in LIRGs and ULIRGs, with a potential interest for high angular resolution observations with ALMA. 
	
	\section{Prospects}

	ALMA (the Atacama Large Millimeter Array) which will become operational by 2012, will dominate the prospects in the field of interstellar molecules in the next decades, by providing orders of magnitude gain in sensitivity through the whole millimetre and submillimetre domain accessible from the ground. In the meantime, the very next years will see significant upgrades of millimetre facilities and the full harvest of results of mid-infrared spectroscopy (H$_2$, PAHs, etc.) with current infrared space observatories; however, this immediate future will be mainly marked by breakthroughs in submillimetre astronomy, with the space observatory Herschel and the new generation of submillimetre cameras. In parallel to ALMA, other breakthroughs are expected in space mid-infrared spectroscopy with the James Webb Space Telescope (JWST), from the ground with the extremely large telescopes, and later in radio with the Square Kilometer Array (SKA). 

\subsection{Waiting for ALMA: ongoing studies and submillimetre breakthroughs}
%%10.1 Pre-ALMA era: completion of the exploration of the mid-IR and submm windows

	The infrared space observatory {\it Spitzer} is expected to remain in full operation until early 2009. Many of its results are still to come out or even to be observed. It will continue to provide exquisite imaging details on mid-infrared properties of local galaxies, calling for new molecular observations at other $\lambda$, especially of starburst regions. It will also discover a large number of new high redshift ULIRGs, and among them, new prominent, rare sources for follow-up millimetre molecular studies. However, as discussed in Sections 8 and 10, the most important results of {\it Spitzer} for extragalactic molecules are produced by its Infra-Red Spectrometer (IRS), mainly on rotational lines of H$_2$ and aromatic bands of PAHs. One may thus expect comprehensive studies on both topics. H$_2$ lines will provide detailed information on the `warm' (a few 10$^2$\,K) molecular gas in local galaxies. They will give information about its distribution in starburst regions, and cast new light on various kinds of shocks and postshock regions, including cooling flows and extra-galactic shocks (Section 10). The amazing sensitivity of the IRS for low resolution spectroscopy will be fully exploited for extensive studies of PAH features, both locally and at high redshift (Section 8). In nearby galaxies, sensitive comparative studies of small variations in PAH features will confirm, refine and exploit their diagnostic power in various conditions of physical parameters - especially UV radiation, interstellar chemistry and metallicity. The Magellanic Clouds and nearby starbursts will be particularly useful to complement Galactic PAH sources with widely different conditions. Similar comparative studies will address the molecular compounds in dust in the Magellanic Clouds and other nearby galaxies with various metallicities. The Japanese infrared space observatory, AKARI (ASTRO-F), will complement and extend {\it Spitzer} results (see e.g.\ Matsuhara et al.\,2006). In parallel, near-IR and redshifted UV molecular observations, mostly of H$_2$, will fully exploit the capabilities of the large park of 8-10\,m optical telescopes. The adaptive optics breakthrough in near-IR high angular resolution should mostly address AGN (see Sections 10  and 11). HST should benefit from the UV spectrometer COS after 2008, adding outstanding UV capabilities for studying molecules (H$_2$, PAHs, etc., Snow 2005) to those of NICMOS in the near-infrared.
	
	Waiting for ALMA and its first antennae by 2010, millimetre and radio observations of interstellar molecules, especially at high redshift, will continue with the current facilities with significant upgrades. Interferometers will mainly benefit from improvements of their receivers and correlators, with wider bandwidths and frequency coverage, and better sensitivities. The Californian interferometers, Owens Valley and BIMA will gain from their merging into CARMA (Scott \& Pound 2006). However, the IRAM interferometer 
	(Fig.\,7a) 
	has still the largest collecting area. It has just increased its receiver capabilities in a significant way\footnote{The new generation of receivers of the IRAM Plateau de Bure interferometer are more sensitive than the previous ones by a factor $\sim$\,5 in the continuum and 1.5-2.0 in the line detection. The available bandwidth is currently 2\,GHz and will be increased to 4\,GHz with a new generation of correlators.}. The extension of VLA capabilities (EVLA, http://www.aoc.nrao.edu/evla/, Butler 2004) will boost observations of low-J lines of high-z CO. Large single dishes will benefit from similar developments of their receiver and frequency coverage, especially the 100\,m Green Bank Telescope (GBT, http://www.gb.nrao.edu/gbt/, Mason 2004), and later the 50\,m LMT-GTM in construction in Mexico (http://www.lmtgtm.org/). One breakthrough soon to be expected is the direct blind determination of redshifts from CO lines of SMGs with spectrometers with very broad bandwidths (Baker et al.\ 2007).

	The observations of high redshift molecules will certainly increase using the expected breakthrough in the identification of large numbers of high-z ULIRGs with the new generation of submillimetre cameras. Arrays, such as SCUBA2 to be soon in operation (e.g.\ Doug 2004), will have  thousands of Transition Edge Superconducting (TES) detectors.
	%, with a SQUID multiplexer. 
	They will increase the submm mapping speed by a factor of $\sim$100. The 2-year survey programme of SCUBA2 aims to observe $\sim$20 square degrees, and to detect $\sim$\,10$^4$ high-z ULIRGs, to be compared with the few hundreds currently known. Such very large samples will change the prospects of millimetre studies of high-z molecules, focussing on prominent exceptional objects, especially lensed ones, allowing deeper chemistry probes, and various statistical, environmental and clustering studies. 
	
	The space submillimetre observatory, Herschel, to be launched in 2008, will extend such surveys to the whole submillimetre and far-infrared range, providing much more detailed information about  far-infrared luminosities, star formation rates and redshifts. More importantly for the exploration of the molecular world, Herschel will really open up the submillimetre window for sensitive  observations in its whole range with its heterodyne instrument HIFI (e.g.\ Lis 2004, Greve et al.\ 2006c). Compared to ALMA, Herschel will be completely unaffected by atmospheric effects, but both its sensitivity and angular resolution will be severely hampered by differences of three orders of magnitude in collecting area and baseline/diameter with the Herschel 3.5\,m mirror. Both handicaps will cumulate for the observation of extragalactic molecular lines. The results to be expected should thus be limited to strong lines of most abundant molecules, such as high-J lines of CO, HCN, HCO$^+$, CS, etc.\  (Greve et al. 2006c, see also Iono et al.\ 2007), lines of H$_2$O and isotopomers including HDO, and of a few other abundant hydrids such as OH, CH, CH$^+$, H$_2$D, NH$_3$, as well as related fine structure lines of \CI~and \CII, in prominent starburst and AGN sources. 
%TBC
The same applies to low resolution spectroscopy with the other instruments PACS and SPIRE of Herschel; and also to the airborne telescope SOFIA (http://sofia.arc.nasa.gov/, Erickson 2005), with its 2.5\,m mirror, whose various instrument capabilities in the mid- and far-IR may well address molecular lines in prominent extragalactic sources.

\subsection{The ALMA revolution}

\subsubsection{Overview of the ALMA project.}  

   The `Atacama Large Millimeter Array', ALMA, now extended into the `Enhanced Atacama Large Millimeter/Submillimeter Array' with Japan and other East Asian parties, is a world-wide project (http://www.eso.org/projects/alma/, http://www.alma.nrao.edu/, http://www.nro.nao.ac.jp/alma/E/ and Conicyt Chile). It will represent a jump of almost two orders of magnitude in sensitivity and angular resolution as compared with present millimetre/submillimetre interferometers, and will thus undoubtedly produce a major step in astrophysics. The main objectives will be the origins of galaxies, stars and planets. ALMA will be able to detect dust-enshrouded star-forming galaxies at redshifts $z$\,$\ge$\,10, both in the emission of  dust and spectral lines (CO and other species, including C$^+$). ALMA will also allow enormous  gains for the observation of the molecular gas in local and intermediate redshift galaxies of various types. 

   	The ALMA interferometer, to be completed by 2012, will be installed in an exceptional site for submillimetre observations, at Chajnantor, Atacama, Chile, at 5000\,m elevation, with baselines up to 14\,km (Fig.\,7b). Together with the ALMA Compact Array (ACA, driven by NAOJ Japan), it will include 54x12m--dishes and 12x7m ACA--dishes (total collecting area 6500\,m$^2$), providing a very good coverage in the {\it uv} interferometric plane, and allowing high sensitivity fast mapping with angular resolution better than 0.1'', and an ultimate angular resolution better than 0.01''. It will eventually operate in at least eight frequency bands, covering the main atmospheric windows between 4\,mm and 0.4\,mm (84 to 720\,GHz), and will be equipped with very broad-band heterodyne receivers close to  quantum-limit sensitivity, and a huge correlator (16\,GHz, 4096\,channels). 

% Fig7 Bure/ALMA
  \begin{figure*}
   \centering 
\caption{
({\it left}) The IRAM interferometer at Plateau de Bure in the French Alps, at 2500\,m elevation  -- 6\,x\,15m--dishes. ({\it right}) 
Artist view of the ALMA interferometer, to be completed by 2012, at Chajnantor, Atacama, Chile, at 5000\,m elevation -- 54\,x\,12m--dishes and 12\,x\,7m--dishes -- with baselines up to 14\,km.}
 \end{figure*}

	Such capabilities will deeply renew all the fields of millimetre astronomy, with a sensitive extension to the submillimetre range (see e.g.\ the ALMA Design Reference Science Plan (DRSP), http://www.strw.leidenuniv.nl/$\sim$alma/drsp11.shtml, and its future updates). As concerns {\it local galaxies}, one may say that many of the goals of current millimetre astronomy in distant sources of the Milky Way, such as the central regions, will become accessible in nearby galaxies. Similarly, detailed studies currently carried out in nearby galaxies will be easy in rare, relatively distant local galaxies such as ULIRGs. It is thus easy to figure out the enormous impact of ALMA for all kinds of detailed studies of the molecular gas in local galaxies, and more generally our global knowledge of the interstellar molecular medium, without being restricted to a single galaxy. Routine high sensitivity imaging of nearby galaxies with a resolution  approaching 1\,pc will well resolve the giant molecular clouds, revealing many details and the dynamics of various kinds of structures, such as condensations, hot spots, outflows, photodissociation regions, shocks, turbulence, etc. The various aspects of star formation in various types of galaxies can then be addressed in much more detail than currently. Comparisons of molecular abundances and interstellar chemistry, both at galactic scales and for specific regions, will be extended to a variety of galaxy types and element abundances. Dedicated observations of isotopomers will address isotope ratios and nucleosynthesis at galactic scales in this rich diversity of environments. A particular interest will focus on the specificity of the chemistry in AGN and under strong X-ray irradiation.

\subsubsection{ALMA capabilities at high redshift}  

 It is well known that the past history of star formation in the Universe, in dusty and molecular starbursts, is one of the two main driver goals for ALMA. The expected leap forward may be summarized as a sensitivity for dust detection at 850\,$\mu$m $\sim$\,50 times better than SCUBA and more than 10 times SCUBA2, no limitation by source confusion, and a sensitivity for lines an order of magnitude better than the upgraded IRAM interferometer. ALMA will be able to detect dust in luminous infrared galaxies (LIRGs with L$_{{\rm FIR}}$ of a few 10$^{11}$\,L$_\odot$, SFR of a few 10\,M$_\odot$/yr) and map CO lines in ULIRGs, at {\it any redshift} where they might exist; i.e. {\it the first dusty starbursts in the Universe}. One may expect that the major breakthroughs of ALMA in this field will include (Blain 1999 \& 2006, Blain et al.\ 2002, Combes 2005, Omont 2004, Wootten 2001, 2004, Takeuchi et al.\ 2001, Walter \& Carilli 2007, Carilli et al.\ 2007, Combes, 2007, papers by Bertoldi, Walter, etc. in Bachiller et al. 2007):
 
 $\bullet$ Comprehensive studies of high-z dusty starbursts in a few {\it ALMA ultra-deep fields}, with detections up to hundreds of them per field of view, determination of their redshift and luminosity distributions by multi-$\lambda$ mm/submm observations, and CO detection in a number of them\footnote{However, because of its small field of view ($\sim$25'' at 1\,mm), ALMA will have a limited speed for wide surveys, an order of magnitude less than SCUBA2, and a larger factor compared with future projects with larger telescope and TES camera such as CCAT  (e.g.\ Blain 2006). But the sensitivity of such instruments will be severely limited by source confusion contrary to ALMA.}. A clever use of 'gravitational telescopes` by clusters will be needed to carry out comprehensive studies of high-z starbursts down to $\sim$10\,M$_\odot$/yr and detect weaker ones; and in particular to address the earliest starbursts in the Universe

 $\bullet$ Complementary observations of deep fields observed by other instruments or with multi-$\lambda$ coverage. In particular, {\it combined projects by ALMA and JWST} should be unique for unveiling the first major starbursts in the Universe, clumping of starburst galaxies in proto-clusters, deepest studies of highly lensed fields, observations of highest-z Gamma-Ray Bursts, their host galaxy and their environments, etc.
 
 $\bullet$ {\it Sensitive mapping of star-forming galaxies at all redshifts}, both in the dust continuum and mainly in CO, C$^+$ and \CI~lines. The combination of sensitivity, high angular resolution and heterodyne velocity profiles will provide a rich information about detailed structure, star formation, mass of molecular gas, ionization, and their spatial distribution; dynamics, rotation, dynamical masses; mergers and companions; outflows; etc. One will derive a complete picture of the properties and evolution of starburst galaxies through the whole history of the Universe. One may also say that we will in particular build a complete coverage of the history of all major phases of star formation in Milky Way-like galaxies in the last ten billion years or so. Many dedicated observations of this type will address single or multiple objects discovered at other wavelengths, peculiar, rare, prominent, gravitationaly lensed, hosts of various types of AGN, etc.  
 
 $\bullet$ Very fast complete frequency coverage of an atmospheric frequency window, allowing {\it blind redshift determination from CO lines}, as well as multi-line detections in strong sources. One may thus expect a full development of {\it interstellar chemistry}, including isotopomers, in prominent sources at intermediate redshifts, and at all redshifts through {\it absorption line} studies with numerous continuum background sources of modest strength.

%\bigskip

\subsection{Accompanying- and post-ALMA: JWST, extremely large telescopes and SKA}

It is certainly hazardous to try to make predictions 15-20\,year from now in a fast moving field which is only 35\,year old. We will thus just concentrate on the three major fields where projects, as costly as or even more costly than ALMA, are decided or in a pre-decision phase in domains of interest for extragalactic molecules: JWST in near/mid-IR, the ground-based 30-40\,m class telescopes in optical/near-IR, and SKA in radio. However, for all of them, despite their importance for molecules, their main drivers are not centred on molecules as for ALMA.

This leaves aside other important fields, such as UV spectroscopy where no major project following FUSE and HST/COS, is yet decided. It is of course impossible to anticipate about the follow-up of more or less serendipitous discoveries that one may expect in the next five years, and then especially with ALMA and JWST. It is even hard to predict the evolution of eventual molecular studies in fields yet unexplored, such as the reionization epoch and the formation of the very first galaxies, as well as of long pending problems such as the diffuse insterstellar bands.

\subsubsection{JWST and ground-based extremely large telescopes}
	  
	  Current instrumentation technology  may already allow a gain of several orders of magnitude with respect to present facilities for spectroscopy in the whole infrared range, with tremendous potential impact on molecular studies,  especially at high redshift. This will be achieved in the near- and mid-infrared up to 27\,$\mu$m, by the 6m James Webb Space Telescope (JWST, Gardner et al.\ 2006). One may thus expect much deeper studies, with better angular resolution, than current ones for vibration lines of hot molecular gas, mostly H$_2$, in various types of local galaxies, mainly starbursts, AGN, cooling flows, etc. For this purpose, JWST will be complemented by the new generation 30-40\,m ground telescopes\footnote{Such projects presently include: the European Extremely Large Telescope (E-ELT, https://www.eso.org/projects/e-elt/); the Giant Magellan Telescope (GMT, http://www.gmto.org/) and the Thirty Meter Telescope (TMT, http://www.tmt.org/). They will allow deeper, higher-resolution spectroscopy in the atmospheric windows with the sensitivity needed for extragalactic observations.}, especially for high angular resolution near-IR spectroscopy of molecules, mainly H$_2$, in various extragalactic shocks and in the galactic nuclei of starburst galaxies and AGN.
	  
	  The Mid-InfraRed Instrument (MIRI) of JWST will have orders of magnitude improvements in sensitivity, spatial and/or spectral resolution compared with other facilities and will be a unique facility for astrochemistry in the next decade. MIRI will expand tremendously the capabilities of other facilities, with e.g.\ two orders of magnitude more sensitive, one order of magnitude larger spatial resolution and a factor of 4-6 higher spectral resolution than {\it Spitzer} (van Dishoeck et al.\ 2005). One may thus expect much more comprehensive studies in local galaxies of warm H$_2$ rotational lines, PAH emission and molecular features in dust. Even more important will be the extensions at high redshift allowing the comparison of the evolution in the history of galaxies of these essential molecular features of the interstellar medium. H$_2$ lines in particular could trace any kind of warm molecular gas in high-z starbursts, AGN, mergers, outflows, and various shocks.
	  
	  The 30-40\,m telescopes will be unique for high-sensitivity spectroscopy of redshifted UV molecular lines, especially H$_2$ in Damped Lyman-$\alpha$ systems (DLAs) on the line of sight of QSOs and GRBs at very high z. This will be a unique tool for studying H$_2$ at high z and the conditions in DLAs, as well as the possible variation of m$_e$/m$_p$, with extensions to other molecules such as CO, and studying H$_2$ in GRBs and their host galaxies (Theuns \& Srianand 2006, Campana et al.\ 2007).
	  
    There are also projects, some of them less advanced, to take advantage of the extraordinary possible sensitivity for far-infrared spectroscopy in space. A first step could be the Japan-led SPICA project with a 3m-class telescope (Matsumoto 2005). It could be followed by a larger far-IR telescope, such as the SAFIR project (Benford et al.\ 2004), and eventually by a far-IR space interferometer, but it is difficult to anticipate what could be the exact time-scale for such projects.

\subsubsection{SKA, mega-masers and cold molecular gas at very large redshift}

 The 'Square Kilometer Array' (SKA) will provide two orders of magnitude increase in collecting area over existing telescopes in the cm range, allowing for study of the \HI~content of galaxies to cosmologically significant distances (i.e. to z\,$\ga$\,2 rather than z\,$\sim$\,0.2). Radio studies of molecular lines, especially at high z and mega-masers, will benefit from the same sensitivity gain which is larger than brought by ALMA for millimetre astronomy. Among the five key science projects of SKA (see e.g.\ Carilli \& Rawlings 2004), observations of high-z molecules are quoted in `Galaxy evolution and cosmology' for the precise measurement of H$_0$ using extragalactic water masers, and in `Probing the dark ages' for the incomparable sensitivity of the SKA enabling studies of the molecular gas, dust, and star formation activity in the first galaxies. As discussed e.g.\ by Carilli \& Blain (2002) and Blain, Carilli \& Darling (2004), the detection of redshifted low-J lines of CO will be carried out by the SKA at z\,$\ga$\,2 at a rate at least comparable to that of ALMA for higher-J lines. SKA will be complementary to ALMA for studying the cold gas and resolving its morphology, and also for detecting OH and H$_2$O mega-masers and redshifted low-J lines of molecules such as HCN, HCO$^+$, CS, CN, etc., whose high-J lines are not easily excited.

\section{General conclusion}

%It is not an easy task to summarize the most salient conclusions of this review dealing with a subject as  vast and rich as molecules in galaxies.

Molecules may be seen as playing two essential roles in the Universe: as the cosmic material which forms stars; and as a first step in building complexity which culminates in life. For the second aspect of astrobiology, it is obvious that, because of the enormous distances, galaxies are not the best hunting ground to try to understand the numerous and complex steps which could lead from the simple interstellar molecules to the genetic code. On the other hand, the panoramic views provided by molecular signals from galaxies are one of the best sources of information to grasp the essential features of {\it star formation} at galactic scales. CO is the best tracer of normal, cold molecular gas. In various types of local galaxies, its millimetre lines reveal the standard giant molecular clouds, home of `normal' massive star formation such as in most of the Milky Way. CO lines also deeply probe the most extreme, massive, dusty starbursts, especially at high redshift, without being blocked and much distorted by extinction. CO and better HCN line intensities yield good estimates of the global star formation rates. The high resolution velocity profiles of CO lines, routinely offered by heterodyne techniques, provide molecular and dynamical masses of starburst galaxies, key parameters for learning about their past and future star formation history. The great news is that such a comprehensive information about star formation is available not only for our local neighbours, but more and more up to the most distant galaxies that we see in their youth and infancy. This will be soon extended by ALMA at z\,$\ga$\,10, practically up to the frontier of the `dark ages' when the first stars and galaxies formed and the Universe was only a few percents of its present age. We are indeed living the dream of the first explorers of stellar populations in the Milky Way, by directly catching the action of successive starbursts which formed most of the stars in Milky Way sisters at various epochs of their past life up to z\,$\ga$\,2. Even farther out in space and time, through molecular lines, we watch and get insight in the most powerful starbursts in the Universe, progenitors of the most massive (elliptical) galaxies today. Disentangling their history is a basic clue for understanding the joint evolution of the most fascinating objects in the Universe: their massive dark matter halos, their stellar spheroids, their central super-massive black-holes, and the massive clusters of galaxies surrounding today the most massive ones. However, the power of molecular lines to trace star formation in galaxies is far from being limited to such climaxes, but it is almost universal even for much milder episodes in various kinds of galaxies. Even much less spectacular, some of these episodes may be of great importance for their hosts, either dwarf such as the Magellanic Clouds, or giant such as elliptical galaxies harbouring cooling flows. While CO traces the cold bulk of molecular gas, observations of rotational lines of H$_2$ itself are spectacularly progressing with the development of space mid-infrared spectroscopy. They trace patches of warm gas in particularly active regions of starbursts, mergers, vicinities of AGN, and various types of shocks even at extra-galactic scales. CO and H$_2$ lines should thus be major tracers of violent events in galaxy lifes, such as mergers, accretion shocks, outflows and other effects of feedback action from supernovae and AGN.

Star formation is not by far the only fascinating question of contemporary astronomy that the young extragalactic molecular astrophysics may address. Large aromatic molecules (PAHs) are recognized as an ubiquitous major component and a reservoir of carbon of the interstellar medium of most types of galaxies. They are now currently studied up to redshift at least $\sim$2 in the most massive galaxies; but we have still to learn about their behaviour in the first billion years of galaxy lives. Similarly, little is known about molecular mantles in strong starbursts and at high redshift. More generally, all processes of dust surface chemistry, including H$_2$ formation, remain challenging. The extragalactic chemistry of the deuterium is still in its infancy, together with the information it can bring about the dust-gas cycle in various galactic environments. If the current results about extragalactic interstellar chemistry appear a bit disapointing as too similar to that of the Milky Way or less rich, it could be due to our lack of sensitivity, and a reassessment with ALMA is obviously necessary. In any case, the progress expected in isotopic ratio determination will be important for a better appreciation of nucleosynthesis and the chemical evolution of various types of galaxies. Tracing molecules in AGN, in the vicinity of the nucleus, up to the molecular torus, is one of the fields which will most benefit from the gains in sensitivity and angular resolution from space or with millimetre and radio interferometry and adaptive optics. This will provide unique opportunities to probe X-ray dominated chemistry and the special properties of the molecular gas in the `molecular torus'. VLBI detections of H$_2$O mega-masers already probe the even inner region of the accretion disk, offering a unique way of accurate determination of the mass of the super-massive black hole. Their expected extension at high redshift is very promising for a direct accurate calibration of the cosmic distance scale, and in particular the precise measurement of H$_0$. More generally extragalactic OH and H$_2$O mega-masers provide
unique examples of radiation amplification with the most extreme cosmic scales and energies. Fully appreciating the properties of mega-masers remains challenging, as well as extending their systematic detection in the most powerful starbursts at high redshift. Molecules, such as H$_2$ and OH, also participate in high precision spectroscopy of absorption lines at high redshift, allowing one to probe possible variations of fundamental constants such as m$_e$/m$_p$ and $\alpha$. The multiplication of high-z background sources, especially gamma-ray bursts, will provide new opportunities, as well as probing molecules at the interface between (proto-) galaxies and the intergalactic medium.

  It is clear that the vast and rich field of molecules in galaxies will see many exciting developments in the next years and importantly contribute to the progress of the exploration of the world of galaxies and their evolution.

\section*{Acknowledgements}
I owe a great debt of gratitude to Al Glassgold and Pierre Cox for their past and present close  collaboration, their careful reading of the whole manuscript and numerous helpful comments and suggestions. I specially thank James Lequeux and Fran{\c c}oise Combes, as well as two anonymous referees, for numerous helpful comments and suggestions. I also want to thank Fran{\c c}ois Boulanger, Asunci\`on Fuente, George Helou, Mich\`ele Leduc, Gary Mamon, David Merritt, Padelis Papadopoulos, Patrick Petitjean, H\'el\`ene Roussel and Phil Solomon for various help and comments.

\section*{References}

\footnotesize
%\scriptsize
{\bf Note.} {\it The references are grouped by sections. However, any reference appears only once, generally in the first section where it is quoted.} {\bf The most important general references and reviews are grouped with Section 1.}

\clearpage

{\scriptsize

\section*{List of Main Abbreviations}

~~~~~~~~\,ACA Atacama Compact Array (NAOJ/ALMA)

AGB Asymptotic Giant Branch (late stage of red giant stars)

AGN Active Galactic Nucleus

AIB Aromatic Infrared Bands

AKARI (Previously known as ASTRO-F, Institute of Space and Astronautical Science [ISAS], Japan)

ALMA Atacama Large Millimeter/submillimeter Array 

APEX Atacama Pathfinder Experiment, is a collaboration between Max Planck Institut f\"ur Radioastronomie (MPIfR), Onsala Space Observatory (OSO), and ESO 

AST/RO Antarctic Submillimeter Telescope and Remote Observatory 
%(Center for Astrophysical Research in Antarctica, CARA)

ATCA Australia Telescope Compact Array 

BIMA Berkeley Illinois Maryland Array 

CARMA Combined Array for Research in Millimeter-wave Astronomy (Caltech, the BIMA partners [Maryland, Illinois, and Berkeley], and the National Science Foundation)

CCAT Cornell Caltech Atacama Telescope (25m-class sub-millimetre radiotelescope project)

CMB Cosmic Microwave Background

CSO Caltech Submillimeter Observatory (operated by Caltech under a contract from the NSF) 

DIB Diffuse Interstellar Band

DLA Damped Lyman-$\alpha$ absorption system 

ESO European Southern Observatory

FUSE Far Ultraviolet Spectroscopic Explorer (NASA)

EVLA Expanded VLA (NRAO)

FCRAO Five College Radio Astronomy Observatory (University of Massachussets and NSF)

FIR Far InfraRed

%FWHM Full Width at Half Maximum
%
GBT Green Bank Telescope (NRAO)

GMC Giant Molecular Cloud 

HST Hubble Space Telescope (NASA)

IRAM Institute for RadioAstronomy in the Mm-range (CNRS Centre National de la Recherche Scientifique France; MPG Max Planck Gesellschaft Germany; IGN Instituto Geográfico Nacional, Spain)

IRS InfraRed Spectrograph ({\it Spitzer}, NASA)

ISO Infrared Space Observatory (European Space Agency [ESA])

ISOCAM The ISO Camera

ISM InterStellar Medium

JCMT James Clark Maxwell Telescope [PPARC (UK), NRC (Canada) and NWO (Netherlands)]

JWST John Webb Space Telescope (NASA)

LIRG Luminous InfraRed Galaxy

LMC Large Magellanic Cloud

LMT-GTM Large Millimeter Telescope/Gran Telescopio Milimetrico (The Instituto Nacional de Astrofisica, Optica y Electronica in Puebla, Mexico, and University of Massachusetts)

LTE Local Thermodynamical Equilibrium

LVG Large Velocity Gradient

MIR Mid-InfraRed

MW Milky Way

NANTEN Southern 4-meter radiotelescope (Nagoya University, Carnegie Inst.\ of Washington) 

NAOJ National Astronomical Observatory of Japan

NRAO National Radio Astronomy Observatory (NSF facility  operated by Associated Universities, Inc.)

NUGA NUclei of GAlaxies: the IRAM Survey of Low Luminosity AGN

OVRO Owens Valley Radio Observatory (Caltech)

PAH Polycyclic Aromatic Hydrocarbon

PDR PhotoDissociation Region

QSO Quasi-Stellar Object (quasar)

SAFIR Single Aperture Far-Infrared observatory (project)

SCUBA (SCUBA2) The Submillimetre Common-User Bolometer Array at JCMT

SED Spectral Energy Distribution

SEST Swedish--ESO Submillimetre Telescope

SFR Star Formation Rate

SKA Square Kilometer Array

SMA Submillimeter Array (the Smithsonian Astrophysical Observatory [SAO] and the Academia Sinica Institute of Astronomy and Astrophysics [ASIAA])

SMC Small Magellanic Cloud

SMG SubMillimetre Galaxy

SOFIA Stratospheric Observatory for Infrared Astronomy (NASA and the DLR, German Aerospace Center)

SPICA Space Infrared Telescope for Cosmology and Astrophysics (Inst. of Space and Astronautical Science, Japan)

%SQUID Supraconductor Quantum Interference Device

SWS Short Wavelength Spectrometer (ISO, ESA)

ULIRG Ultra-Luminous InfraRed Galaxy

VLA Very Large Array (NRAO)

VLBI Very Long Baseline Interferometry

VLT Very Large Telescope (ESO)

}

\end{document}